\documentclass[11pt,twoside,a4paper,cmspaper,final,collab]{cms-tdr}
\def\svnVersion{d7d7b22-D}\def\svnDate{2026/05/18}\def\cmsCernNoTag{CERN-EP-2026-133}\def\cmsCernDate{\today}\def\cmsMessage{Submitted to the Journal of High Energy Physics}
\begin{document}\cmsNoteHeader{NPS-25-003}

\providecommand{\PGgst}{\HepParticle{\PGg}{}{\ast}\xspace}
\newcommand{\ee}{{\ensuremath{\Pe\Pe}}\xspace}
\newcommand{\mumu}{{\ensuremath{\PGm\PGm}}\xspace}
\newcommand{\emu}{{\ensuremath{\Pe\PGm}}\xspace}
\newcommand{\etau}{{\ensuremath{\Pe\tauh}}\xspace}
\newcommand{\mutau}{{\ensuremath{\PGm\tauh}}\xspace}
\newcommand{\wjets}{\ensuremath{\PW+\text{jets}}\xspace}
\newcommand{\il}{138\fbinv}
\newcommand{\Paa}{\ensuremath{\PGf_{1}}\xspace}
\newcommand{\Pab}{\ensuremath{\PGf_{2}}\xspace}
\newcommand{\maa}{\ensuremath{m_{\Paa}}\xspace}
\newcommand{\mab}{\ensuremath{m_{\Pab}}\xspace}
\newcommand{\mtt}{\ensuremath{m_{\PGt\PGt}}\xspace}
\newcommand{\mbb}{\ensuremath{m_{\PQb\PQb}}\xspace}
\newcommand{\mellell}{\ensuremath{m_{\Pell\Pell}}\xspace}
\newcommand{\dzeta}{\ensuremath{D_{\zeta}}\xspace}
\newcommand{\ttjets}{\ensuremath{\PQt\PAQt+\text{jets}}\xspace}
\newlength\cmsTabSkip\setlength{\cmsTabSkip}{1ex}

\cmsNoteHeader{NPS-25-003}
\title{Search for Higgs boson decays into two neutral scalars with unequal masses in final states with \texorpdfstring{\PQb}{b} quarks and tau leptons in proton-proton collisions at \texorpdfstring{$\sqrt{s}=13\TeV$}{sqrt(s)=13 TeV}}

\date{\today}

\abstract{
	A search for Higgs boson (\PH) decays into a pair of neutral scalars \Paa and \Pab, with \Pab heavier than \Paa, is performed in final states with \PQb quarks and tau leptons. Depending on the masses of the neutral scalars, \Pab can undergo a cascade decay into \Paa{}\Paa. For both the cascade and non-cascade scenarios, one \Paa is required to decay to a pair of tau leptons. Proton-proton collision data corresponding to an integrated luminosity of \il collected with the CMS detector at the LHC at $\sqrt{s}=13\TeV$ are analyzed. No statistically significant excess over the standard model expectation is observed. Upper limits are set on the products $\sigma \mathcal{B}(\PH \to \Paa \Pab \to 3\Paa \to 2\PGt 4\PQb)$ and $\sigma \mathcal{B}(\PH \to \Paa \Pab)\ \mathcal{B}(\Paa \to 2\PGt)\ \mathcal{B}(\Pab \to 2\PQb)$ where $\sigma$ is the Higgs boson production cross section. The observed upper limits range between 0.9 and 36.8\unit{pb} at 95\% confidence level, depending on the mass hypothesis and decay scenario.
}

\hypersetup{
pdfauthor={CMS Collaboration},
pdftitle={Search for Higgs boson decays into two neutral scalars with unequal masses in final states with b quarks and tau leptons in proton-proton collisions at sqrt(s) = 13 TeV},
pdfsubject={CMS},
pdfkeywords={CMS, Higgs, exotics}}

\maketitle 
\section{Introduction}
In the standard model (SM), the Brout--Englert--Higgs mechanism~\cite{PhysRevLett.19.1264,Salam:1968rm} facilitates electroweak (EW) symmetry breaking and predicts the existence of a scalar particle---the Higgs boson (\PH)~\cite{PhysRevLett.13.321, HIGGS1964132, PhysRevLett.13.508, PhysRevLett.13.585, PhysRev.145.1156, PhysRev.155.1554}.
In 2012, the ATLAS and CMS Collaborations discovered a scalar particle with a mass of 125\GeV, exhibiting properties consistent with those predicted for the SM Higgs boson~\cite{Aad_2012, Chatrchyan_2012, chatrchyan_observation_2013}.
However, many theories beyond the SM can accommodate a scalar particle of this mass with properties consistent with current experimental constraints. These models address known shortcomings of the SM.
In particular, extended scalar sector models can explain the existence of dark matter and the observed matter-antimatter asymmetry in the universe~\cite{Grzadkowski_2010, Drozd_2014}.
Such models often predict that the Higgs boson decays into additional light scalar particles, which decay into SM fermions.
The latest constraints on the Higgs boson decay into undetected particles by the CMS and ATLAS Collaborations are currently rather loose: ${<}16\%$~\cite{CMS:2022dwd}, and ${<}12\%$~\cite{ATLAS:2022vkf} respectively, at 95\% confidence level (\CL), motivating searches for exotic decays of the Higgs boson.

Searches for exotic decays of the SM Higgs boson have focused on its decay to light pseudoscalar particles, \Pa, using $\PH \to \Pa \Pa$ and $\PH \to \PZ \Pa$ processes.
A wide range of searches for these decay modes to various final states involving SM fermions have been performed by the ATLAS and CMS Collaborations~\cite{ATLAS:2015hpr, CMS:2012qms, CMS:2015nay, CMS:2017dmg, CMS:2015twz, ATLAS:2015rsn, ATLAS:2018coo, CMS:2018jid, ATLAS:2015unc, CMS:2018qvj, CMS:2020ffa, ATLAS:2018emt, CMS:2024uru, ATLAS:2018jnf, CMS:2019spf, CMS:2024zfv, CMS:2022xxa, CMS:2022fyt, CMS:2025hjt, CMS:2024jyb, ATLAS:2025qyn}.
Each search targeting a final state with SM fermions provides a model-independent upper limit on the associated Higgs boson branching fraction ($\mathcal{B}$).
In addition, some results have been interpreted in two-Higgs-doublet models with an additional scalar singlet (2HDM+S)~\cite{Branco_2012}, next-to-minimal supersymmetric SM (NMSSM)~\cite{Fayet1975104, Kaul198236, Barbieri1982343, Nilles1983346, Frere198311, Derendinger1984307, Drees:1988fc, Maniatis:2009re, Ellwanger:2009dp}, models with axion-like-particles~\cite{Kim:1979if, AxionsDM}, and dark supersymmetry~\cite{ArkaniHamed:2008qn, Baumgart:2009tn, Falkowski:2010cm}.
These studies have not revealed any significant deviation from the SM.

In addition to the $\PH \to \Pa \Pa$ decay, exotic Higgs boson decays to light neutral scalars with unequal mass, $\PH \to \Paa \Pab$, may also occur in some extended scalar sector models.
For example, the two-real-singlet model (TRSM)~\cite{Robens_2020}, where the SM is extended by two real scalar singlets, predicts three neutral scalar bosons and proposes several benchmark scenarios based on the scalar masses~\cite{Robens:2023oyz}.
Previous searches for heavy resonances decaying to the SM Higgs boson have been interpreted in terms of these benchmark scenarios by the CMS Collaboration~\cite{CMS:2022suh, CMS:2023boe, CMS:2024phk}.
In this paper, we explore the possibility that the heaviest of the three scalar particles is the observed Higgs boson, which decays to a pair of lighter scalars with non-degenerate masses via $\PH \to \Paa \Pab$, where $\mab > \maa$.
The branching fraction $\mathcal{B}(\PH \to \Paa \Pab)$ is a function of the scalar masses $\maa$ and $\mab$ and, in the TRSM, reaches up to 7--8\% in the intermediate mass range $\mab \approx$ 60--80\GeV~\cite{Robens_2020}.

If the decay channel $\Pab \to \Paa \Paa$ is kinematically open ($\mab \geq 2\maa$), this cascade decay is the dominant decay mode [$\mathcal{B}(\Pab \to \Paa \Paa)\approx 100\%$] with a very sharp transition to the non-cascade decays at the $\mab = 2\maa$ threshold.
For the non-cascade scenario, \ie, $\mab \leq 2\maa$, both scalars \Paa and \Pab decay directly to SM particles, with branching fractions identical to that of an SM-like Higgs boson with the corresponding mass.
The ATLAS experiment has used the \PZ boson-associated Higgs boson production mode to probe the scenario in which all scalar particles decay hadronically, the final states having four or six b quarks depending on whether the decay includes a cascade~\cite{ATLAS:2025rfm}.
Upper limits on the branching fractions at 95\% \CL are in the ranges 10--20\% and 24--38\% for the cascade and non-cascade decay scenarios, respectively.

At lighter scalar masses, final states with tau leptons become more relevant.
The study of final states with \PQb quarks and tau leptons benefits from relatively large branching fractions.
Leptonic triggers also provide higher signal purity in event selection, whereas the relatively higher \pt thresholds for hadronic tau lepton decays, and lower identification efficiencies of the jets from the fragmentation of b quarks ($\PQb$ jets) and tau leptons, pose a challenge.

\begin{figure}[ht!]
    \centering
        \includegraphics[width=0.42\textwidth]{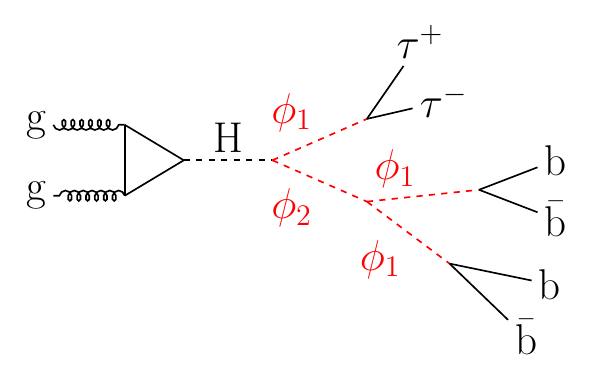}
        \includegraphics[width=0.42\textwidth]{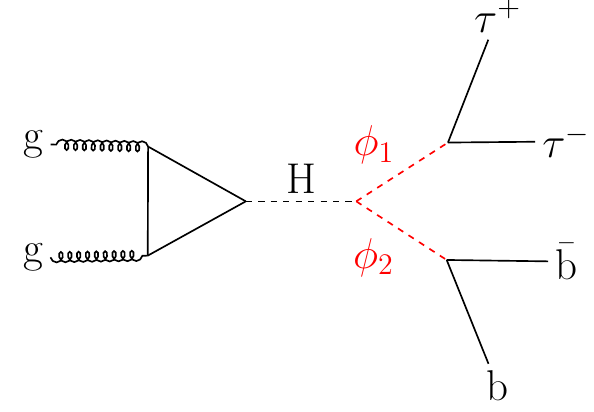} \\
        \includegraphics[width=0.42\textwidth]{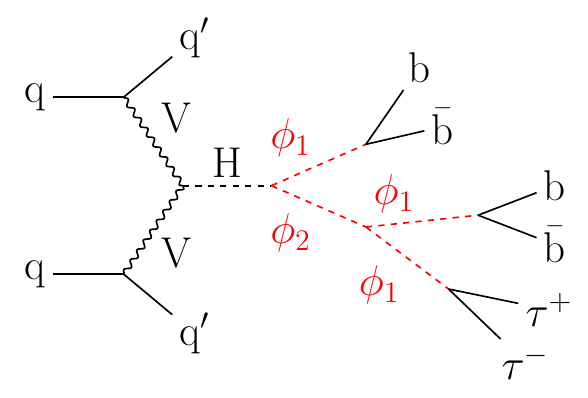}
        \includegraphics[width=0.42\textwidth]{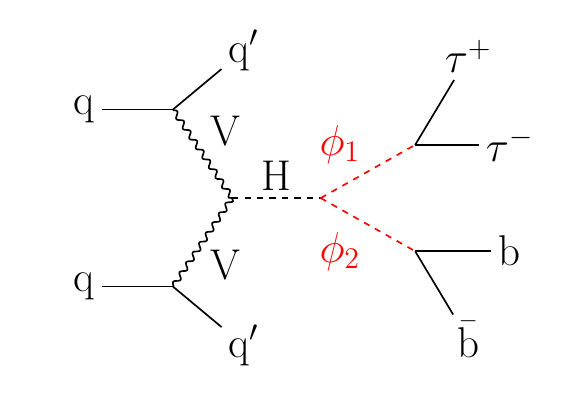}
    
    \caption{Representative schematic diagrams showing the ${\Pg\Pg}$F (upper) and VBF (lower) production of the Higgs boson decaying to $2\PGt4\PQb$ (cascade, left) and $2\PGt2\PQb$ (non-cascade, right) final states.}
    \label{fig:diagrams}
\end{figure}

This is the first search using CMS data for the asymmetric decays $\PH \to \Paa \Pab$, previous searches having focused on symmetric decays.
The gluon-gluon fusion (${\Pg\Pg}$F) and vector-boson fusion (VBF) production of the Higgs boson are targeted.
The search is based on \il of proton-proton collision data collected during 2016--2018 at a center-of-mass energy of 13\TeV.
The decay chains considered are illustrated in Fig.~\ref{fig:diagrams}.

One of the \Paa scalars must decay to a pair of \PGt leptons, used for the ``online'' selection of events at the trigger level, and in the case of cascade decays, the other \Paa scalars must each decay to a pair of \PQb quarks. For non-cascade decays, the \Pab scalar must decay into a pair of \PQb quarks.
Depending on the tau lepton decay, three final states with electrons, muons, and hadronically decaying tau leptons (\tauh) are considered: \mutau, \etau, and \emu.
Light dilepton (\ee and \mumu) final states suffer from a large, irreducible Drell--Yan background, \ensuremath{{\PZ}/\PGgst+\text{jets}} production (DY), whereas the final state with two \tauh has low signal sensitivity as a result of high trigger \pt thresholds.
Further requirements on the presence of $\PQb$ jets are applied for preselecting signal-like events.
Finally, a signal categorization based on a boosted decision tree (BDT) is used, which provides powerful background reduction for all \Paa and \Pab mass hypotheses considered.
An alternative method based on applying selection criteria (cuts) on variables to define signal-enriched regions, referred to as the ``cut-based'' method, is explored as a cross-check, and the results are included in Appendix~\ref{appendix}. 

The paper is organized as follows. 
Sections 2 and 3 provide a description of the CMS detector and the simulated data samples.
Final-state object selections and event categorization procedures are presented in Sections 4 and 5, respectively.
In Section 6, the background estimation methods are described.
Section 7 presents the sources of systematic uncertainties in the search, while results are detailed in Section 8.
Finally, a summary is provided in Section 9.
Tabulated results are provided in the HEPData record for this analysis~\cite{hepdata}.

\section{The CMS experiment}
The CMS apparatus~\cite{CMS:2008xjf,CMS:2023gfb} is a multipurpose, nearly hermetic detector, designed to trigger on~\cite{CMS:2020cmk,CMS:2016ngn,CMS:2024aqx} and identify electrons, muons, photons, and (charged and neutral) hadrons~\cite{CMS:2020uim,CMS:2018rym,CMS:2014pgm}. Its central feature is a superconducting solenoid of 6\unit{m} internal diameter, providing a magnetic field of 3.8\unit{T}. Within the solenoid volume are a silicon pixel and strip tracker, a lead tungstate crystal electromagnetic calorimeter (ECAL), and a brass and scintillator hadron calorimeter (HCAL), each composed of a barrel and two endcap sections. Forward calorimeters extend the pseudorapidity coverage provided by the barrel and endcap detectors. Muons are reconstructed using gas-ionization detectors interleaved with the layers of the steel flux-return yoke outside the solenoid. 

Events of interest are selected using a two-tiered trigger system. The first level, composed of custom hardware processors, uses information from the calorimeters and muon detectors to select events at a rate of around 100\unit{kHz} within a fixed latency of 4\mus~\cite{CMS:2020cmk}. The second level, known as the high-level trigger, consists of a farm of processors running a version of the full event reconstruction software optimized for fast processing, and reduces the event rate to a few kHz before data storage~\cite{CMS:2016ngn,CMS:2024aqx}. More detailed descriptions of the CMS detector, together with a definition of the coordinate system used and the relevant kinematic variables, can be found in Refs.~\cite{CMS:2008xjf,CMS:2023gfb}.

\section{Simulated samples}
A number of Monte Carlo (MC) event generators are used to produce events using either leading-order (LO) or next-to-LO (NLO) matrix element calculations.
In all cases, parton showering and fragmentation are implemented using \PYTHIA (version 8.212) with the CP5 event tune~\cite{Sjostrand:2014zea}. The CMS detector is simulated using the \GEANTfour package~\cite{Agostinelli:2002hh}. To model the effect of additional proton-proton collisions within the same or adjacent bunch crossings (pileup), minimum-bias interactions are simulated and superimposed on the hard-scattering events. Simulated events are then reweighted to reproduce the pileup distribution in data. 

{\tolerance=1600
The signal samples are generated in the ${\Pg\Pg}$F and VBF production modes using \MGvATNLO (version 2.6.5) at LO~\cite{Alwall:2007fs}.
All SM backgrounds containing the Higgs boson are generated using \POWHEG v2.0 at NLO~\cite{Alioli:2008tz, Bagnaschi:2011tu, Nason:2009ai, Luisoni:2013cuh, Hartanto:2015uka}.
The ${\Pg\Pg}$F production processes are normalized to the theoretical cross section  $\sigma_\mathrm{ggF}^{13\TeV}=48.58\pm 1.56\unit{pb}$~\cite{deFlorian:2016spz} at next-to-next-to-NLO accuracy in perturbative quantum chromodynamics (QCD) and NLO in EW corrections.
The VBF production processes use a theoretical cross section of $\sigma_{\mathrm{VBF}}^{13 \TeV}=3.78\pm 0.08\unit{pb}$~\cite{deFlorian:2016spz}, which includes NLO QCD and EW corrections.
\par}

Cascade and non-cascade signal mass points are chosen to cover the available parameter space following the benchmark scenarios of Ref.~\cite{Robens_2020}.
In the cascade scenario, one of the lower-mass \Paa decays to tau leptons, while the other two \Paa scalars decay to $\PQb$ quarks.
In contrast, for the non-cascade scenario, we require that the lower-mass \Paa decays to tau leptons and \Pab to $\PQb$ quarks.
This is a choice made at the analysis level to target a single invariant mass peak from the tau leptons in the final fit of the simulated distribution to the observed data.
Thus, the final-state signature is $\PH \to \Paa \Pab \to 3\Paa \to 2\PGt4\PQb$ for the cascade and $\PH \to \Paa \Pab,\ \Paa \to 2\PGt,\ \Pab \to 2\PQb$, \ie, $\PH \to \Paa \Pab \to 2\PGt2\PQb$ for the non-cascade scenario, as shown in Fig.~\ref{fig:diagrams}.
The corresponding Higgs boson branching fractions are $\mathrm{B_{C}} = \mathcal{B}(\PH \to \Paa \Pab \to 3 \Paa \to 2\PGt4\PQb)$ and $\mathrm{B_{NC}} = \mathcal{B}(\PH \to \Paa \Pab)\ \mathcal{B}(\Paa \to 2\PGt)\ \mathcal{B}(\Pab \to 2\PQb)$ for the cascade and non-cascade decay scenarios, respectively.

The dominant sources of background are the DY process, the production of a top quark-\hspace{0pt}antiquark pair with additional jets (\ttjets), single top quark production, and massive vector boson pair production (diboson).
The ${\PZ}/\PGgst\to\PGt\PGt$ background is estimated using the embedding method described in Section~\ref{sec:background}~\cite{embedding}.
Events with jets misidentified as $\tauh$ candidates ($\text{jet}\to\tauh$) in the \mutau and \etau channels, and the QCD multijet contribution in the \emu channel, are estimated using control regions (CRs) in data.
These background estimation methods are discussed in detail in Section~\ref{sec:background}.

The DY process in the dilepton final state (\Pell{}\Pell, with \Pell being \Pe or \PGm) is modeled using \MGvATNLO at LO.
The \POWHEG \textsc{box} v2.0 framework~\cite{Nason:2004rx, Frixione:2007vw, Alioli:2010xd, Alioli:2010xa} event generator is used to produce \ttjets and single top events at NLO in QCD.
The simulated \ttjets events are reweighted to match the top quark \pt distribution at next-to-NLO (NNLO) QCD and NLO EW~\cite{Czakon_2017} precision.
Diboson and \wjets\ events are generated by \MGvATNLO at LO.
The \ttjets, DY, and \wjets\ samples are normalized to theoretical cross section values accurate to NNLO in QCD~\cite{Czakon:2011xx, Ball:2014uwa, Martin:2009bu, Gao:2013xoa, Ball:2012cx, Campbell:2020fhf, PDF4LHCWorkingGroup:2022cjn, Melnikov:2006kv, Martin:2009iq}.
In addition, for events from simulated DY samples, a correction to the Z boson \pt distribution is applied by reweighting the simulated events to match the data distribution in bins of the dilepton invariant mass \mellell and the \pt of the dilepton system following Ref.~\cite{CMS:2022kdi}. 

\section{Object reconstruction}
A particle-flow algorithm~\cite{CMS:2017yfk} aims to reconstruct and identify each individual particle in an event, with an optimized combination of information from the various elements of the CMS detector. The energy of a photon is obtained from the ECAL measurement. The energy of an electron is determined from a combination of its momentum at the primary interaction vertex as determined by the tracker, the energy of the corresponding ECAL cluster, and the energy sum of all bremsstrahlung photons spatially compatible with originating from the electron track. The energy of a muon is obtained from the curvature of the corresponding track. The energy of a charged hadron is determined from a combination of the momentum measured in the tracker and the matching ECAL and HCAL energy deposits, corrected for the response function of the calorimeters to hadronic showers. Finally, the energy of neutral hadrons is obtained from the corresponding corrected ECAL and HCAL energies.

For each event, hadronic jets are clustered from these reconstructed particles using the infrared and collinear safe anti-\kt algorithm~\cite{Cacciari:2008gp, Cacciari:2011ma} with a distance parameter of 0.4. Jet momentum is determined as the vectorial sum of all particle momenta in the jet, and is found from simulation to be, on average, within 5 to 10\% of the true momentum over the whole \pt spectrum and detector acceptance. Pileup can contribute additional tracks and calorimetric energy depositions, increasing the apparent jet momentum. To mitigate this effect, tracks identified to be originating from pileup vertices are discarded and an offset correction is applied to correct for remaining contributions~\cite{CMS:2020ebo}. Jet energy corrections are derived from simulation studies so that the average measured energy of jets becomes identical to that of particle-level jets. In situ measurements of the momentum balance in dijet, $\PGg + \text{jet}$, DY, and multijet events are used to determine any residual differences between the jet energy scale in data and in simulation, and appropriate corrections are made~\cite{CMS:2016lmd}. Additional selection criteria are applied to each jet to remove jets potentially dominated by instrumental effects or reconstruction failures~\cite{CMS:2020ebo}.

The \textsc{DeepJet} flavor classification algorithm~\cite{Bols:2020bkb,Sirunyan:2017ezt} is used to determine (tag) which are the \PQb jets.
The medium working point of the \PQb tagging discriminator is used to identify \PQb jets, which corresponds to an 80\% tagging efficiency and a 1\% probability to mistag a light-flavor quark or a gluon jet as a \PQb jet.

The missing transverse momentum vector \ptvecmiss is computed as the negative vector sum of the \pt of all the particle-flow candidates in an event, and its magnitude is denoted as \ptmiss~\cite{CMS:2019ctu}. The value of \ptvecmiss is modified to account for corrections to the energy scale of the reconstructed jets in the event. Anomalous high-\ptmiss events can be due to a variety of reconstruction failures, detector malfunctions, or noncollision backgrounds. Such events are rejected by event filters that are designed to identify more than 85--90\% of the spurious high-\ptmiss events with a false positive rate less than 0.1\%~\cite{CMS:2019ctu}.

The \tauh decays are reconstructed from jets, using the hadrons-plus-strips algorithm~\cite{CMS:2018jrd}, which combines 1 or 3 tracks with energy deposits in the calorimeters to identify the tau lepton decay modes.
Neutral pions are reconstructed using strips with dynamic size in $\eta$-$\phi$ space from reconstructed electrons and photons, where the strip size varies as a function of the \pt of the \Pe or \PGg candidate.
To distinguish genuine \tauh decays from jets originating from the hadronization of quarks or gluons, and from \Pe or \Pgm, the \textsc{DeepTau} algorithm is used~\cite{CMS:2022prd}.
Information from all individual reconstructed particles near the \tauh axis is combined with properties of the \tauh candidate and the overall event.
The probability of a jet to be misidentified as \tauh by the \textsc{DeepTau} algorithm depends on the \pt and quark flavor of the jet.
In simulated events of \PW boson production in association with jets, the rate is found to be 0.43\%, with an identification efficiency of 70\% for genuine \tauh decays.
The rate for an \Pe (\Pgm) to be misidentified as a \tauh is 2.6 (0.3)\% for a genuine \tauh identification efficiency of 80 ($>$99)\%.

\section{Event selection}
\label{sec:eventcat}
A combination of single-lepton triggers and cross triggers requiring a \tauh and an electron or muon is used to identify \mutau and \etau events, while only triggers requiring both an electron and a muon are used for the \emu channel at the online reconstruction level.
If multiple single-lepton and cross-trigger decisions are available, the event is selected using a logical ``OR'' operation.
Offline \pt thresholds are 1\GeV greater than the online \pt thresholds for \Pe and \Pgm, and 5\GeV greater in the case of \tauh.
The trigger \pt thresholds changed over the data-taking periods, as tabulated in Table~\ref{tab:triggers}.

\begin{table*}[ht!]
	\centering
	\topcaption{The \pt thresholds for \Pe, \Pgm, and \tauh at the trigger level for the three data-taking periods. For the \etau and \mutau channels, the \pt thresholds for the \Pe and \Pgm are dependent on the specific high-level trigger used, \ie, the single-lepton or the cross trigger. Thresholds are stated as follows: the single-lepton and cross-trigger \pt thresholds, and thresholds for the highest \pt (leading) and second-highest \pt (subleading) leptons. Multiple single-lepton triggers are listed as comma-separated \pt thresholds. For cross triggers, the leading and subleading \pt thresholds are applied simultaneously to the lepton pair. }
	\begin{tabular}{lccccccc}
		 & & \multicolumn{2}{c}{\emu} & \multicolumn{2}{c}{\etau} & \multicolumn{2}{c}{\mutau}  \\
		 & & \Pe & \PGm & \Pe & \tauh & \PGm & \tauh \\
		\hline
		\multirow{3}{*}{ 2016 } & Single-lepton \pt         &        \NA       &       \NA      &         25            &          \NA         &          22          &         \NA    \\
		                        & Leading-lepton \pt        &        23        &       23       &  \multirow{2}{*}{\NA} & \multirow{2}{*}{\NA} &          \NA         &         20     \\
		                        & Subleading-lepton \pt     &        12        &       8        &                       &                      &          19          &         \NA    \\
		[\cmsTabSkip]
		\multirow{3}{*}{ 2017 } & Single-lepton \pt         &        \NA       &       \NA      &       27, 32          &          \NA         &       24, 27         &         \NA    \\
		                        & Leading-lepton \pt        &        23        &       23       &         \NA           &          30          &          \NA         &         27     \\
		                        & Subleading-lepton \pt     &        12        &       8        &         24            &          \NA         &          20          &         \NA    \\
		[\cmsTabSkip]
		\multirow{3}{*}{ 2018 } & Single-lepton \pt         &        \NA       &       \NA      &       32, 35          &          \NA         &       24, 27         &         \NA    \\
		                        & Leading-lepton \pt        &        23        &       23       &         \NA           &          30          &          \NA         &         27     \\
		                        & Subleading-lepton \pt     &        12        &       8        &         24            &          \NA         &          20          &         \NA    \\
	\end{tabular}
	\label{tab:triggers}
\end{table*}
Events passing the trigger requirements are subsequently processed to select offline reconstructed objects.
First, all \Pgm, \Pe, \tauh, and jets passing loose identification and isolation requirements are selected.
Muons are selected within $\abs{\eta} < 2.4$ using the medium identification criteria of Ref.~\cite{CMS:2018rym}, which has an overall selection efficiency of 99.5\%. 
Electrons are identified within $\abs{\eta} < 2.5$ using a multivariate analysis discriminant~\cite{CMS:2020uim} having a 90\% selection efficiency for the chosen working point.
The relative isolation for a muon (electron), defined as the ratio of the sum of \pt deposited by other particles within an angular distance of $\Delta R=\sqrt{\smash[b]{(\Delta\phi)^2+(\Delta\eta)^2}} < 0.4\,(0.3)$ to the muon (electron) \pt, is required to be less than 0.15.
The \tauh decays are selected for $\abs{\eta} < 2.3$ using the very loose working points of the \textsc{DeepTau} discriminator against jets, electrons, and muons~\cite{CMS:2022prd}.

Events are required to contain two tau lepton candidates, sorted into three exclusive channels:
\begin{itemize}
	\item \mutau: at least one \Pgm, exactly zero \Pe, and at least one \tauh;
	\item \etau: exactly zero \Pgm, at least one \Pe, and at least one \tauh;
	\item \emu: at least one \Pgm and at least one \Pe.
\end{itemize}

The two lepton candidates are required to be spatially separated, satisfying $\Delta R > 0.4 (0.3)$ for the \mutau and \etau (\emu) channels, and all jets are equired to be separated by $\Delta R > 0.4$ from both lepton candidates.
Once events are sorted into channels, the \pt thresholds and isolation criteria are tightened.
For the \mutau and \etau events, the \tauh must pass the medium working point of the \textsc{DeepTau} discriminator against jets and have $\pt > 20\GeV$ in the absence of a cross trigger.
The \tauh \textsc{DeepTau} discriminator against muons (electrons) must satisfy the tight (loose) working point in the \mutau channel.
Similarly, the \tauh discriminator against electrons (muons) must satisfy the tight (loose) working point in the \etau channel.
For the \emu events, the relative electron isolation is tightened to be ${<}0.10$.
These requirements are optimized from simulation to yield better sensitivity to signal events where the \PGt lepton decay products are close to each other.

A single \PGt{}\PGt candidate per event is then selected from all possible combinations of leptons in the event.
Multiple candidates arise in ${<}5\%$ of events; in such cases, the pair with higher \pt and tighter lepton isolations is preferred.
The two leptons of the \PGt{}\PGt candidate must have opposite charge and also match the online reconstructed leptons in position within $\Delta R < 0.5$ for that event to be considered.
For the \tauh case, the charge is determined by the constituent charged tracks from the hadrons-plus-strips reconstruction.
Finally, the selected event is required to have at least one $\PQb$ jet.
After the preselection of events, signal region (SR) categories are defined using a BDT discriminator.
The invariant mass of the \PGt{}\PGt candidate (\mtt), including neutrino energies, is reconstructed with the SVfit algorithm~\cite{svfit, Kalinowski:2025vci}.
The signal is searched for as a peak in the \mtt distribution, centered at \maa.
The CRs with larger contributions from background processes are also defined to validate the background estimation methods, though not used for signal extraction.

Each final state is subdivided into two categories based on the presence of exactly one \PQb jet or two or more \PQb jets in the event.
In total, six different BDT discriminators are utilized, according to the number of \PQb jets and the decay modes of the \PGt leptons.
For the signal simulation, all the different mass hypotheses are incorporated in the BDT training.
Further, only events from the three major background processes for each channel are included in the training.
Two sets of events, one for training and another for testing, are prepared by mixing signal and background events passing the preselection criteria.
The mixture of background events employed in the BDT training is based on their yields in the preselected region with at least one \PQb jet.
The models are trained using XGBoost~\cite{Chen:2016:XGBoost}, and the number of trees or evaluators range between 300 and 500.
Since the BDT is utilized for a binary classification between signal and background events, the ``binary logistic'' option is applied, to set the learning objective, and a negative log-likelihood function is used for the evaluation of these models.
The shape of the \mtt distribution is verified to remain largely unchanged after the BDT selection is applied.

Among the important variables included in the training are the visible invariant mass of the combined leading \PQb-tagged jet and \PGt{}\PGt candidate, $m^{\text{vis}}(\PGt\PGt\PQb_1)$, the transverse invariant mass between the \ptvecmiss and the leading constituent of the \PGt{}\PGt, and an alignment variable called \dzeta.
The variable \dzeta is defined following Ref.~\cite{2016bbtautau} as
        \begin{equation} \dzeta \equiv p_{\zeta} - 0.85 p_{\zeta}^{\text{vis}},     \label{eqn:D_zeta} \end{equation}
where we define the $\zeta$ axis along the bisector of the directions of the visible products of the two tau leptons of the \PGt{}\PGt pair transverse to the beam direction.
This axis approximates the direction of the \PGt{}\PGt system.
We define $p_{\zeta}$ to be the component of \ptvecmiss along the $\zeta$ axis, and $p_{\zeta}^{\text{vis}}$ to be the sum of projections onto the $\zeta$ axis of the \pt of the visible \PGt decay products~\cite{Abulencia:2005kq}.
The variable \dzeta provides good discrimination against the \ttjets background, which tends to have small values because of the large component of \ptvecmiss that is not aligned with the \PGt{}\PGt system. 
The ${\PZ}/\PGgst\to\PGt\PGt$ background has large \dzeta values due to \ptmiss being approximately collinear to the \PGt{}\PGt system.
The signal has intermediate \dzeta values since \ptvecmiss is approximately aligned with the \PGt{}\PGt system, but its magnitude is small.

The transverse mass (\mT) of a visible particle $i$ combined with \ptmiss is defined as
        \begin{equation} \mT (i, \ptmiss) \equiv \sqrt{2 \pt^{i} \ptmiss\left[1 - \cos(\Delta\phi)\right]},     \label{eqn:MT} \end{equation}
where $\pt^{i}$ is the transverse momentum of the particle $i$, and $\Delta\phi$ is the azimuthal angle between its direction and \ptvecmiss.
Events from \ttjets and misidentified \tauh background are characterized by large \ptmiss and thus have higher \mT than signal events.
In addition, the angular separations between various object pairs are particularly discriminating and are therefore fed into the BDT.
Another important feature used is the number of jets in the event.
Further, for events containing at least two \PQb jets, a variable $\Delta m$ is defined to measure the normalized difference between the invariant masses of the two \PQb jets (\mbb) and the \PGt{}\PGt candidate (\mtt):
\begin{equation} \Delta m \equiv \frac{\mbb - \mtt }{\mtt}.     \end{equation}

\begin{figure}[ht!]
    \centering
        \includegraphics[width=0.44\textwidth]{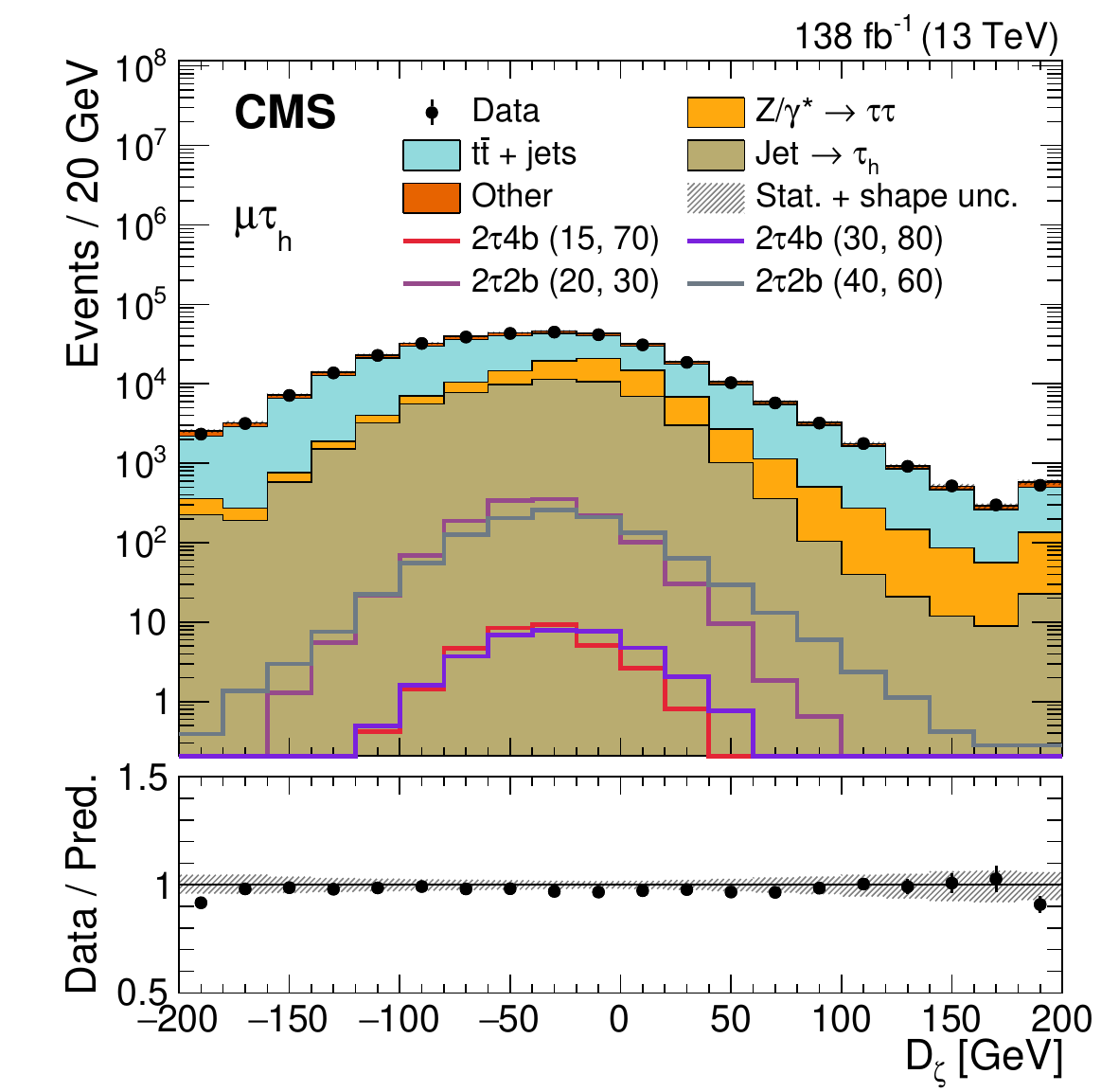}
        \includegraphics[width=0.44\textwidth]{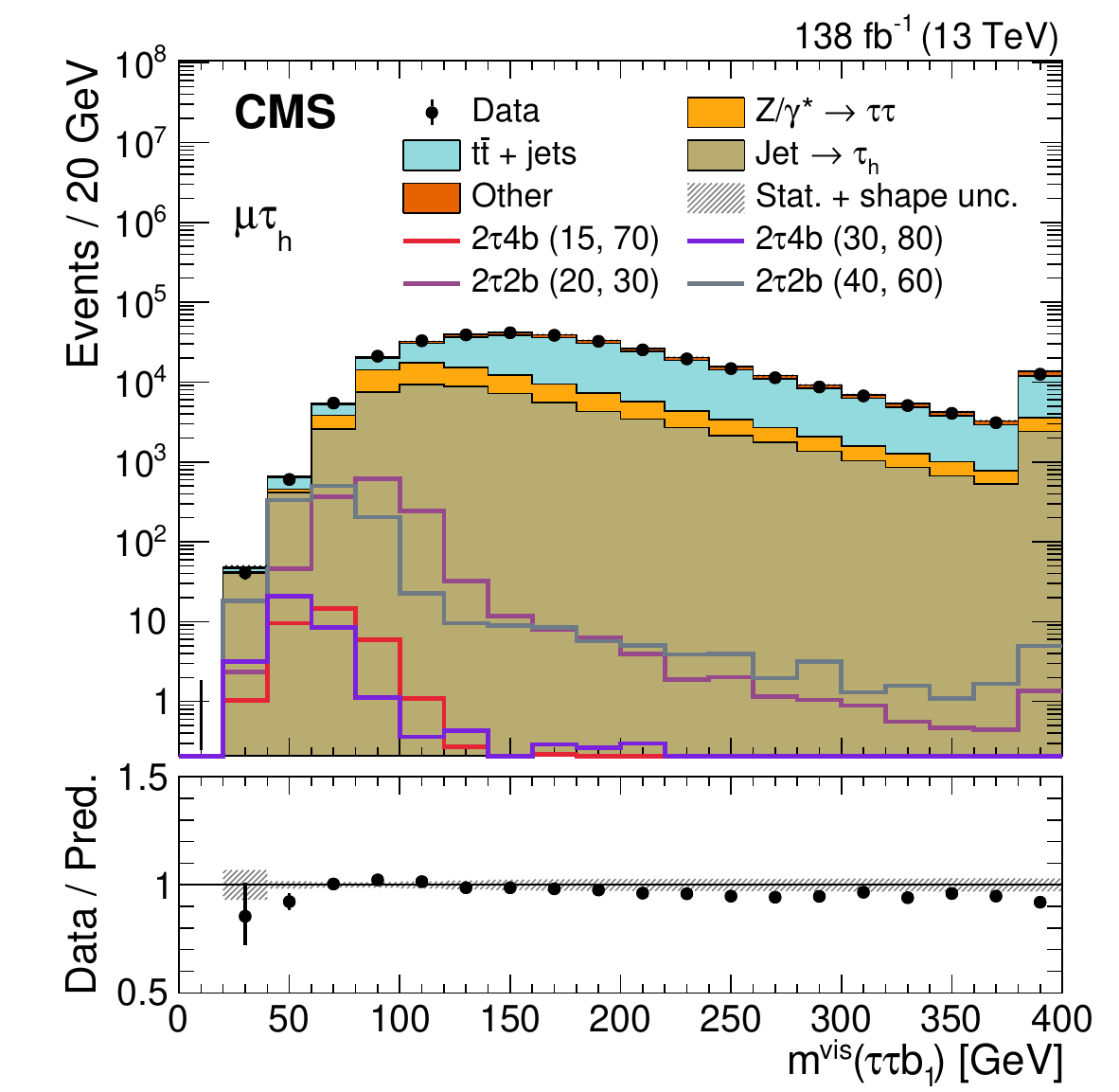} \\
        \includegraphics[width=0.44\textwidth]{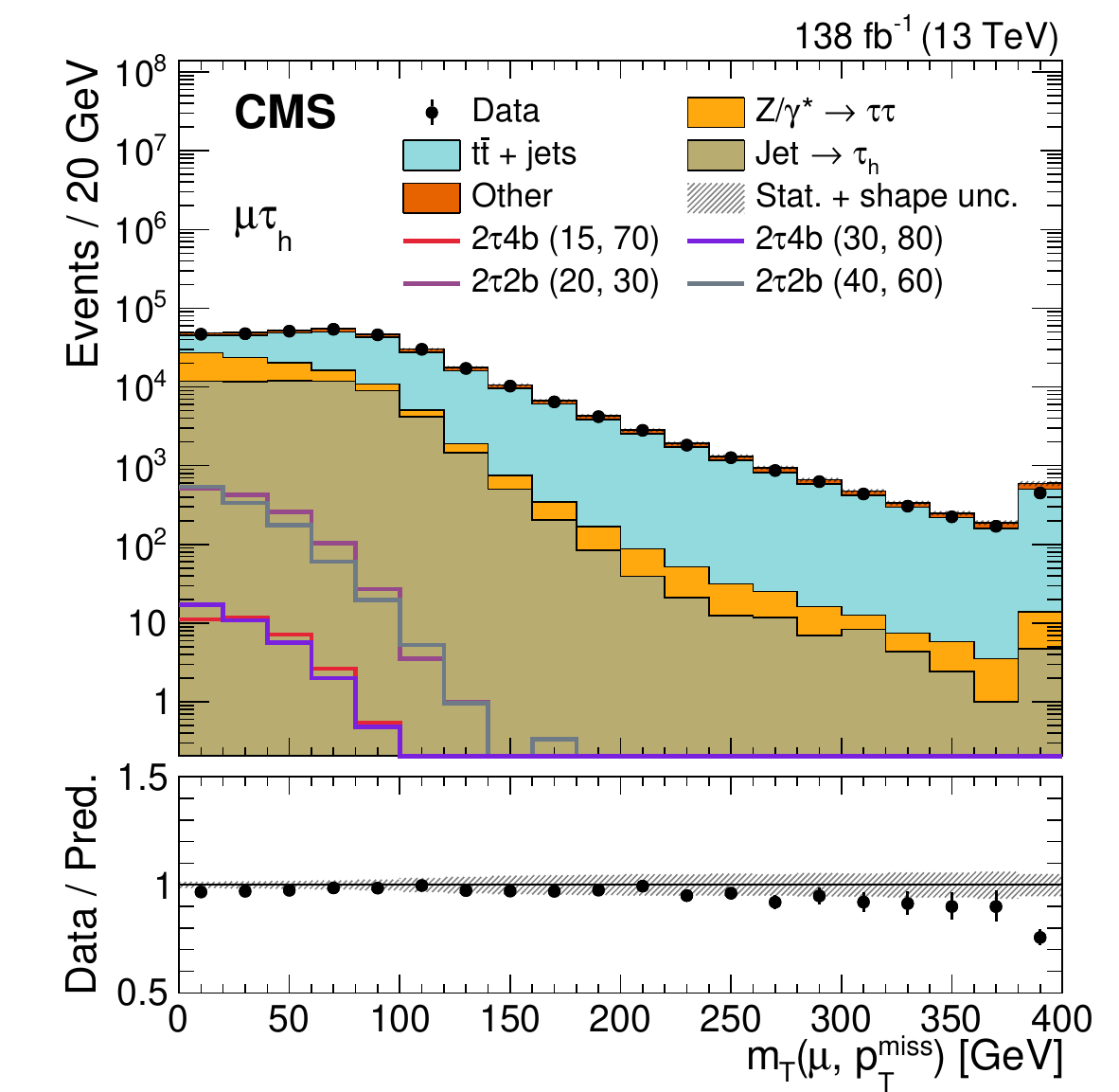}
        \includegraphics[width=0.44\textwidth]{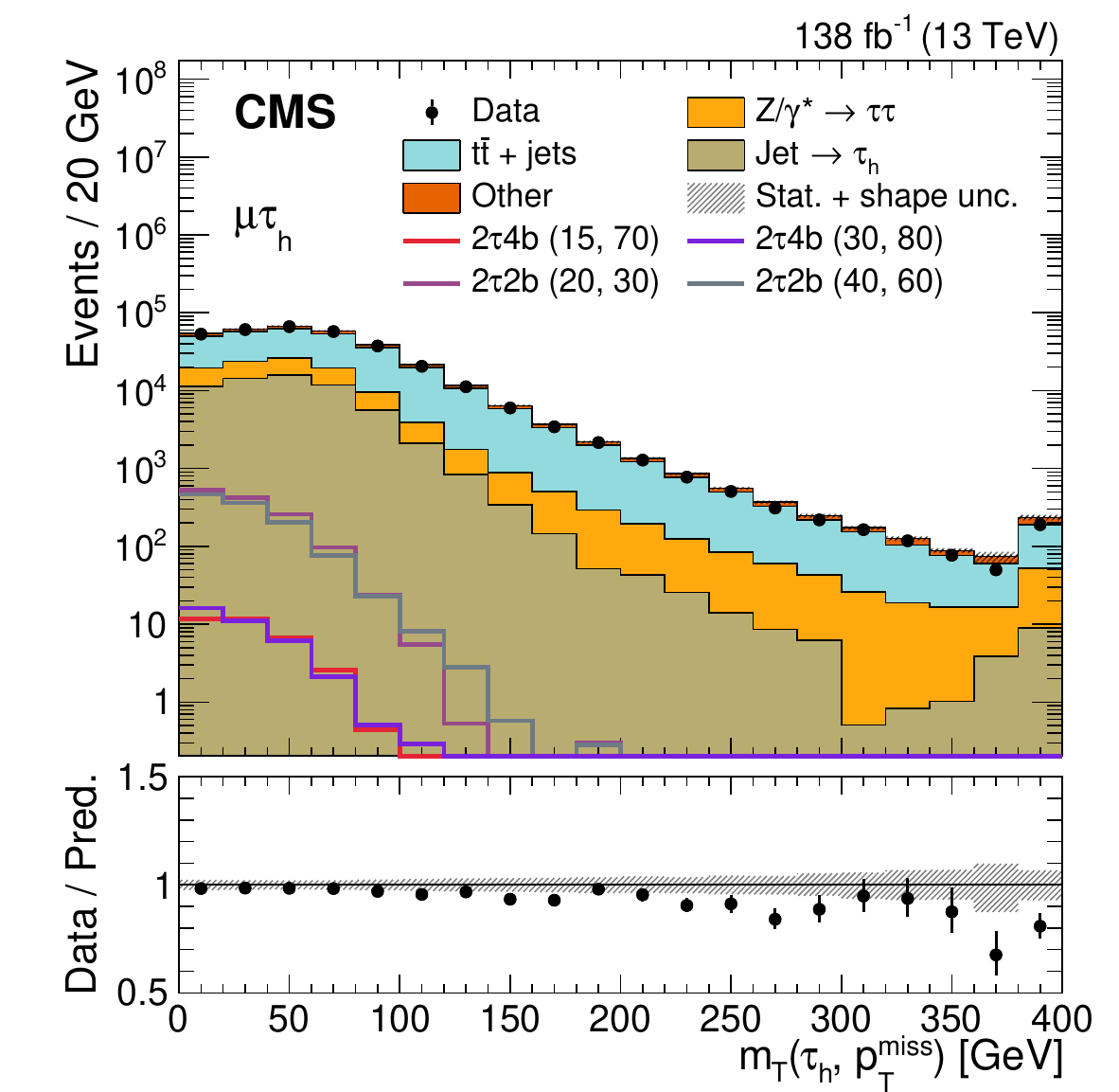}
    
    \caption{Pre-fit distributions of \dzeta (upper left), $m^{\text{vis}}(\PGt\PGt\PQb_1)$ (upper right), $\mT(\PGm, \ptmiss)$ (lower left), and $\mT(\tauh, \ptmiss)$ (lower right), including underflow and overflow bins, for preselected events with at least one \PQb-tagged jet for the \mutau channel, without any SR requirements. The data are shown by the markers with vertical bars and various backgrounds by the colored histograms. The combination of statistical and shape systematic uncertainties is displayed with the hatched areas. The colored open histograms display the predicted signal distribution for two cascade decays and two non-cascade decays, with four different values of \Paa and \Pab masses, for an assumed branching fraction of 100\%. The lower panel of each plot shows the ratio of the data to the sum of the predicted number of background events. The vertical bars on the points show the statistical uncertainty in the ratio.}
    \label{fig:vars_mutau}
\end{figure}

\begin{figure}[ht!]
    \centering
        \includegraphics[width=0.44\textwidth]{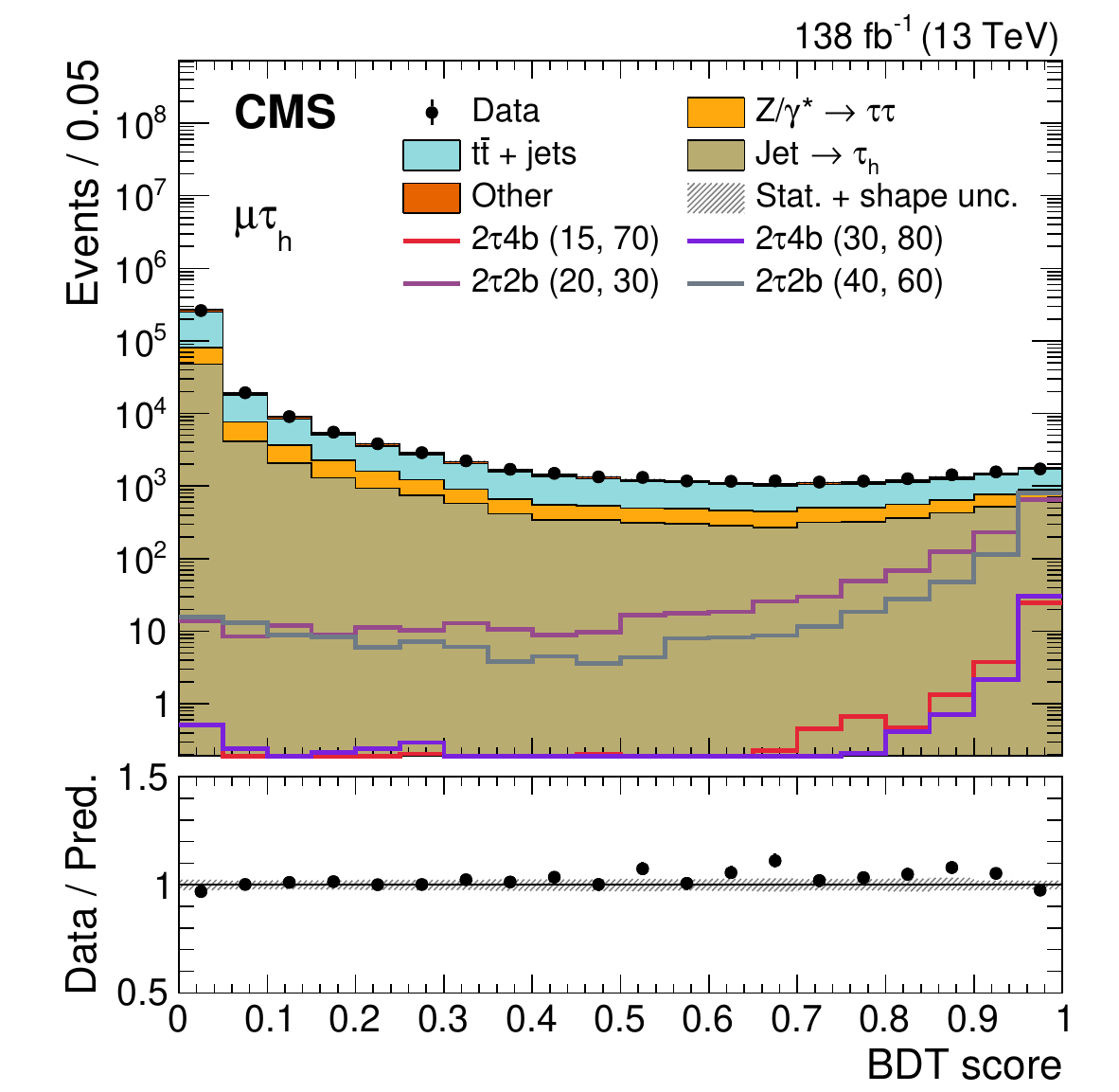}
        \includegraphics[width=0.44\textwidth]{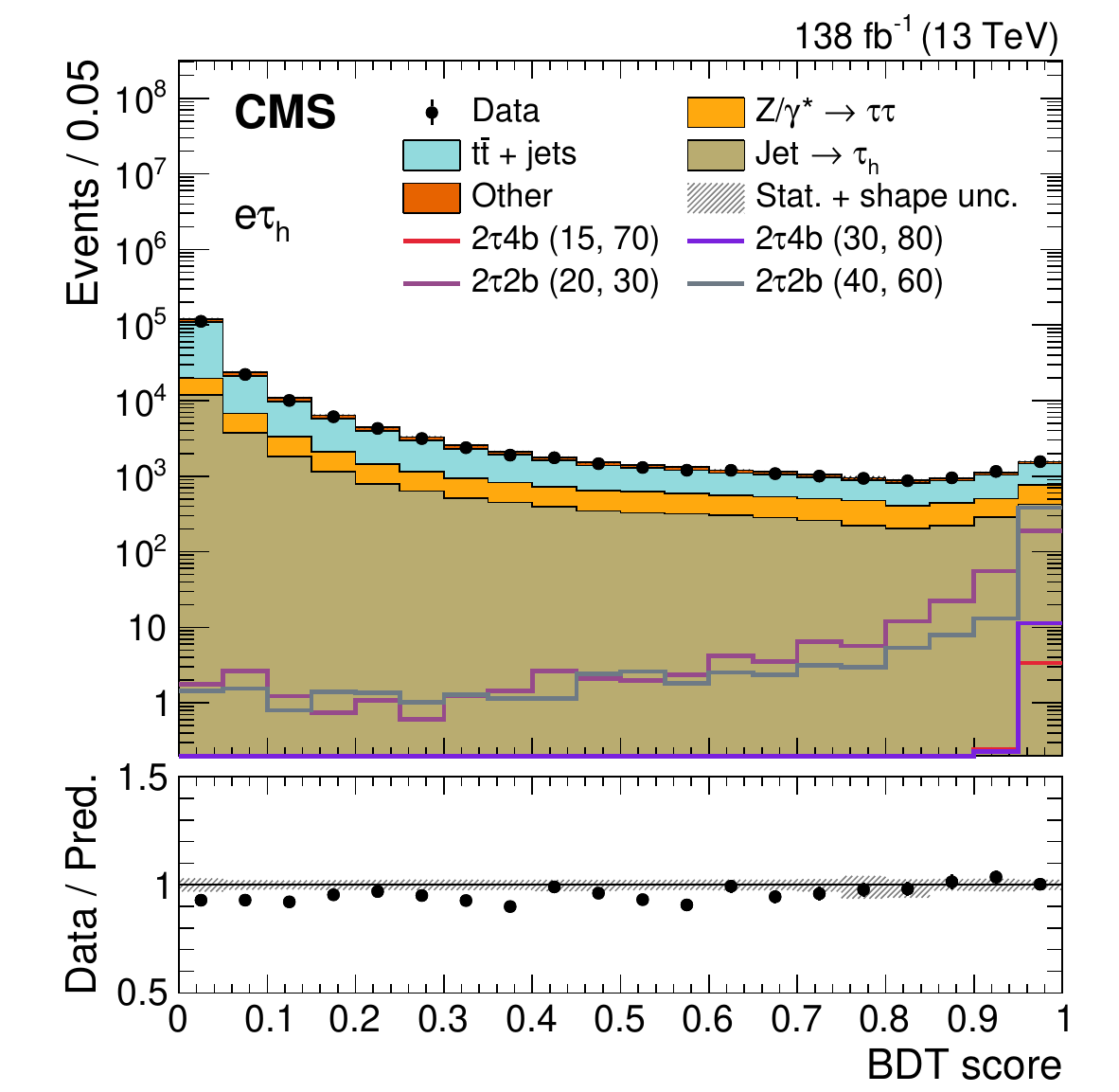}
        \includegraphics[width=0.44\textwidth]{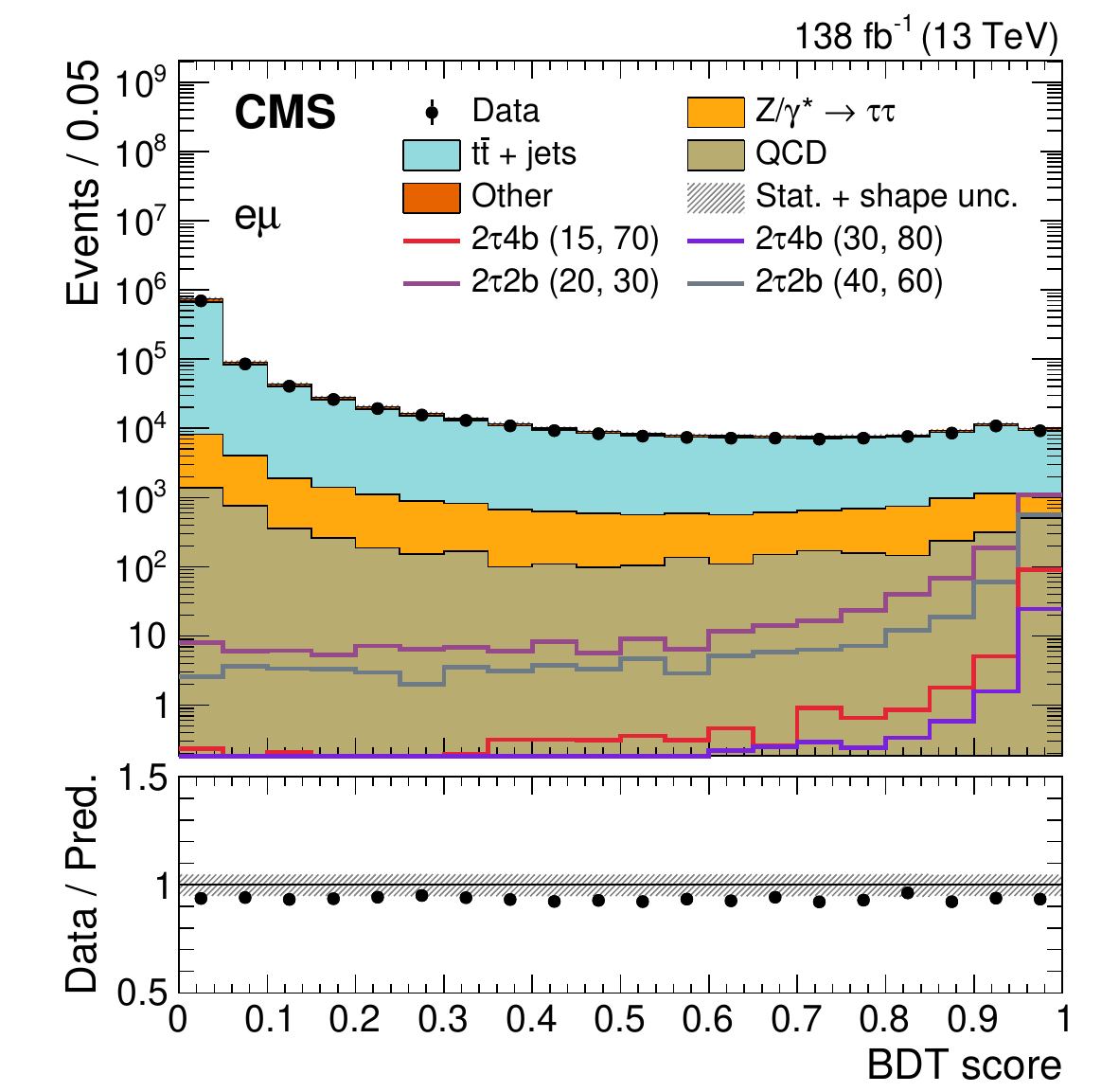}
    
    \caption{Pre-fit BDT score distribution for preselected events with at least one \PQb-tagged jet for the \mutau (upper left), \etau (upper right), and \emu (lower) channels, without any SR requirements. The data are shown by the markers with vertical bars and various backgrounds by the colored histograms. The combination of statistical and shape systematic uncertainties is displayed with the hatched areas. The colored open histograms display the predicted signal distribution for two cascade decays and two non-cascade decays, with four different values of \Paa and \Pab masses, for an assumed branching fraction of 100\%. The lower panel of each plot shows the ratio of the data to the sum of the predicted number of background events. The vertical bars on the points show the statistical uncertainty in the ratio.}
    \label{fig:bdtscores}
\end{figure}

These variables can be used to discriminate signal from background directly using a cut-based method (see Appendix~\ref{appendix}); however, a BDT is used for further optimization, yielding better sensitivity across the mass range.
Figure~\ref{fig:vars_mutau} shows the background and signal distributions of a few important discriminating variables in events with at least one \PQb-tagged jet for the \mutau channel, before (``pre-fit'') the maximum likelihood fit described in Section~\ref{sec:results}.
For all following figures, ``Other'' indicates the grouped background contribution from ${\PZ}/\PGgst\to\Pe\Pe$ or $\PGm\PGm$, single top quark, diboson, and SM Higgs boson production processes.
In Figs.~\ref{fig:vars_mutau}--\ref{fig:results_postfit_emu_bdt} and \ref{fig:results_postfit_mutau}--\ref{fig:results_postfit_emu}, and in the text, the masses \maa and \mab corresponding to a signal hypothesis are indicated in parentheses, in units of \GeV.

The kinematic variables used as input to the BDT discriminants are listed in Table~\ref{tab:BDT-variables}.
They are optimized for each channel based on discriminator performance parameters, such as the logarithmic loss, misclassification error fraction, and the area under the receiver operating characteristic curve.
The hypothesis of the heavier scalar \Pab decaying to \PQb jets in this analysis broadens the $\Delta m$ distribution.
In addition, the angular separation $\Delta R$ variables from the asymmetric signal models have differences in shape compared to the symmetric signal models.
Therefore, the BDTs that are trained and implemented in this search are optimized for the asymmetric signal scenario, and not for $\PH \to \Pa \Pa \to 2\PGt 2\PQb$.
In the following, events containing exactly one \PQb jet are denoted by 1\PQb, and events containing at least two \PQb jets are denoted by 2\PQb.
Figure~\ref{fig:bdtscores} shows the BDT score distributions in events with at least one \PQb jet in the three channels.

\begin{table*}[ht!]
        \centering
        \topcaption{List of kinematic variables used as inputs to the BDT discriminator, with their importance rankings. The leading and subleading \PQb-tagged jets are denoted by $\PQb_1$ and $\PQb_2$, respectively. Depending on the channel, the importance of an input variable in the performance of the BDT varies. The increasing importance of a variable in the training is denoted by decreasing numerical values.}
        \begin{tabular}{lcccccc}
		Variable                                 & \mutau, 1\PQb & \mutau, 2\PQb & \etau, 1\PQb & \etau, 2\PQb & \emu, 1\PQb & \emu, 2\PQb  \\
                \hline
                \dzeta                                   & 7    & \NA  & 8    & \NA   & 7     & \NA   \\
                $\pt^{\text{vis}}(\PGt\PGt)$             & 2    & \NA  & 1    & \NA   & 4     & 4     \\
                $\pt^{\Pe}$                              & \NA  & \NA  & \NA  & \NA   & 5     & \NA   \\
                $m^{\text{vis}}(\PGt\PGt\PQb_1)$         & 1    & \NA  & 4    & 1     & 1     & \NA   \\
                $m^{\text{vis}}(\PGt\PGt\PQb_2)$         & \NA  & 1    & \NA  & \NA   & \NA   & \NA   \\
                $m^{\text{vis}}(\PGt\PGt\PQb\PQb)$       & \NA  & 3    & \NA  & \NA   & \NA   & \NA   \\
                $m_{\PQb\PQb}$                           & \NA  & \NA  & \NA  & 6     & \NA   & \NA   \\
                $\mT(\PGm,\ \ptmiss)$                    & 4    & 4    & \NA  & \NA   & \NA   & \NA   \\
                $\mT(\Pe,\ \ptmiss)$                     & \NA  & \NA  & 2    & 2     & 2     & 1     \\
                $\mT(\tauh,\ \ptmiss)$                   & 5    & \NA  & 5    & \NA   & \NA   & \NA   \\
                $\mT(\PQb_1,\ \ptmiss)$                  & \NA  & \NA  & \NA  & \NA   & 3     & 3     \\
                $\Delta  R(\tauh,\ \PQb_1)$      & 6    & \NA  & 7    & 3     & \NA   & \NA   \\
                $\Delta  R(\tauh,\ \PQb_2)$      & \NA  & 5    & \NA  & 4     & \NA   & \NA   \\
                $\Delta  R(\Pe,\ \PQb_1)$        & \NA  & \NA  & 3    & 5     & \NA   & \NA   \\
                $\Delta  R(\PGt\PGt,\ \PQb_1)$   & 8    & 6    & \NA  & \NA   & 6     & \NA   \\
                $\Delta  R(\PGt\PGt,\ \PQb_2)$   & \NA  & \NA  & \NA  & \NA   & \NA   & 5     \\
                $\Delta m$                               & \NA  & 2    & \NA  & \NA   & \NA   & 2     \\
                Jet multiplicity                         & 3    & \NA  & 6    & \NA   & 8     & \NA   \\
        \end{tabular}
        \label{tab:BDT-variables}
\end{table*}

Events in each channel are divided into separate SR categories depending on their BDT score, as shown in Table~\ref{tab:BDT-categories}.
The \mutau and \etau channels include four SRs, namely SR1, SR2, SR3, and SR4 in the 1\PQb events.
There are three SRs: SR1, SR2, and SR3 in the \emu channel 1\PQb events.
All three channels have two categories in 2\PQb events, SR1 and SR2.
In total, there are 17 SRs in the BDT-based categorization. 

\begin{table*}[ht!]
        \centering
        \topcaption{The BDT discriminator score ranges defining the different SRs and CRs.}
        \begin{tabular}{lccccc}
		& SR1 & SR2 & SR3 & SR4 & CR \\
		\hline
		\mutau, 1\PQb   & [0.99, 1] & [0.98, 0.99) & [0.95, 0.98) & [0.80, 0.95) & [0, 0.80) \\
		\mutau, 2\PQb   & [0.99, 1] & [0.84, 0.99) & \NA          & \NA          & [0, 0.80) \\
		\etau,  1\PQb   & [0.99, 1] & [0.98, 0.99) & [0.95, 0.98) & [0.80, 0.95) & [0, 0.80) \\
		\etau,  2\PQb   & [0.99, 1] & [0.80, 0.99) & \NA          & \NA          & [0, 0.80) \\
		\emu,   1\PQb   & [0.99, 1] & [0.97, 0.99) & [0.90, 0.97) & \NA          & [0, 0.80) \\
		\emu,   2\PQb   & [0.99, 1] & [0.90, 0.99) & \NA          & \NA          & [0, 0.80) \\
        \end{tabular}
	\label{tab:BDT-categories}
\end{table*}

\section{Background estimation}
\label{sec:background}
The DY ${\PZ}/\PGgst\to\PGt\PGt$ process is one of the three major backgrounds and is estimated using embedded samples for all three channels. 
These samples are produced by first selecting reconstructed ${\PZ}/\PGgst\to\PGm\PGm$ events in data with high purity, requiring a 20\GeV lower bound on the dimuon invariant mass~\cite{embedding}.
The two muon tracks and their energy deposits in both calorimeters are removed to produce a ``cleaned'' event.
Tau leptons and their decays are simulated and merged with the cleaned event in a way that preserves the original \Pgm kinematic distributions, resulting in an embedded event. The primary advantage is that event characteristics that are difficult to simulate, e.g., associated jet production, pileup, detector noise, and resolution effects, are taken directly from data.
As a result, fewer corrections are required than those needed for fully simulated samples, and thus smaller systematic uncertainties need to be taken into account for embedded samples.

The QCD multijet contribution to the \emu final state is estimated using the data in a CR with same-sign \emu pairs.
The event selection in the CR is otherwise identical to that in the \emu SRs.
The contributions of other processes in the CR are taken from simulation and subtracted from the data.
The resulting number of data events in the CR is scaled by the ratio of the expected multijet contribution in the SR to the expected multijet contribution in the CR.
The scale factors are calculated in data orthogonal to the SR, as functions of the jet multiplicity and the $\Delta R$ separation between the \Pe and \Pgm, in order to account for possible kinematic differences between the two regions.

Backgrounds with hadronic jets that are misidentified as \tauh candidates contribute significantly to the \etau and \mutau final states and are estimated from the data.
This background is dominated by \wjets, QCD multijets, and \ttjets processes with at least one top quark decaying to hadrons.
A background-enriched sideband (SB) region is created by requiring that the events pass all the baseline \etau or \mutau selection criteria, but the \tauh\ fails the \textsc{DeepTau} discrimination working point against jets.
The data in this SB are reweighted with a factor $f/(1-f)$, where $f$ is the probability for a jet to be misidentified as a \tauh candidate and is evaluated as a function of $\pt(\tauh)$.
A selected data sample enriched in ${\PZ}/\PGgst\to\PGm\PGm+\text{jets}$ events is used to measure this misidentification probability.
The final state must contain a \Pgm{}\Pgm compatible with the decay of the \PZ boson, as well as a \tauh candidate.
The contribution of simulated diboson events containing a genuine \tauh is subtracted from data to avoid contamination in the measurement of $f$.
The measurement is done separately for the \etau\ and \mutau final states.
This is because the antilepton discrimination in the \tauh identification changes depending on the flavor of the lepton (\Pe or \Pgm)~\cite{CMS:2022prd}.
The difference between the two misidentification probability measurements is observed to be around 10\%.
The misidentification probability also depends on the jet multiplicity, which characterizes the hadronic activity in the event.

All other backgrounds, including \ttjets with decays to \Pe or \PGm, are estimated from simulation.
However, the contribution with two genuine tau leptons in the final state is already accounted for by the embedded sample described above, so the background estimate from simulation does not include these events.

\section{Sources of systematic effects}
The systematic uncertainties included in the analysis are as follows:

\begin{description}

 \item[Integrated luminosity:] 
the integrated luminosities for the 2016, 2017, and 2018 data-taking years have 1.2--2.5\% individual uncertainties~\cite{CMS-LUM-17-003,CMS-PAS-LUM-17-004,CMS-PAS-LUM-18-002}, while the overall uncertainty for the full 2016--2018 period is 1.6\%.
The difference in measured integrated luminosity using different methods is included as a normalization uncertainty.

 \item[Trigger:]
an uncertainty of 2\% is applied per single-lepton trigger, and the uncertainty in the lepton and \tauh cross trigger depends on $\pt(\tauh)$, ranging between 5 and 10\%~\cite{CMS:2020uim,CMS:2018rym,CMS:2018jrd}.
Uncertainties associated with trigger efficiencies affect both the shape and normalization of the \mtt distributions.

 \item[Energy scales of genuine and misidentified tau leptons:] 
in the \mutau and \etau final states, uncertainties associated with \tauh energy scale corrections are 3--5\% and 0.2--1.1\%, respectively, depending on the \pt and decay mode of the \tauh~\cite{CMS:2018jrd}.
For genuine \tauh, a shape uncertainty in the \tauh energy scale is applied, binned in the tau reconstruction decay mode.
The uncertainties in the energy scales for \Pe and \Pgm misidentified as \tauh are also included for each tau lepton reconstruction decay mode.
For \Pgm misidentified as \tauh, an uncertainty of about 1\% is taken, while the uncertainty for \Pe misidentified as \tauh is 1.2\%~\cite{embedding}.

 \item[Identification efficiencies of genuine and misidentified tau leptons:]
for genuine \tauh passing the medium working point of the \textsc{DeepTau} against jet algorithm, a scale factor binned in $\pt(\tauh)$ is applied for the \textsc{DeepTau} against jet identification efficiency. 
The uncertainties in the corrections are propagated to the simulated events and affect both the shape and normalization of the distributions.
These uncertainties are taken as uncorrelated across $\pt(\tauh)$ bins and data-taking years.
In the \etau channel, there is an additional uncertainty in the above discrimination efficiency, since the analysis uses a looser antilepton working point than the one used to measure the misidentification probabilities.
Uncertainties in the efficiencies of the \textsc{DeepTau} identification against electrons and muons are applied to events with \Pe and \Pgm being misidentified as \tauh, respectively.
The uncertainties in the correction factors for the \textsc{DeepTau} discriminant identification efficiencies are on the order of 2\% ~\cite{embedding}.

 \item[Energy scales and identification efficiencies of electrons and muons:] 
the uncertainties in the \Pe and \Pgm energy scales vary with $\eta$, and are typically within 3\%~\cite{CMS:2020uim,CMS:2018rym}.
They are considered as normalization uncertainties.
The uncertainties associated with \Pe and \Pgm identification of 2\% each~\cite{CMS:2020uim,CMS:2018rym} are also taken as normalization uncertainties.

 \item[Jet energy corrections:]
the jet energy scale and resolution uncertainties include several sources parameterized as functions of the jet \pt and $\eta$~\cite{CMS:2016lmd}.
Those variations can modify the selected event sample and are considered as shape uncertainties, which can vary between 10 and 20\% and are propagated to the \ptmiss calculation. 

 \item[Background estimation methods:]
the \PZ boson \pt reweighting uncertainty in DY samples, which amounts to 10\% of the correction, is taken as an \mtt shape uncertainty.
The embedded samples include a 4\% normalization uncertainty because of the insufficient knowledge of the unfolding corrections to the initially selected muons~\cite{embedding}.
Moreover, shape uncertainties related to tracking efficiencies and contamination from non-DY events in the embedded sample are considered.
Since the contribution of the QCD multijet background in the \emu\ channel is obtained from a same-sign CR with a limited number of events, the validity of the method is tested in independent same-sign CRs.
This test leads to a 20\% normalization uncertainty.
The uncertainty in the scale factor between the same-sign CRs and opposite-sign SRs is modeled using shape variations in the fit utilized to obtain the nominal values.
The misidentification probability, $f$, for a jet to be misidentified as a \tauh candidate depends on the jet multiplicity.
A normalization uncertainty is applied to the estimate of the events where hadronic jets are misidentified as \tauh because $f$ is measured in ${\PZ}/\PGgst\to\PGm\PGm$ events with different jet multiplicities and jet flavor composition.
Since the relative contribution of misidentified jets from \wjets and \ttjets processes varies from one SR to another, the normalization uncertainties are considered uncorrelated between SRs, and vary between 20 and 30\%.
In addition, the uncertainties in the measured $f$ values are considered as shape variations, as a function of the $\pt(\tauh)$.

 \item[Modeling uncertainties:]
there is a 3.6\% normalization uncertainty in the signal to account for PDF, strong coupling constant (\alpS), QCD scale, and renormalization and factorization scale uncertainties~\cite{deFlorian:2016spz}.
The PDF and \alpS uncertainties are included in the simulated backgrounds and consist of 4.2\% for \ttjets, 5\% for diboson production, and 5\% for single top quark processes.
In the case of \ttjets, besides the normalization uncertainty, there are additional acceptance uncertainties arising from renormalization and factorization scale uncertainties. 
The uncertainties are determined from the variations in the factorization and renormalization scales and are taken as shape variations.

 \item[Limited size of the samples:]
to account for the limited size of the simulated samples and the data CRs and SBs used to estimate backgrounds, a bin-by-bin statistical uncertainty estimated using a Poisson nuisance parameter per bin is assigned to distributions in those samples~\cite{Barlow:249779}.

\end{description}

The impact of different groups of uncertainties in the observed limits on the products $\sigma \mathrm{B_{C}}$ and $\sigma \mathrm{B_{NC}}$, where $\sigma$ is the Higgs boson production cross section, is shown in Table~\ref{tab:groups-of-NPs} for two representative mass points of cascade and non-cascade decays, (30, 80) and (40, 60)\GeV, respectively.
They are obtained by successively freezing different groups of nuisance parameters, and therefore represent their relative contributions to the total uncertainty in the measured signal strength.
The bin-by-bin statistical uncertainties are included together in the ``Statistical'' category.
The category ``Theoretical'' contains modeling uncertainties from the PDF, \alpS, and QCD scale in both signal and SM Higgs boson processes, as well as the uncertainty in the cross sections of \ttjets, single top, and diboson processes.
The category ``$\text{jet}\to\tauh$ normalization'' includes normalization uncertainties in the jet misidentified as \tauh background for the \mutau and \etau channels.
Other systematic uncertainties, including variations in final-state object identification, energy scale, etc., are grouped in the ``Other experimental'' category.
The dominant uncertainty stems from the limited size of the data in the SRs, followed by the normalization of the jet misidentified as \tauh background.

\begin{table}[ht]
	\centering
	\topcaption{Impact of different groups of uncertainties in the observed limits on the products $\sigma \mathrm{B_{C}}$ and $\sigma \mathrm{B_{NC}}$ using BDT-based SR categories. The contribution of each uncertainty group in the total signal strength uncertainty is given as a percentage. Once cascade and one non-cascade example is shown, where the \Paa and \Pab mass values are given in \GeV.}
	\begin{tabular}{lcc}
		Uncertainty group & \multicolumn{2}{c}{Impact on $\sigma \mathrm{B_{C}}$ and $\sigma \mathrm{B_{NC}}$ upper limits (in \%)} \\ 
		& Cascade (30, 80) & Non-cascade (40, 60) \\
	\hline
		Statistical                                  & 83           & 85          \\
		Systematic                                   & 56           & 53          \\
		\hspace{3mm}Theoretical                      & 10           & 18          \\
		\hspace{3mm}Jet$\to\tauh$ normalization      & 33           & 26          \\
		\hspace{3mm}Other experimental               & 45           & 42          \\
	\end{tabular}
	\label{tab:groups-of-NPs}
\end{table}

\section{Results}
\label{sec:results}
A binned maximum likelihood fit is performed on the \mtt distribution.
The SR yields are treated as Poisson distributed, while the systematic uncertainties with their correlations are introduced as nuisance parameters with log-normal or Gaussian function constraints.
No significant excess compatible with the probed signal model is observed over the predicted SM background.
Upper limits are set on the product of the cross section and branching fraction of the examined final states using the CMS statistical analysis tool \textsc{Combine}~\cite{CMS:2024onh}.
The ``LHC profile likelihood ratio''~\cite{CMS:2024onh} is taken as the test statistic, which is assumed to be distributed as in the asymptotic approximation~\cite{Cowan:2010js}.
The upper limits are dominated by the uncertainties in the normalization of the misidentified \tauh background and statistical fluctuations in the data.

\begin{figure}[ht!]
    \centering
    \includegraphics[width=0.42\textwidth]{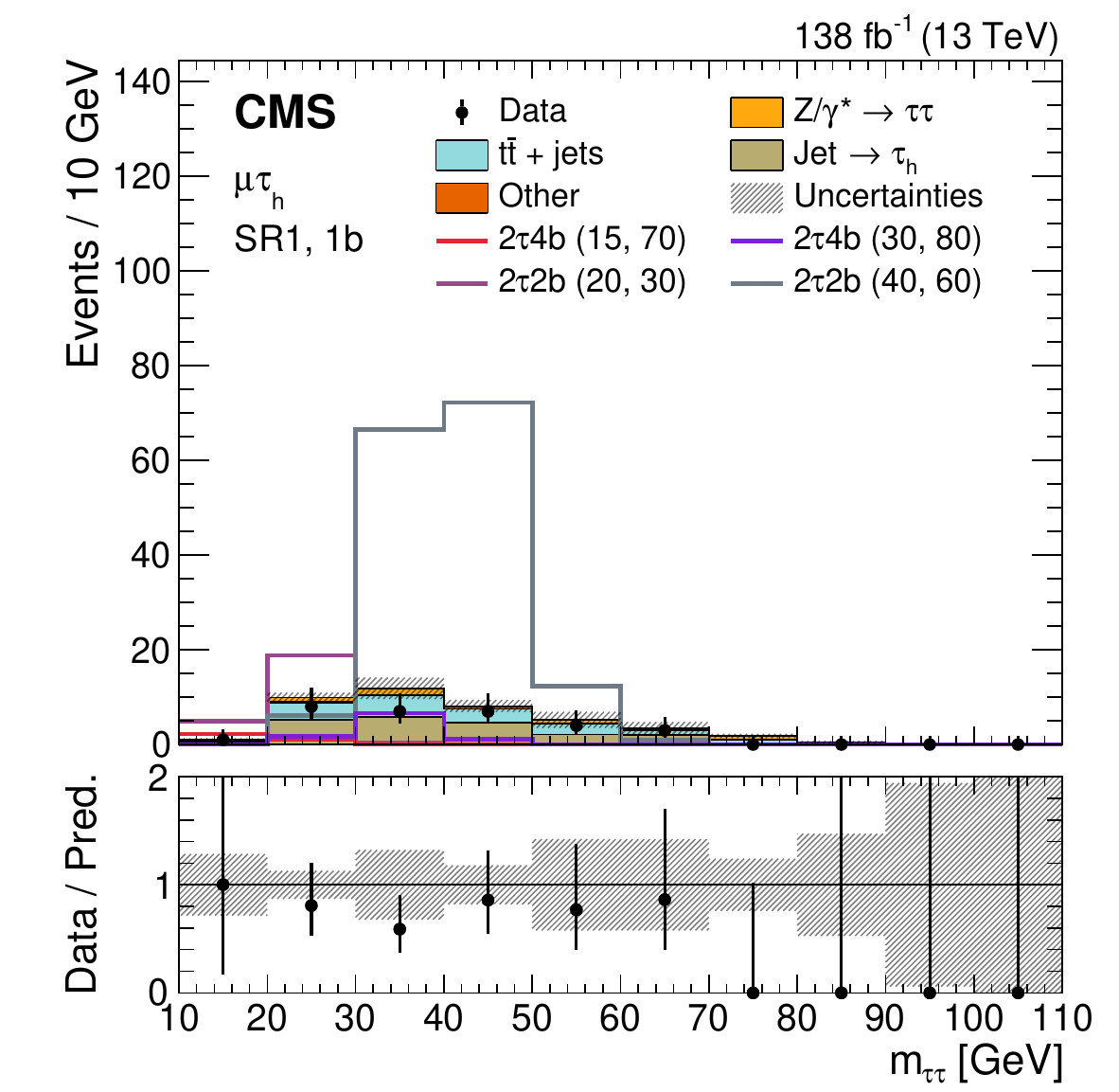}
    \includegraphics[width=0.42\textwidth]{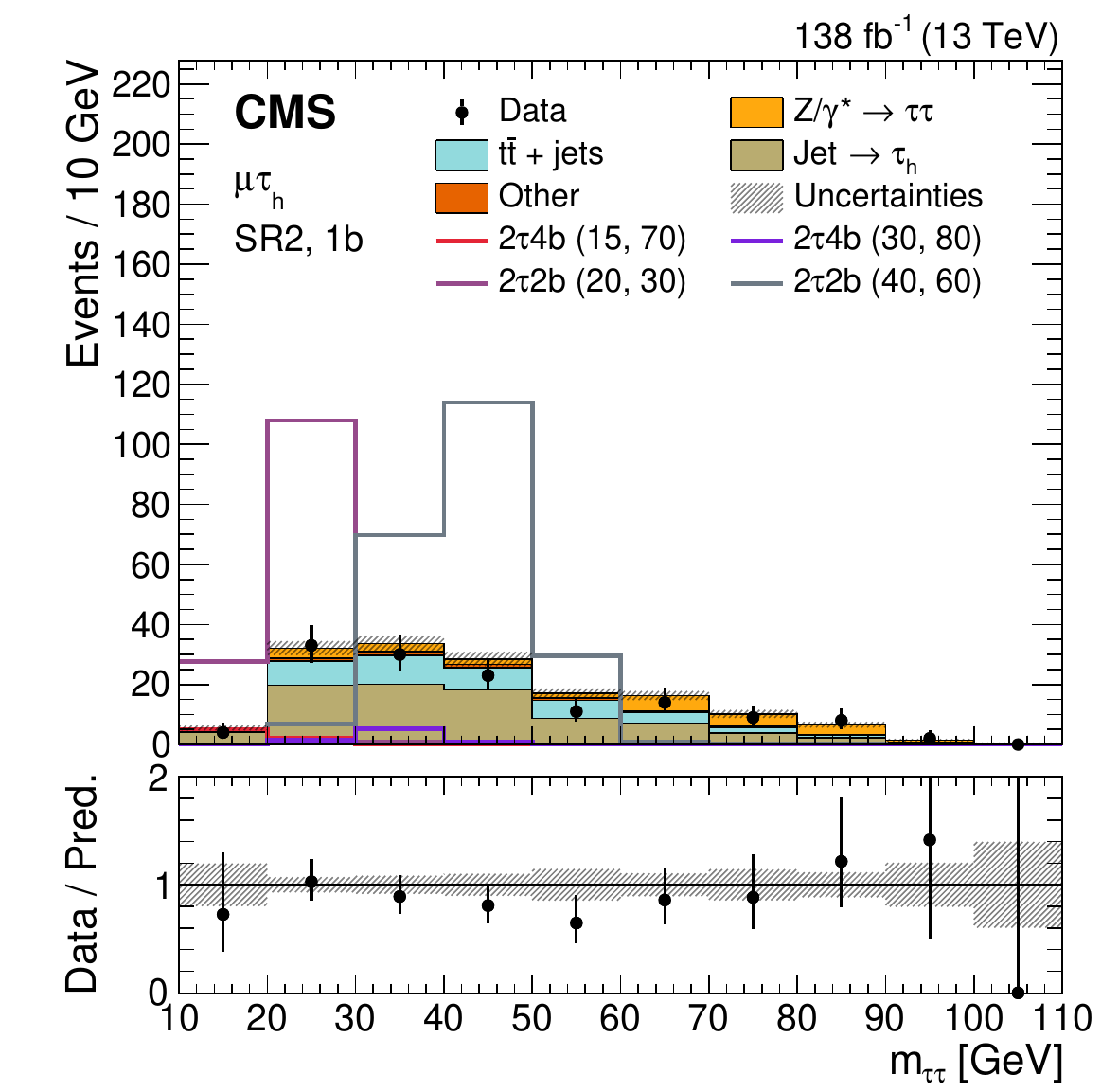} \\
    \includegraphics[width=0.42\textwidth]{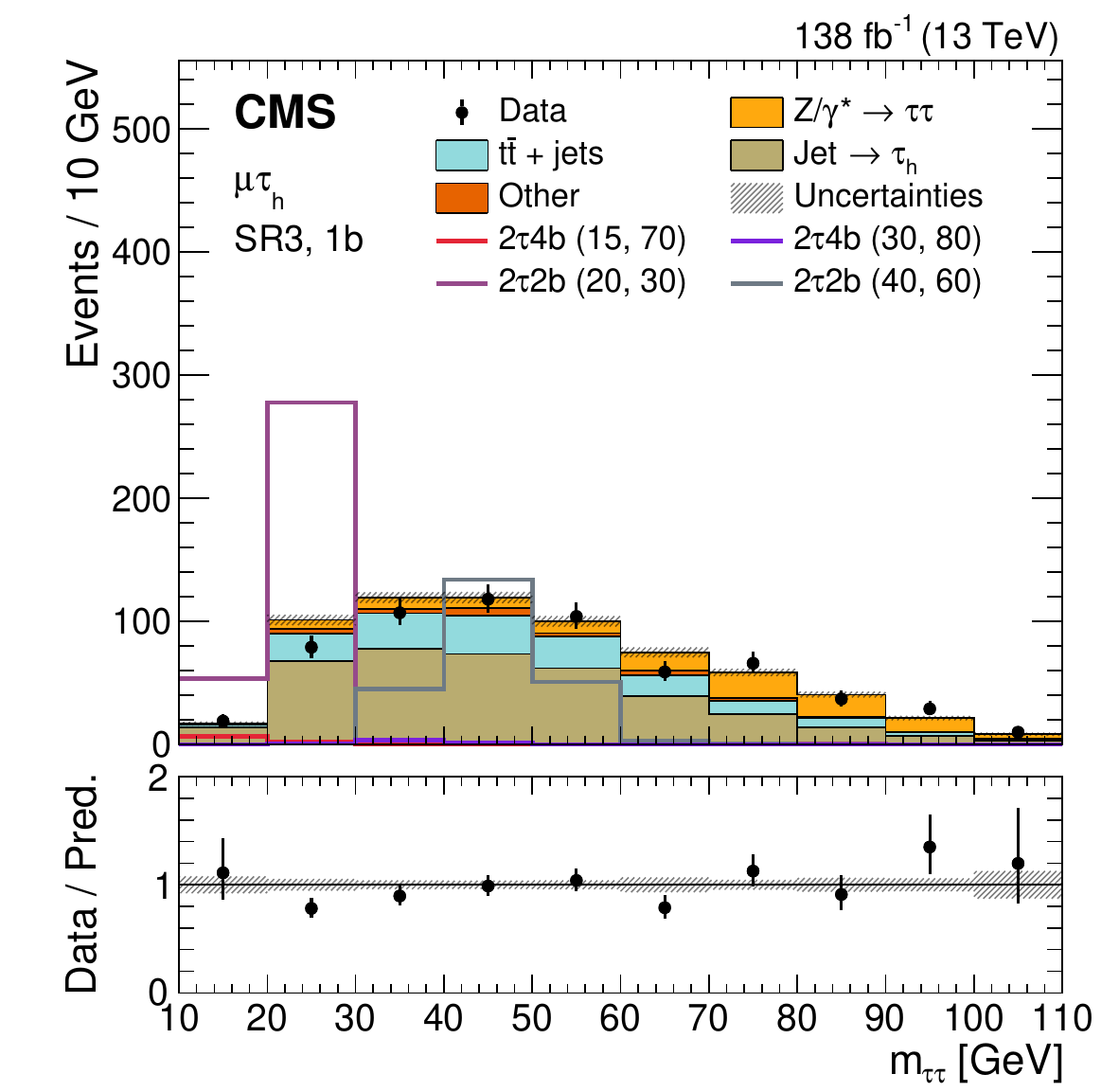}
    \includegraphics[width=0.42\textwidth]{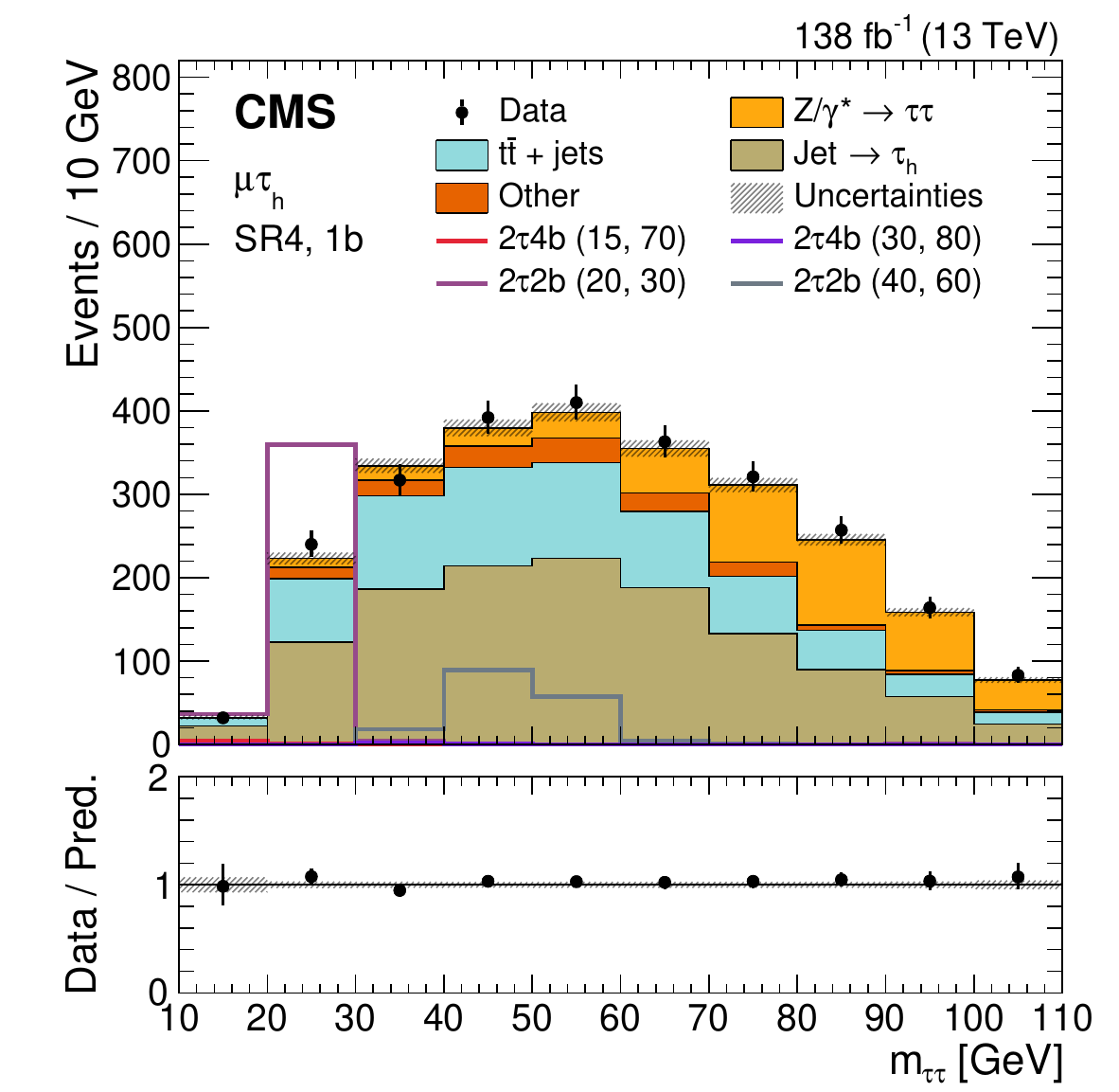} \\
    \includegraphics[width=0.42\textwidth]{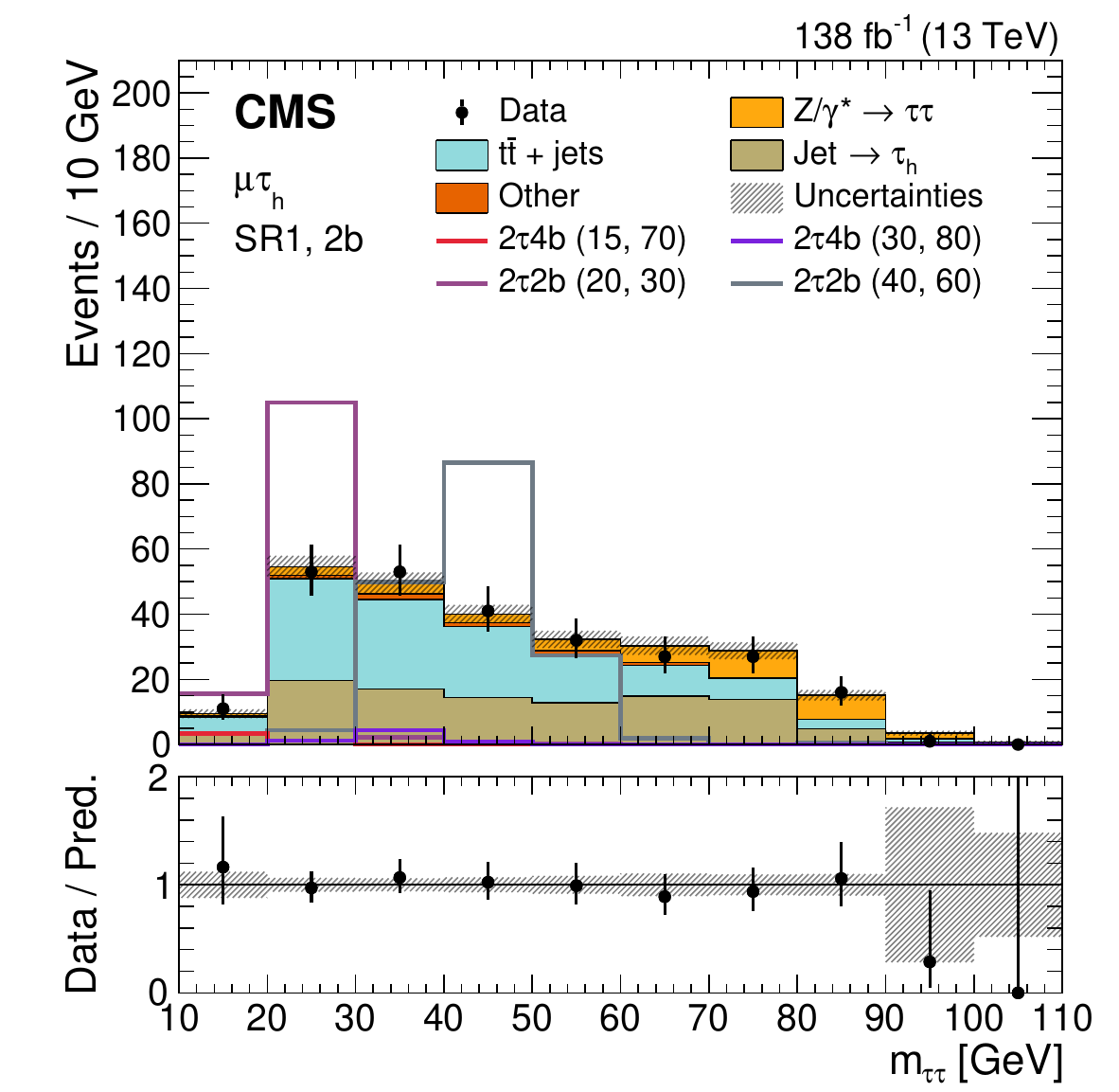}
    \includegraphics[width=0.42\textwidth]{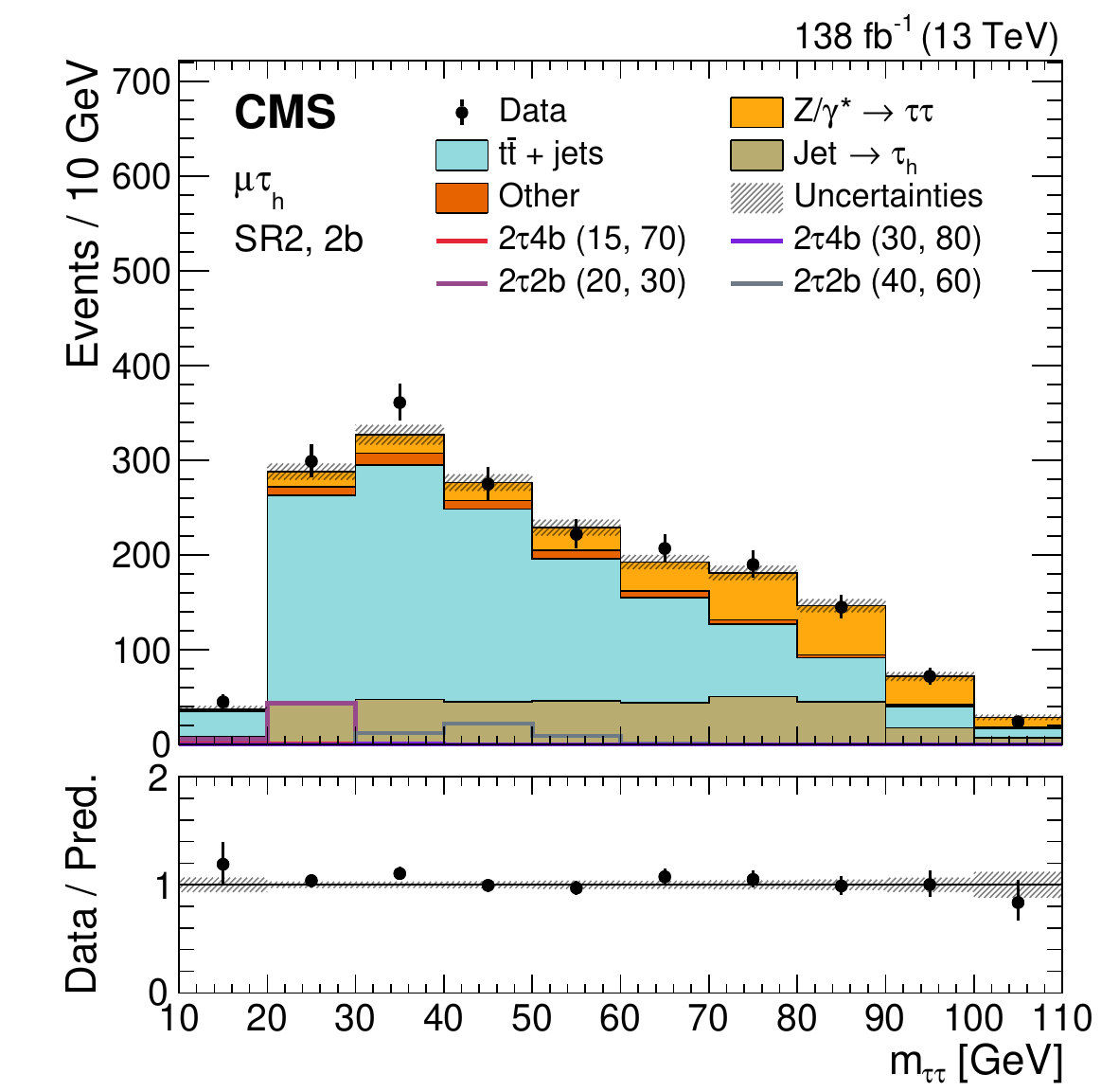}
    \caption{Background only, post-fit \mtt distributions for the \mutau channel, in events with exactly one \PQb-tagged jet: SR1 (upper left), SR2 (upper right), SR3 (middle left), and SR4 (middle right), and in events with at least two \PQb-tagged jets: SR1 (lower left) and SR2 (lower right). The data are shown by the markers with vertical bars and various backgrounds by the colored histograms. The total systematic uncertainty is shown by the hatched area. The colored open histograms display the predicted signal distribution for two cascade decays and two non-cascade decays, with four different values of \Paa and \Pab masses, for an assumed branching fraction of 100\%. The lower plot of each panel gives the ratio of the data to the sum of the predicted number of background events. The vertical bars display the statistical uncertainty in the ratio.}
    \label{fig:results_postfit_mutau_bdt}
\end{figure}

\begin{figure}[ht!]
    \centering
    \includegraphics[width=0.42\textwidth]{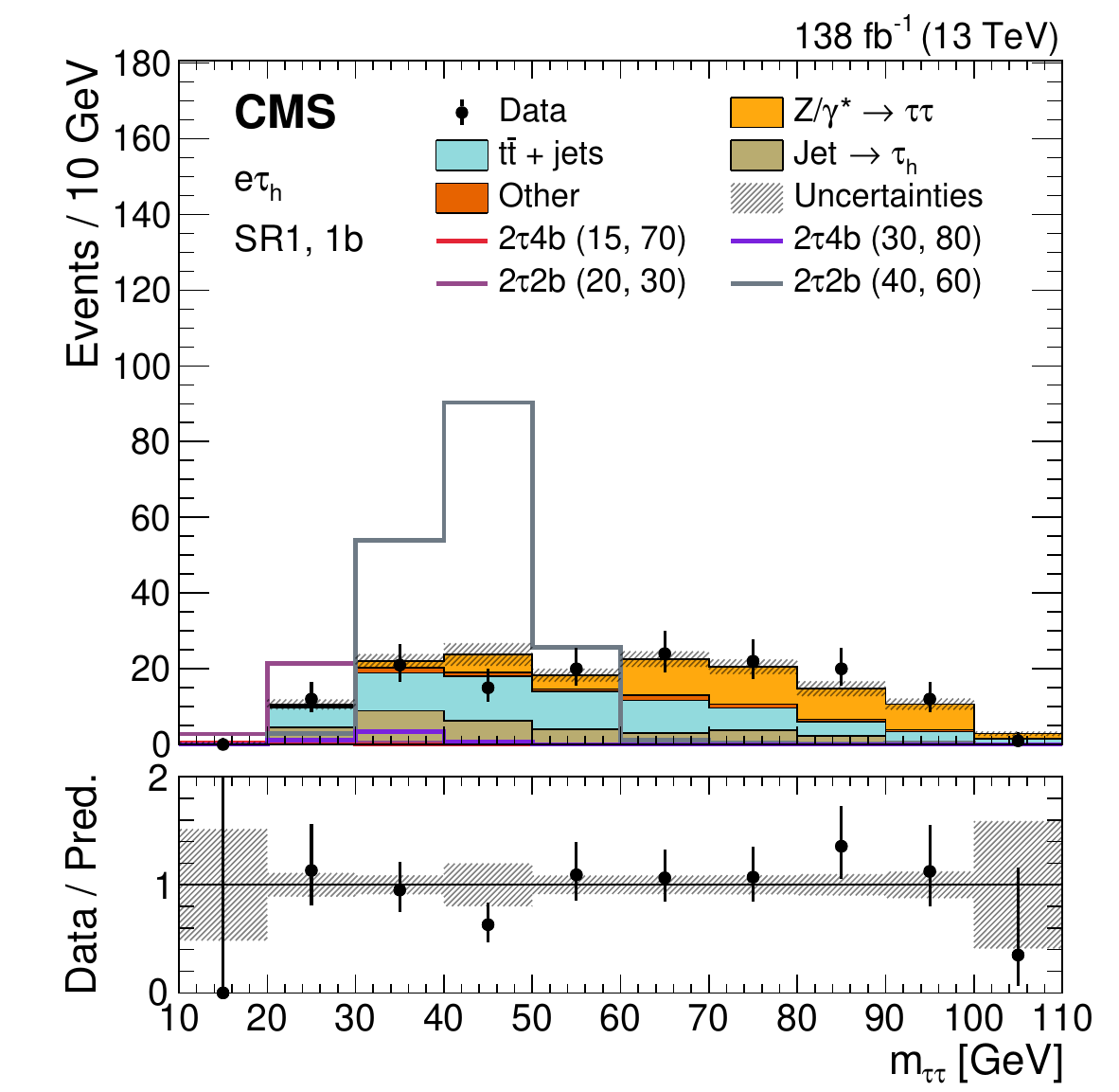}
    \includegraphics[width=0.42\textwidth]{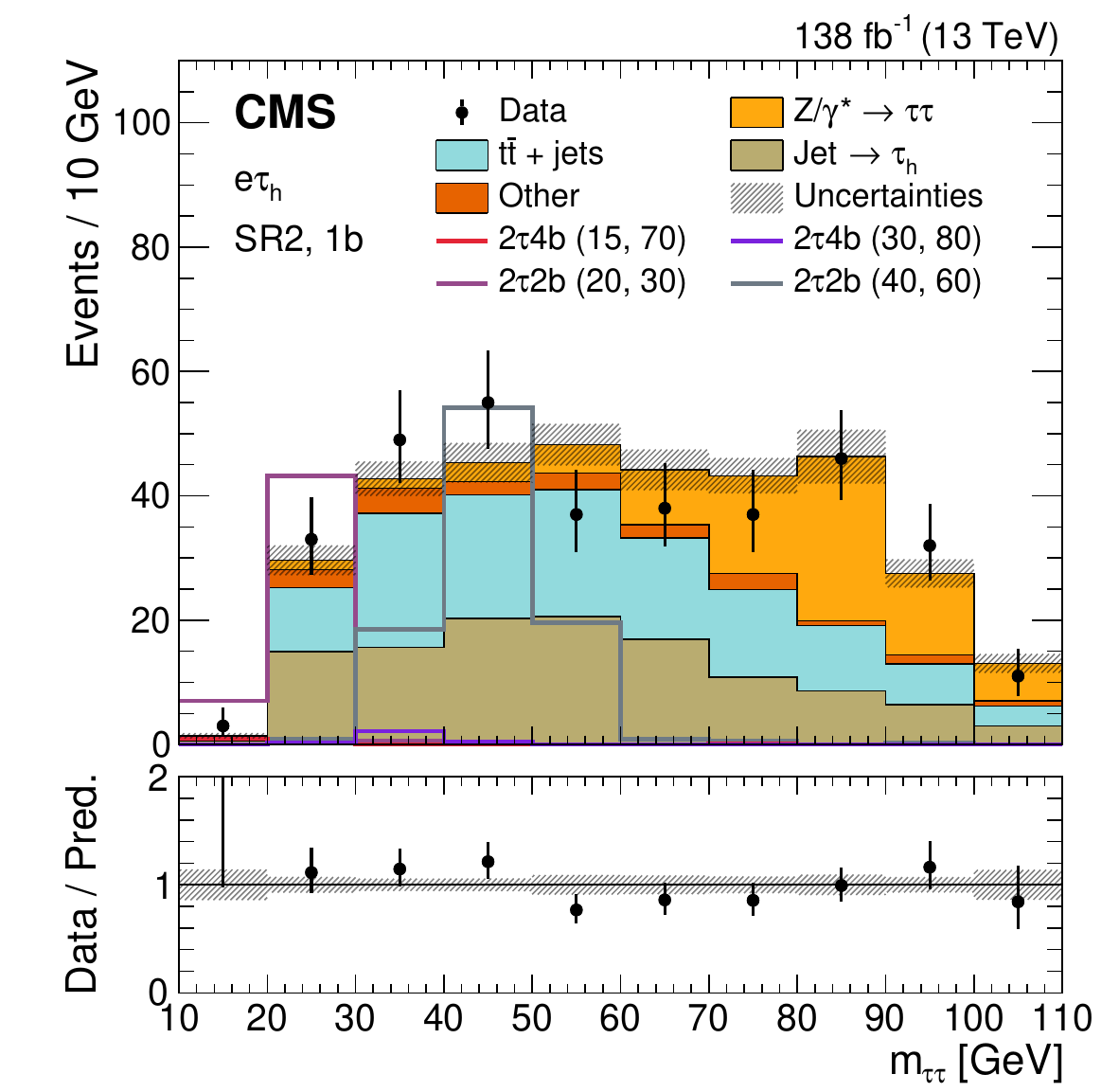} \\
    \includegraphics[width=0.42\textwidth]{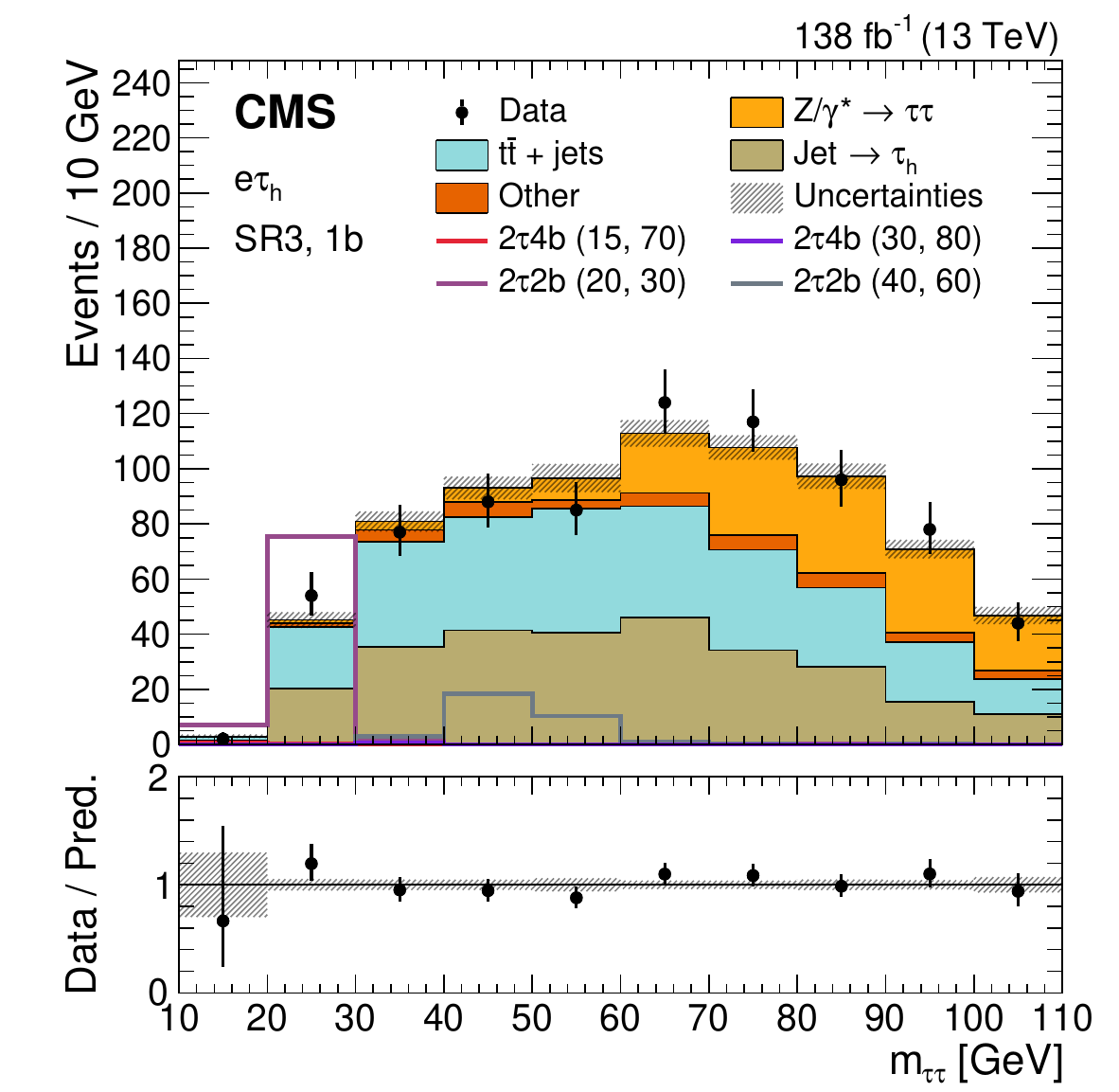}
    \includegraphics[width=0.42\textwidth]{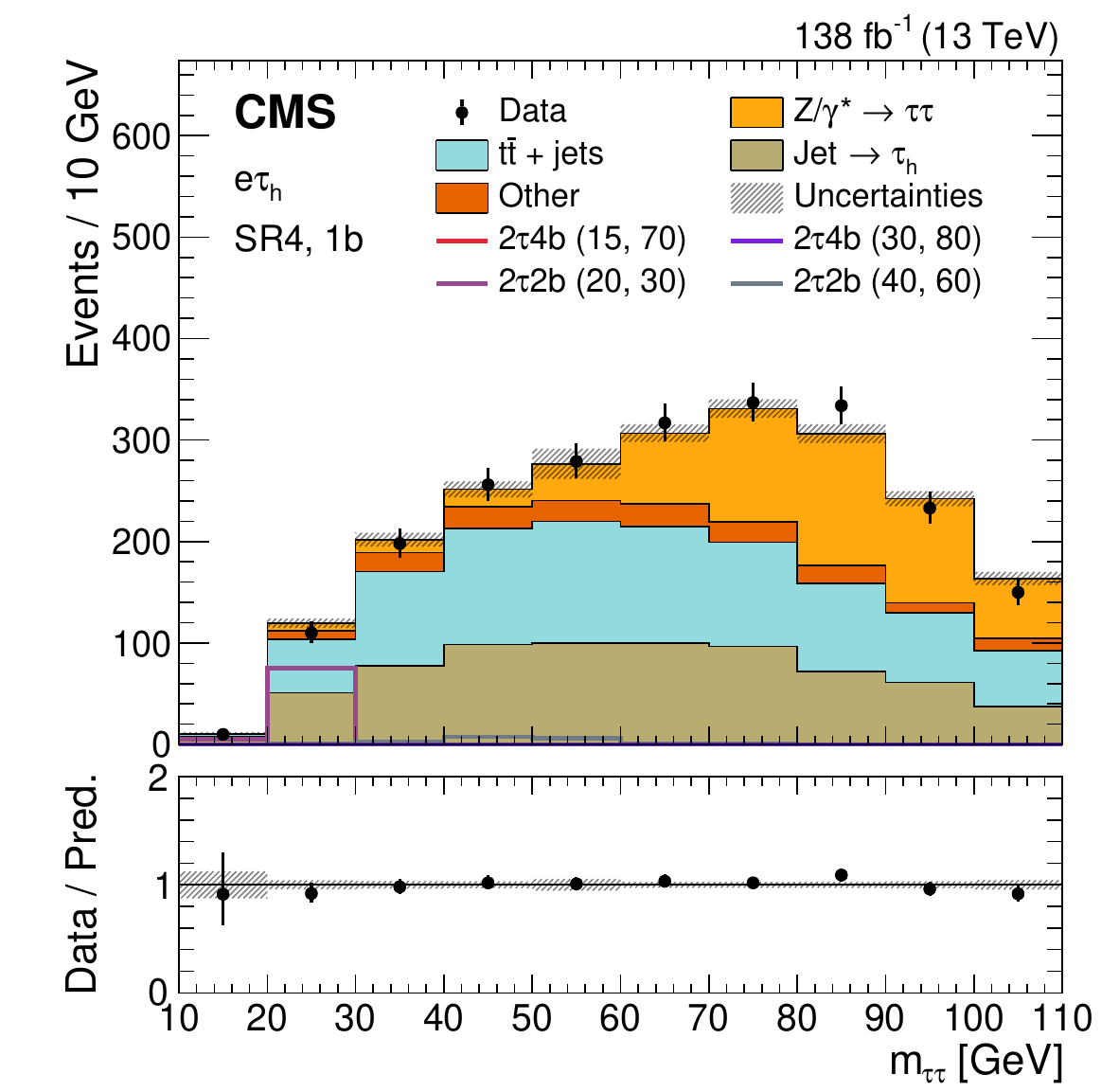} \\
    \includegraphics[width=0.42\textwidth]{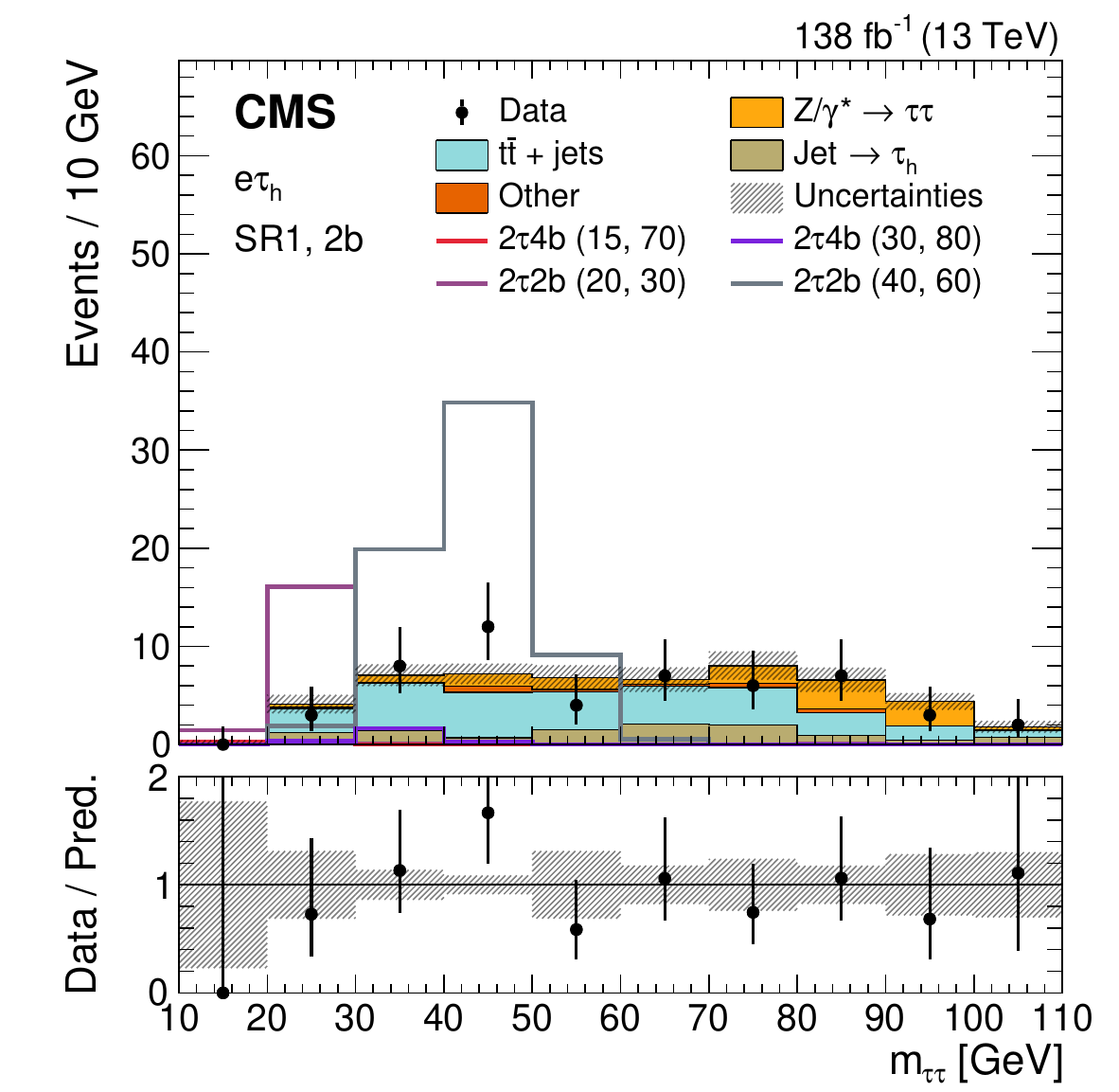}
    \includegraphics[width=0.42\textwidth]{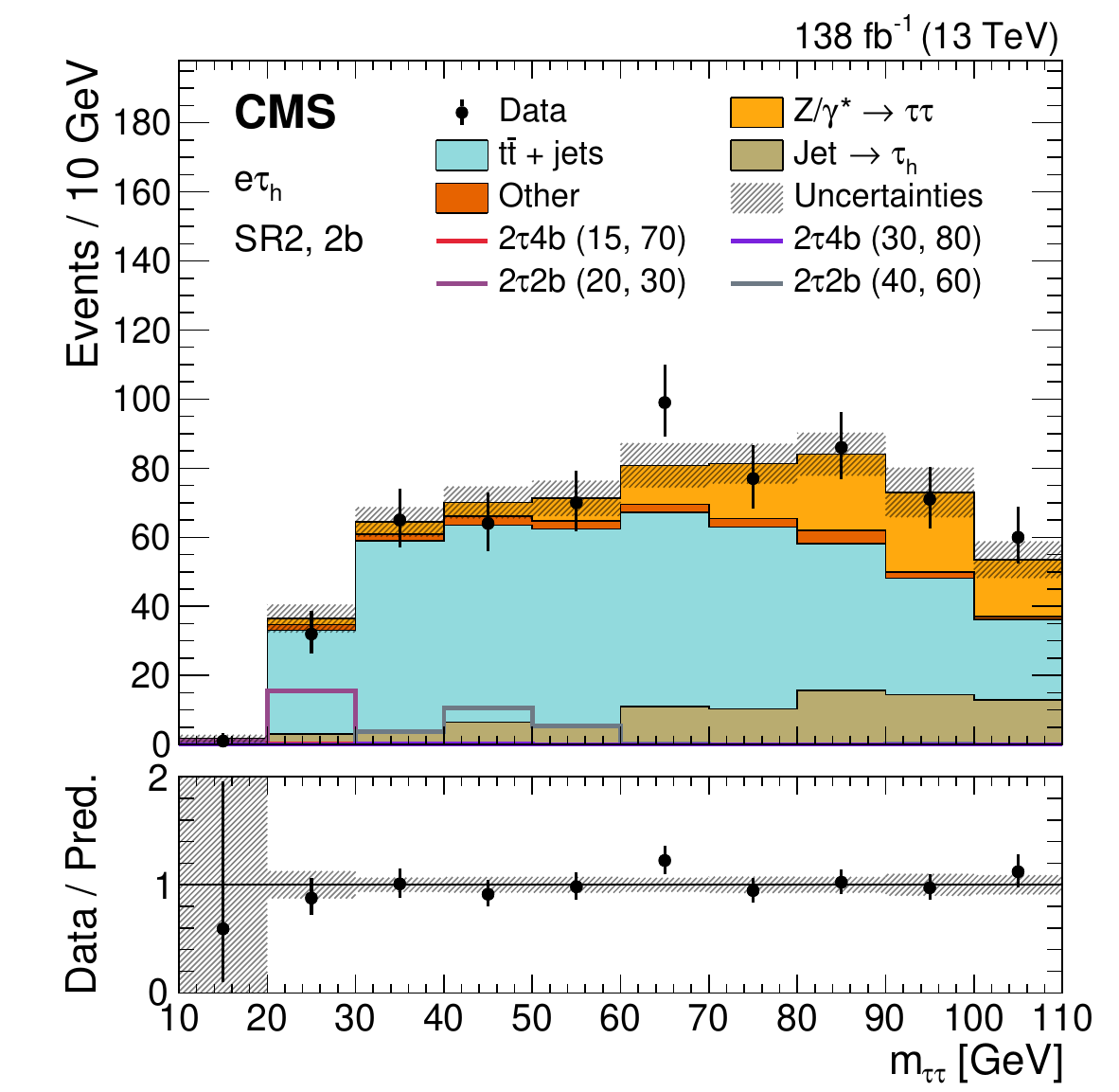} 
    \caption{Background only, post-fit \mtt distributions for the \etau channel, in events with exactly one \PQb-tagged jet: SR1 (upper left), SR2 (upper right), SR3 (middle left), and SR4 (middle right), and in events with at least two \PQb-tagged jets: SR1 (lower left) and SR2 (lower right). The data are shown by the markers with vertical bars and various backgrounds by the colored histograms. The total systematic uncertainty is shown by the hatched area. The colored open histograms display the predicted signal distribution for two cascade decays and two non-cascade decays, with four different values of \Paa and \Pab masses, for an assumed branching fraction of 100\%. The lower plot of each panel gives the ratio of the data to the sum of the predicted number of background events. The vertical bars display the statistical uncertainty in the ratio.}
    \label{fig:results_postfit_etau_bdt}
\end{figure}

\begin{figure}[ht!]
    \centering
    \includegraphics[width=0.42\textwidth]{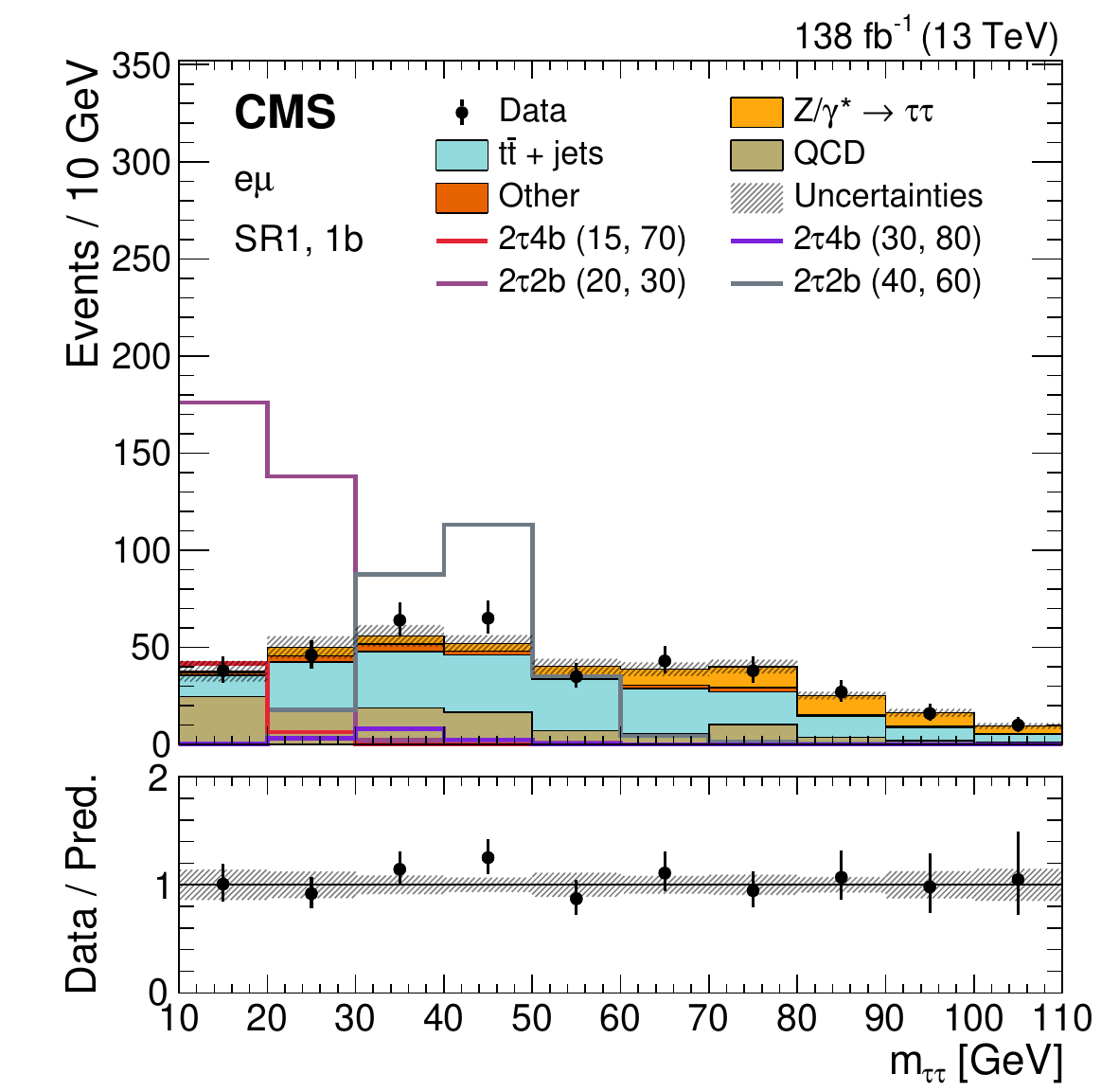}
    \includegraphics[width=0.42\textwidth]{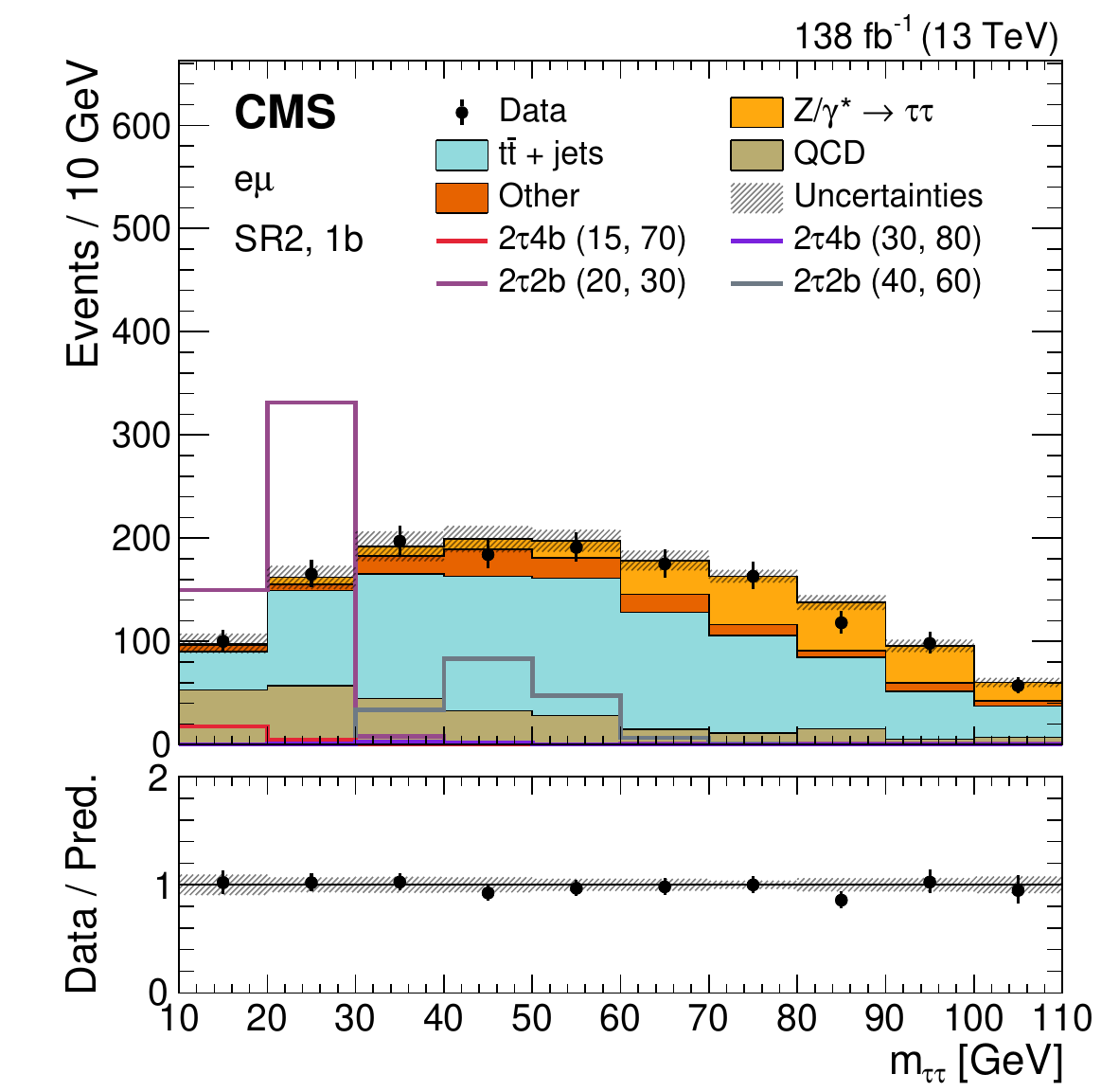} \\
    \includegraphics[width=0.42\textwidth]{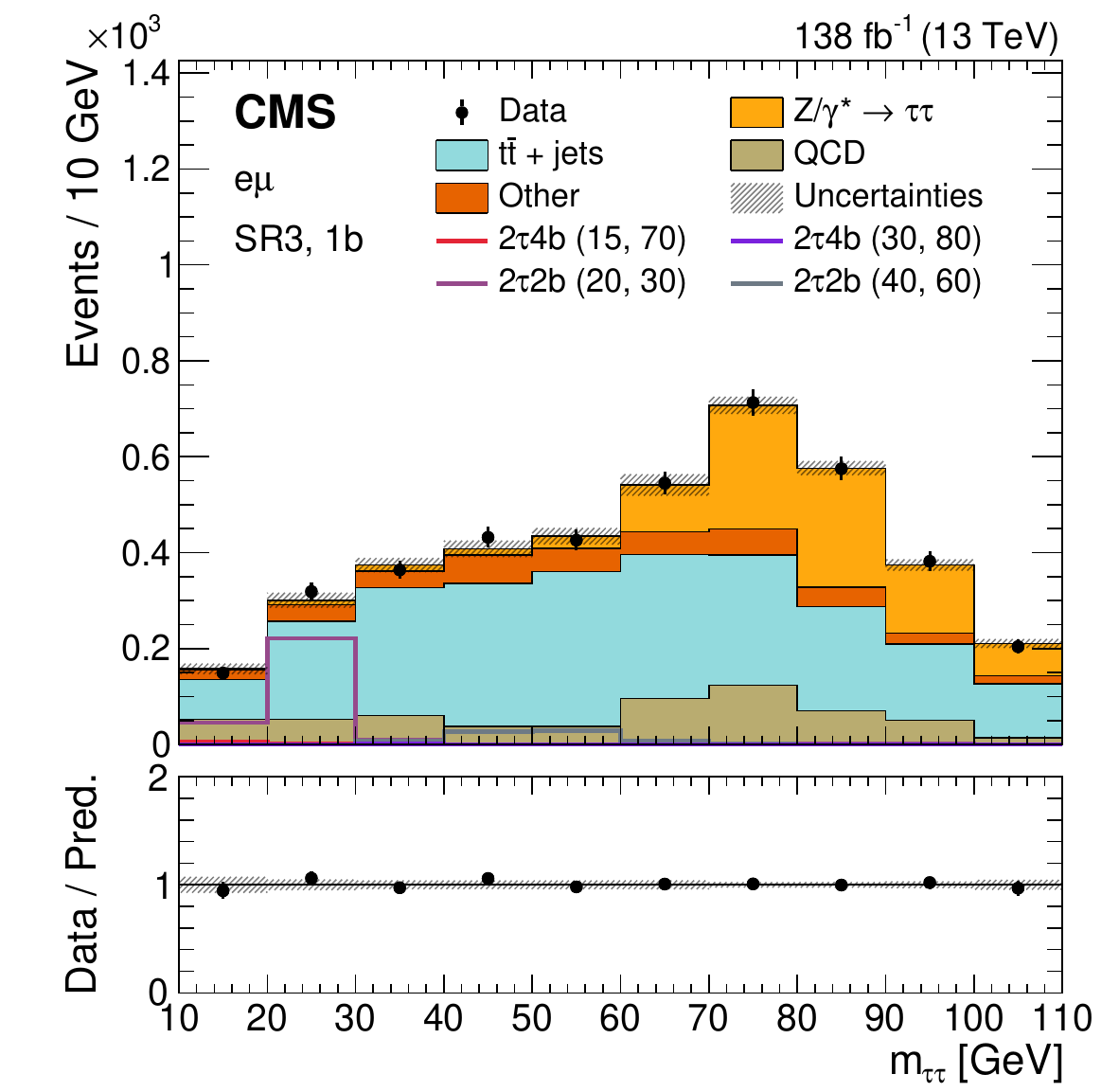}
    \includegraphics[width=0.42\textwidth]{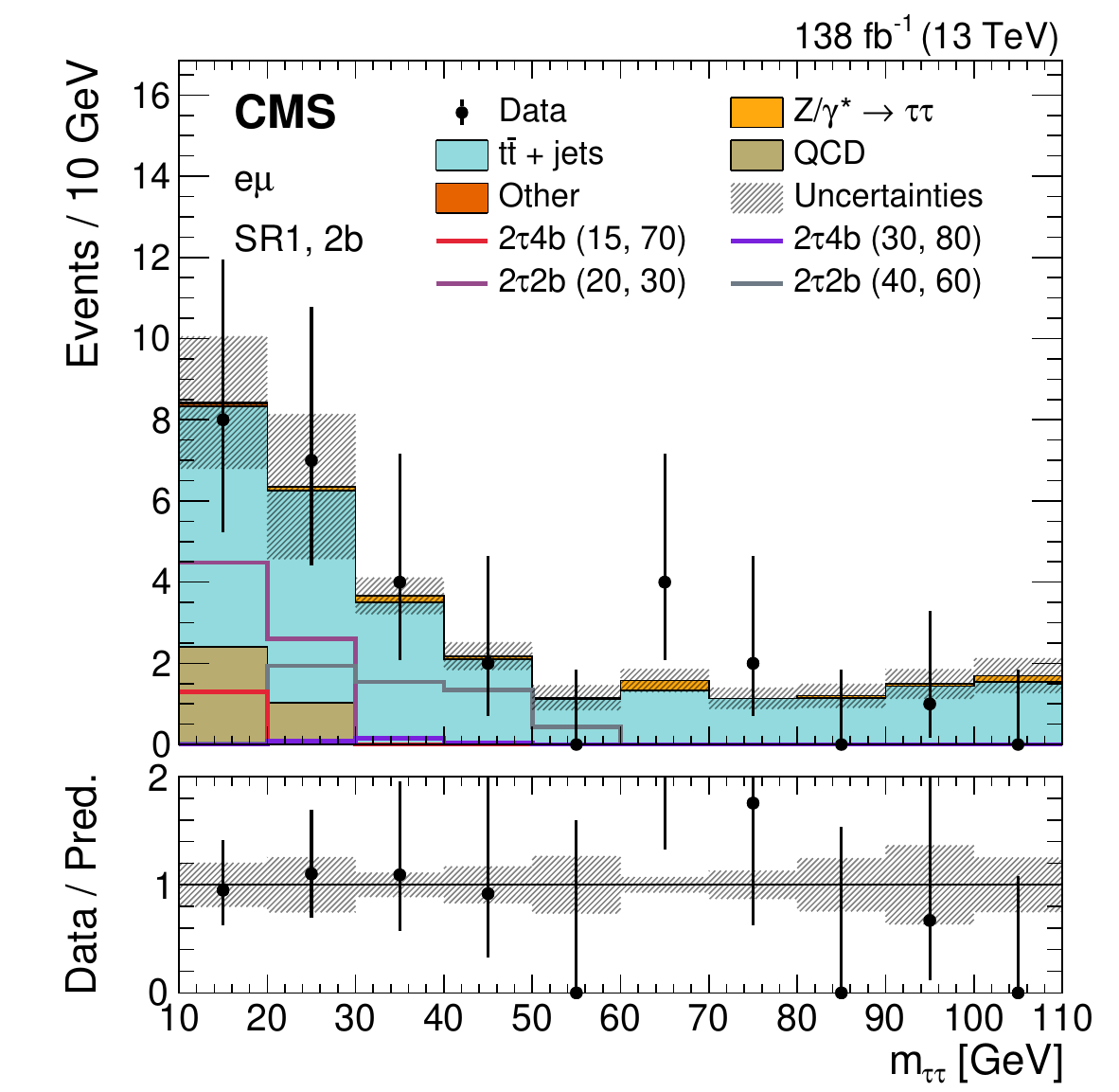} \\
    \includegraphics[width=0.42\textwidth]{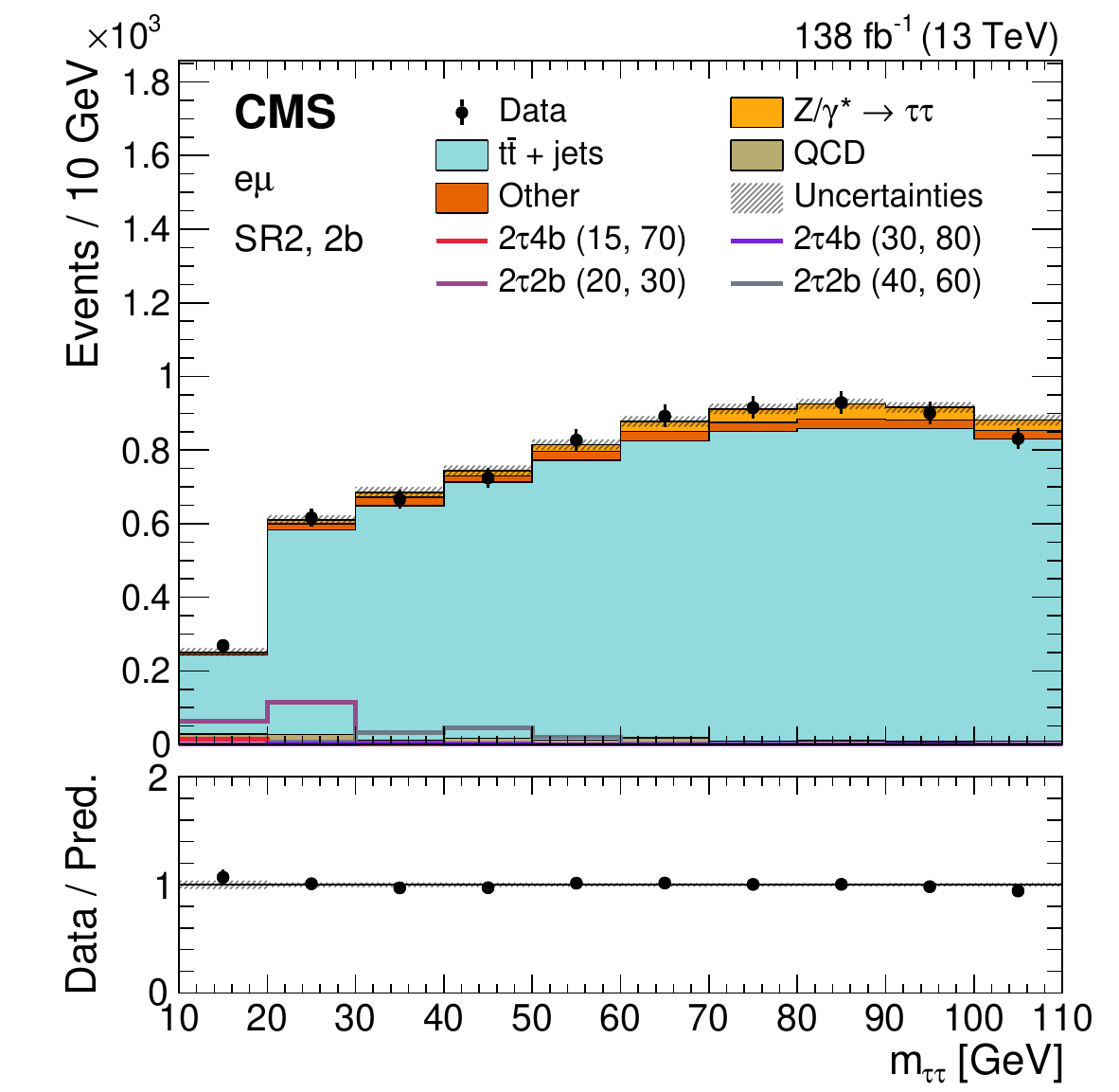}
    \caption{Background only, post-fit \mtt distributions for the \emu channel, in events with exactly one \PQb-tagged jet: SR1 (upper left), SR2 (upper right), and SR3 (middle left), and in events with at least two \PQb-tagged jets: SR1 (middle right) and SR2 (lower). The data are shown by the markers with vertical bars and various backgrounds by the colored histograms. The total systematic uncertainty is shown by the hatched area. The colored open histograms display the predicted signal distribution for two cascade decays and two non-cascade decays, with four different values of \Paa and \Pab masses, for an assumed branching fraction of 100\%. The lower plot of each panel gives the ratio of the data to the sum of the predicted number of background events. The vertical bars display the statistical uncertainty in the ratio.}
    \label{fig:results_postfit_emu_bdt}
\end{figure}

Figures~\ref{fig:results_postfit_mutau_bdt}--\ref{fig:results_postfit_emu_bdt} show the \mtt distributions in the BDT-based SR categories after performing a background-only fit (``post-fit'').
Representative signal distributions are overlaid assuming a 100\% branching fraction for the Higgs boson decay into $2\PGt4\PQb$ for cascade mass hypotheses (15, 70) and (30, 80)\GeV, and into $2\PGt2\PQb$ for non-cascade mass hypotheses (20, 30) and (40, 60)\GeV.

\clearpage

\begin{figure}[ht!]
    \centering
    \includegraphics[width=\textwidth]{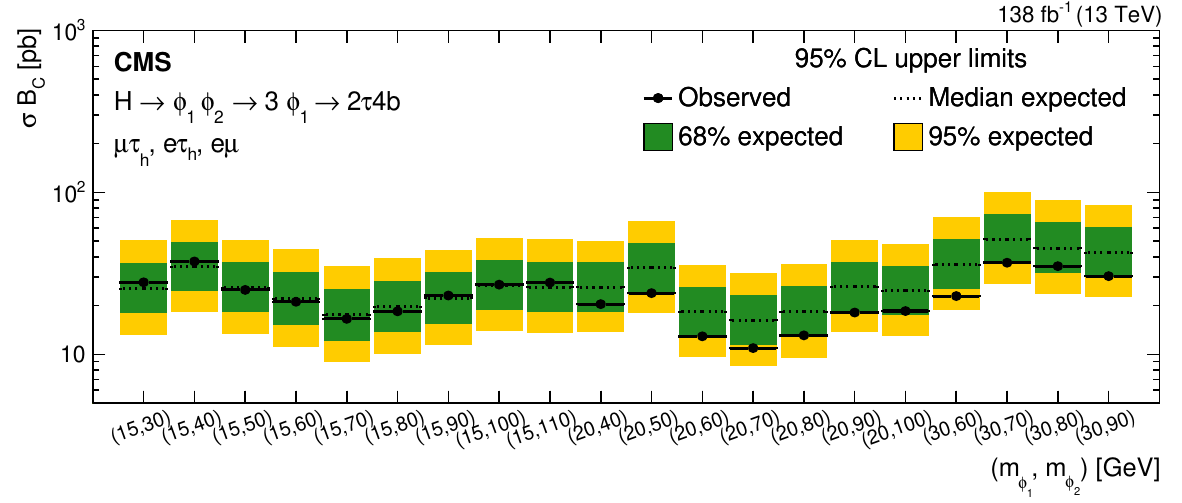}   
    \caption{The observed (points) and median expected (dotted line) 95\% \CL upper limits on the product $\sigma \mathrm{B_{C}}$ for the cascade scenario using the BDT-based event categorization and the fit to the \mtt distribution, for different mass hypotheses (\maa, \mab). The horizontal bars on the points are for better legibility only. The green and yellow regions show the 68 and 95\% expected range for the median value, respectively.}
    \label{fig:bdt_limits_cascade_run2}
\end{figure}

\begin{figure}[ht!]
    \centering
    \includegraphics[width=\textwidth]{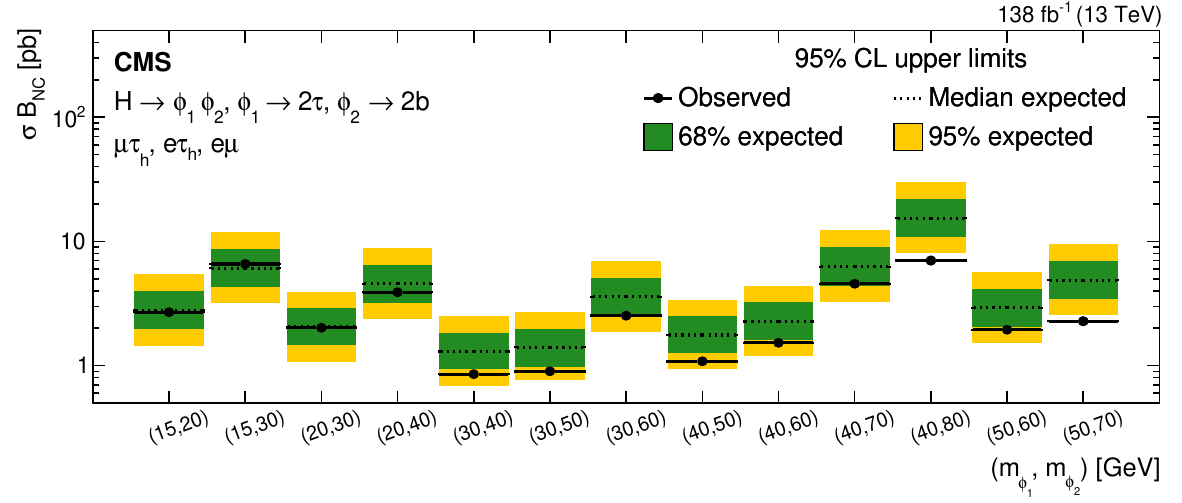}   
    \caption{The observed (points) and median expected (dotted line) 95\% \CL upper limits on the product $\sigma \mathrm{B_{NC}}$ for the non-cascade scenario using the BDT-based event categorization and the fit to the \mtt distribution, for different mass hypotheses (\maa, \mab). The horizontal bars on the points are for better legibility only. The green and yellow regions show the 68 and 95\% expected range for the median value, respectively.}
    \label{fig:bdt_limits_noncascade_run2}
\end{figure}

The 95\% \CL upper limits on the product $\sigma \mathrm{B_{C}}$ for cascade decays and $\sigma \mathrm{B_{NC}}$ for non-cascade decays are shown by the points in Figs.~\ref{fig:bdt_limits_cascade_run2} and \ref{fig:bdt_limits_noncascade_run2}, respectively, for all the considered mass hypotheses.
Figures~\ref{fig:bdt_limits_cascade_run2} and \ref{fig:bdt_limits_noncascade_run2} also display the median expected limits from simulation and the 68 and 95\% \CL regions for the median expected limits.
The decay kinematic distributions and the background shapes depend on the \Paa and \Pab masses, leading to the observed variations in the sensitivity for different mass hypotheses.

For some of the probed mass hypotheses, such as (50, 70) and (40, 80)\GeV in the non-cascade scenario, the observed upper limits are respectively 2.9 and 2.8 standard deviations lower than the expected limit using the BDT-based results.
This is attributed to the deficit in data in the SR bin where a peak in \mtt is expected, as shown in Fig.~\ref{fig:results_postfit_mutau_bdt}.

\begin{figure}[ht!]
    \centering
    \includegraphics[width=0.49\textwidth]{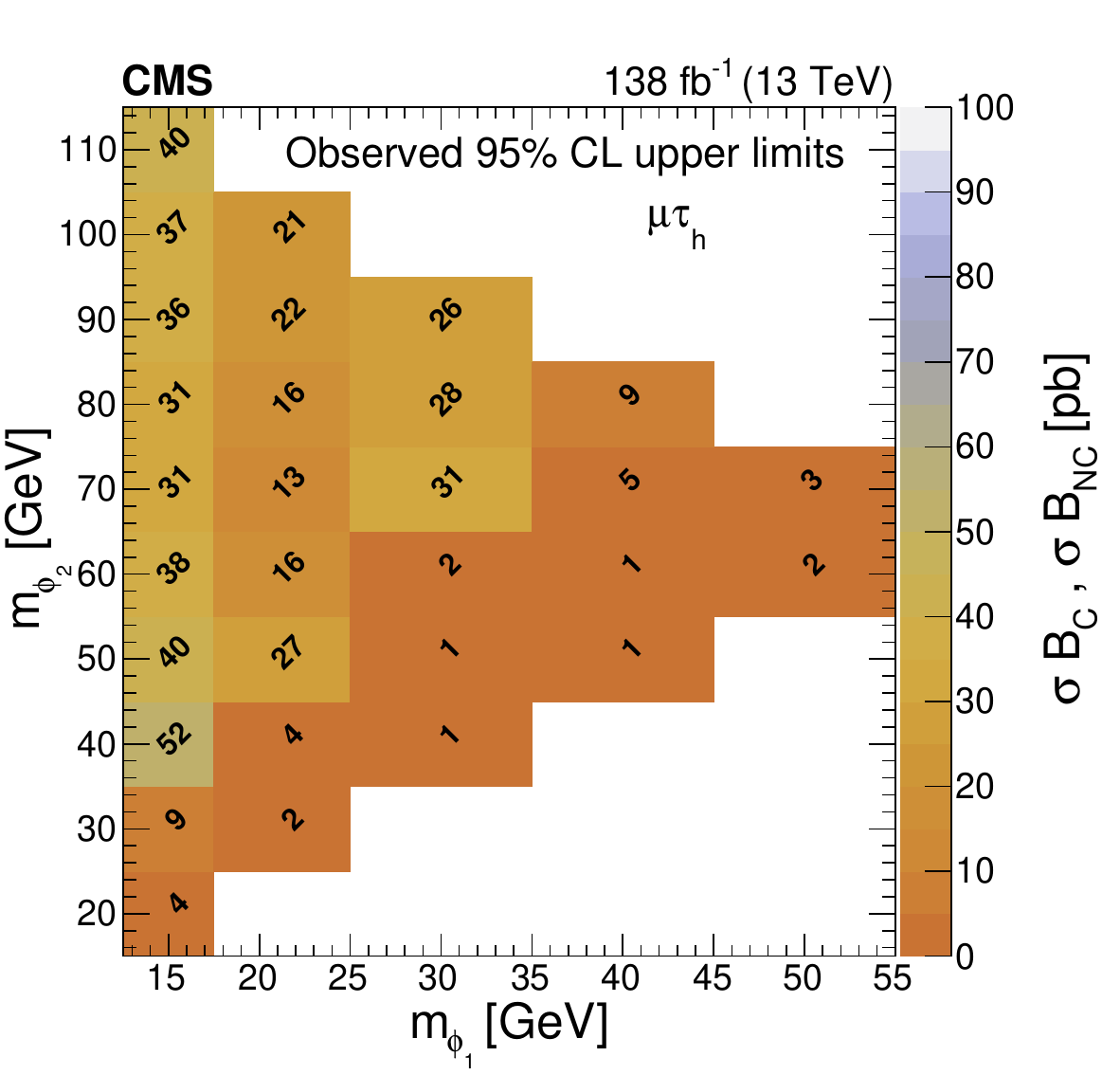}
    \includegraphics[width=0.49\textwidth]{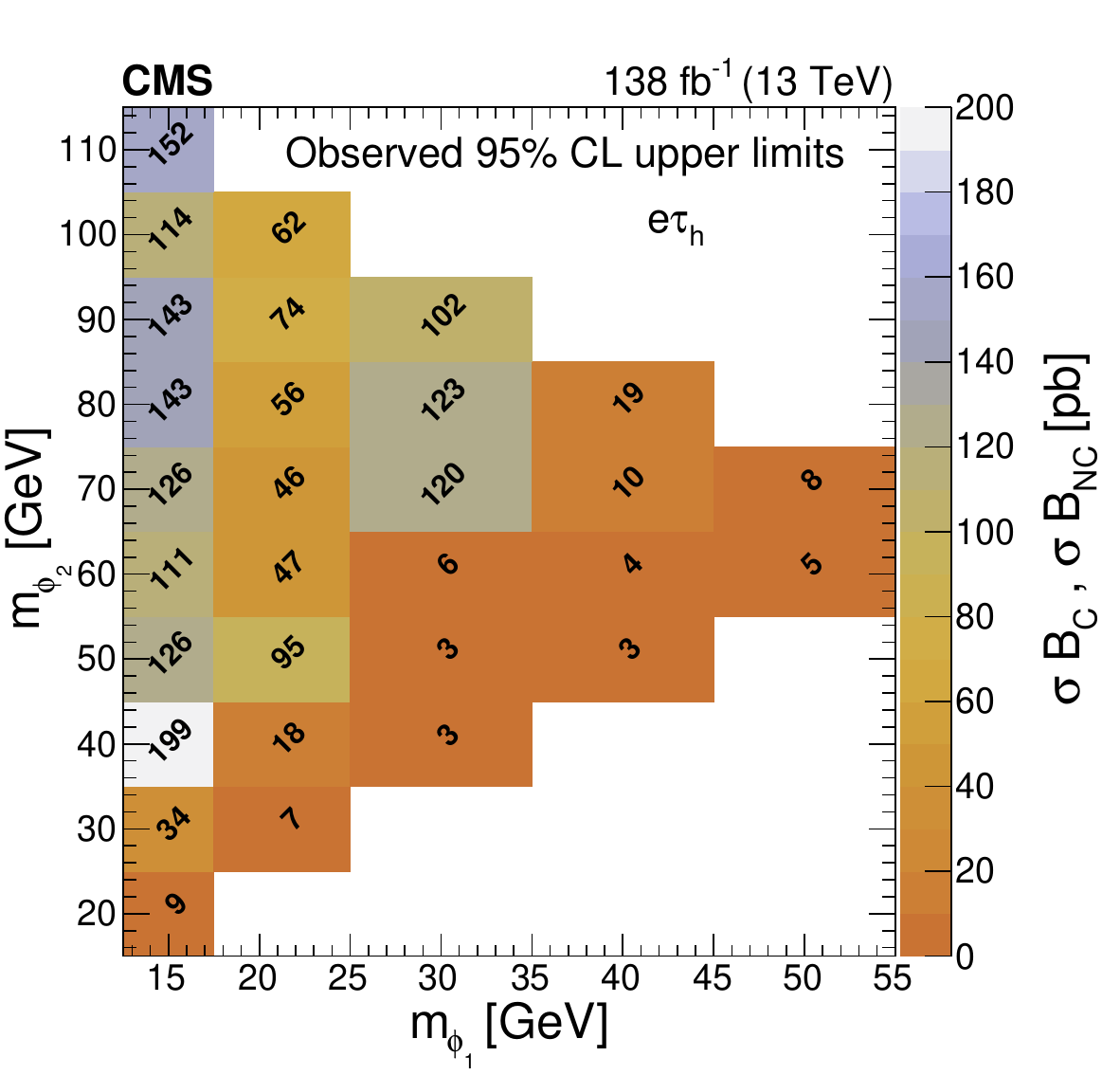} \\
    \includegraphics[width=0.49\textwidth]{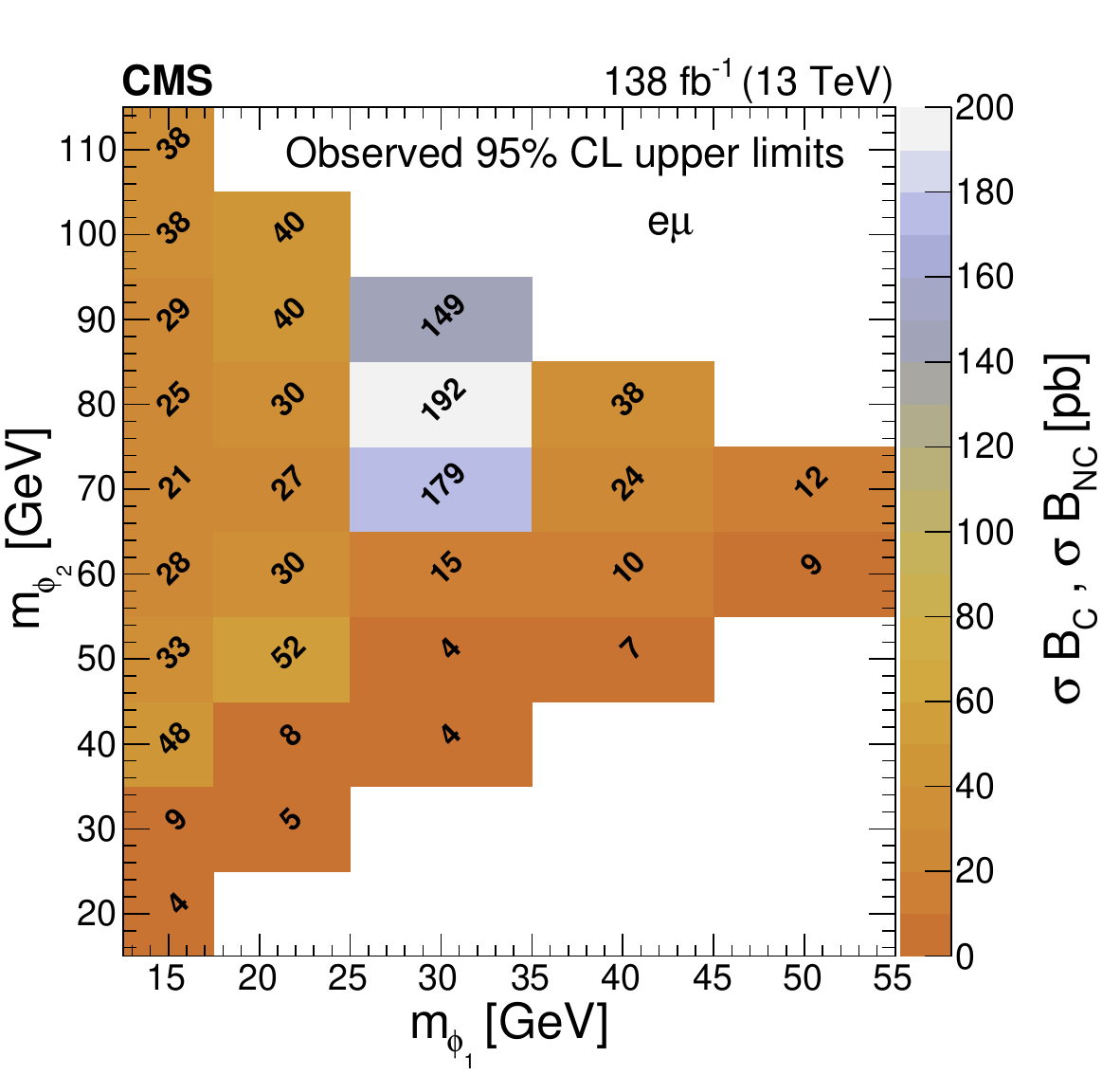}
    \includegraphics[width=0.49\textwidth]{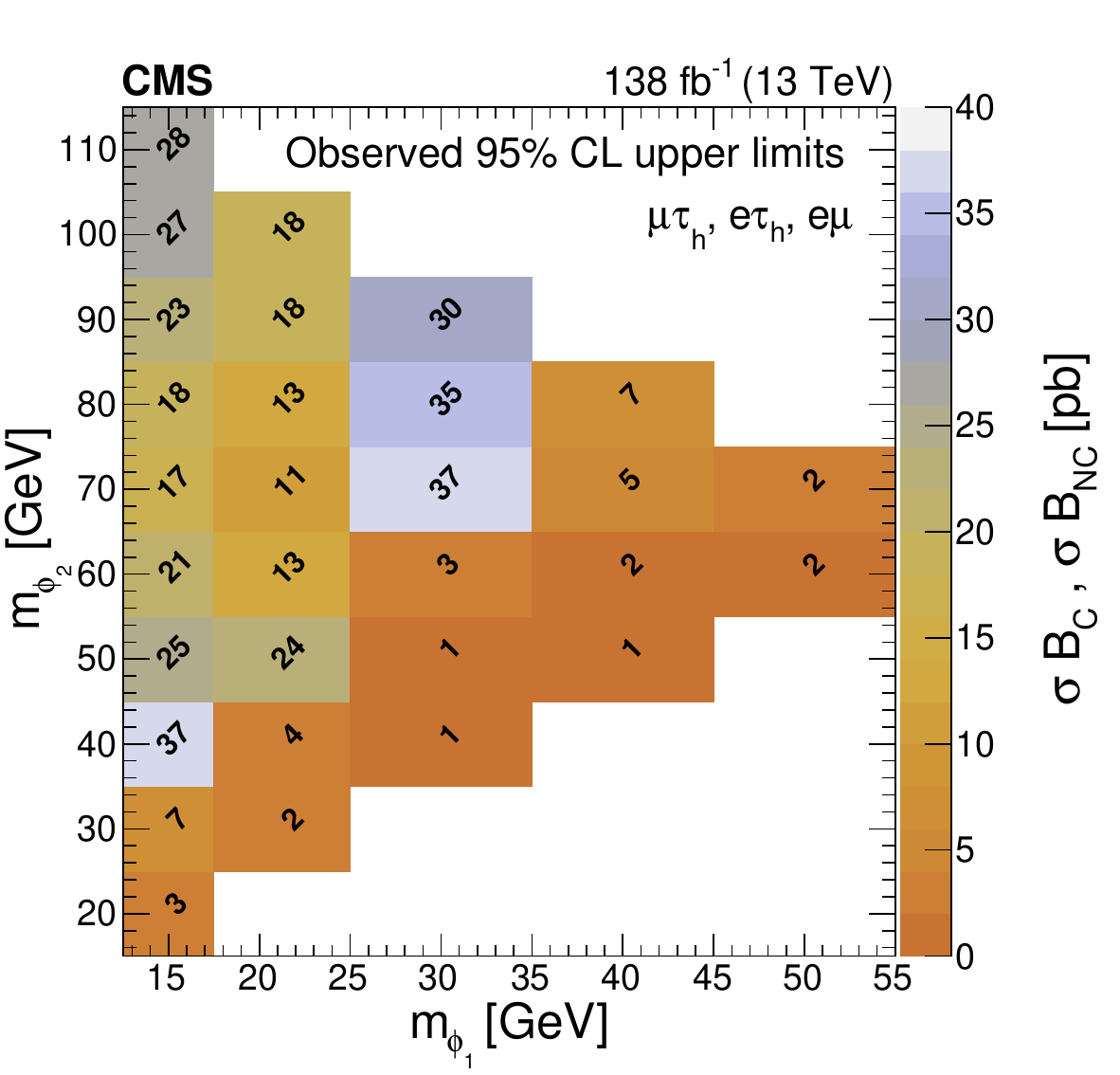}
    \caption{The 95\% \CL upper limits on the products $\sigma \mathrm{B_{C}}$ and $\sigma \mathrm{B_{NC}}$, obtained using the BDT-based event categorization, as a function of the scalar masses $\maa$ and $\mab$. For the (\Paa, \Pab) mass hypotheses (15, 30), (20, 40), and (30, 60)\GeV, only the non-cascade limits are shown. For all other mass hypotheses, either cascade or non-cascade limits are presented, depending on whether the cascade decay is kinematically allowed ($m_{\Pab} \geq 2 m_{\Paa}$). The numbers displayed in the plot are in \unit{pb}.}
    \label{fig:limits_bdtbased}
\end{figure}

Figure~\ref{fig:limits_bdtbased} illustrates the observed 95\% \CL upper limit on the products $\sigma \mathrm{B_{C}}$ and $\sigma \mathrm{B_{NC}}$ for the different mass hypotheses, found from the combination of the three individual channels.
The specific (\Paa, \Pab) mass hypotheses (15, 30), (20, 40), and (30, 60)\GeV are evaluated for both cascade and non-cascade scenarios; however, Fig.~\ref{fig:limits_bdtbased} depicts only the non-cascade limits for these three mass hypotheses as they are lower than the cascade limits.

The loss in sensitivity for certain mass hypotheses is a result of using an admixture of signal samples having different \maa and \mab for the BDT training, which has weaker discriminating power for signals with kinematic distributions closer to the backgrounds.
The analysis is sensitive to non-cascade decays of the neutral scalars, with the upper limits on the branching fraction reaching ${\approx}1$\% for some of the mass hypotheses.
The observed limits shown in Fig.~\ref{fig:bdt_limits_noncascade_run2} correspond to constraints on the product $\sigma \mathcal{B}(\PH \to \Paa \Pab)\ \mathcal{B}(\Paa \to \PQb\PQb)\ \mathcal{B}(\Pab \to \PGt\PGt)$, and can be interpreted as sensitive to scenarios where $\mathcal{B}(\Paa \to \PQb\PQb)$ and $\mathcal{B}(\Pab \to \PGt\PGt)$ are large.
For the most sensitive non-cascade mass hypothesis (30, 40)\GeV, the observed upper limit on $\mathrm{B_{NC}}$ is $1.6\%$, assuming the theoretical SM Higgs boson production cross section in the ${\Pg\Pg}$F and VBF processes of 52.36\unit{pb}~\cite{deFlorian:2016spz}.
The TRSM BP1~\cite{Robens_2020, Robens:2023oyz} predicts $\mathcal{B}(\Paa \to \PQb\PQb) = 86.6\%$ and $\mathcal{B}(\Pab \to \PGt\PGt) = 6.9\%$.
Using these branching fractions, the upper limit on $\mathcal{B}(\PH \to \Paa \Pab)$ for this mass point is 27.3\%.
The corresponding upper limit on the product $\sigma \mathrm{B_{NC}}$ is 0.9\unit{pb}, compared to the TRSM BP1 prediction of 0.18\unit{pb}.
While we cannot rule out the full TRSM BP1 parameter space, we exclude $\mathcal{B}(\PH \to \Paa \Pab)$ values above 27.3\% for this mass hypothesis.
The analysis is also sensitive to non-cascade mass hypotheses (15, 20), (20, 30), (30, 40), (30, 50), (30, 60), (40, 50), (40, 60), and (50, 70)\GeV.
These results are model independent, and specific models can be tested by comparing their predicted branching fractions to these limits.

Results obtained using the cut-based categorization can be found in Appendix~\ref{appendix}.
The BDT-based analysis improves the sensitivity with respect to the cut-based approach by about 20--30\%, depending on the signal model and mass hypothesis.

\section{Summary}
A search for exotic decays of the Higgs boson (\PH) into a pair of light neutral scalars \Paa and \Pab with masses $\mab > \maa$, has been presented.
Final states with at least one jet coming from the fragmentation of a \PQb quark and two tau leptons are studied.
The search utilizes a data sample of proton-proton collisions corresponding to an integrated luminosity of \il, accumulated by the CMS experiment at the LHC during 2016--2018 at a center-of-mass energy of 13\TeV.
The results are dominated by the statistical uncertainties and the systematic uncertainties arising from the normalizations of the backgrounds.
Overall, no statistically significant deviation from the expected standard model background prediction is observed, and upper limits are set on the products $\sigma \mathrm{B_{C}}$ and $\sigma \mathrm{B_{NC}}$ for the cascade and non-cascade decay scenarios, respectively, where $\sigma$ is the Higgs boson production cross section, $\mathrm{B_{C}}=\mathcal{B}(\PH \to \Paa \Pab \to 3 \Paa \to 2\PGt4\PQb)$, and $\mathrm{B_{NC}} = \mathcal{B}(\PH \to \Paa \Pab)\ \mathcal{B}(\Paa \to 2\PGt)\ \mathcal{B}(\Pab \to 2\PQb)$.
This analysis is sensitive to enhanced branching fractions of the neutral scalars into \PQb quarks and tau leptons for some of the mass hypotheses where $\mab \leq 2\maa$.
The observed upper limits at 95\% confidence level on the combined measurement of $\sigma \mathrm{B_{C}}$ and $\sigma \mathrm{B_{NC}}$ range between 0.9 and 36.8\unit{pb}, depending on the mass hypothesis and decay scenario.
These are the first limits using 13\TeV data from CMS on exotic Higgs boson decays into two light neutral scalar particles of unequal masses in final states involving \PQb quarks and tau leptons.

\begin{acknowledgments}
We congratulate our colleagues in the CERN accelerator departments for the excellent performance of the LHC and thank the technical and administrative staffs at CERN and at other CMS institutes for their contributions to the success of the CMS effort. In addition, we gratefully acknowledge the computing centers and personnel of the Worldwide LHC Computing Grid and other centers for delivering so effectively the computing infrastructure essential to our analyses. Finally, we acknowledge the enduring support for the construction and operation of the LHC, the CMS detector, and the supporting computing infrastructure provided by the following funding agencies: SC (Armenia), BMBWF and FWF (Austria); FNRS and FWO (Belgium); CNPq, CAPES, FAPERJ, FAPERGS, and FAPESP (Brazil); MES and BNSF (Bulgaria); CERN; CAS, MoST, and NSFC (China); MINCIENCIAS (Colombia); MSES and CSF (Croatia); RIF (Cyprus); SENESCYT (Ecuador); ERC PRG and PSG, TARISTU24-TK10 and MoER TK202 (Estonia); Academy of Finland, MEC, and HIP (Finland); CEA and CNRS/IN2P3 (France); SRNSF (Georgia); BMFTR, DFG, and HGF (Germany); GSRI (Greece); MATE and NKFIH (Hungary); DAE and DST (India); IPM (Iran); SFI (Ireland); INFN (Italy); MSIT and NRF (Republic of Korea); MES (Latvia); LMTLT (Lithuania); MOE and UM (Malaysia); BUAP, CINVESTAV, CONACYT, LNS, SEP, and UASLP-FAI (Mexico); MOS (Montenegro); MBIE (New Zealand); PAEC (Pakistan); MSHE, NSC, and NAWA (Poland); FCT (Portugal); MESTD (Serbia); MICIU/AEI and PCTI (Spain); MOSTR (Sri Lanka); Swiss Funding Agencies (Switzerland); MST (Taipei); MHESI (Thailand); TUBITAK and TENMAK (T\"{u}rkiye); NASU (Ukraine); STFC (United Kingdom); DOE and NSF (USA).

\hyphenation{Rachada-pisek} Individuals have received support from the Marie-Curie program and the European Research Council and Horizon 2020 Grant, contract Nos.\ 675440, 724704, 752730, 758316, 765710, 824093, 101115353, 101002207, 101001205, and COST Action CA16108 (European Union); the Leventis Foundation; the Alfred P.\ Sloan Foundation; the Alexander von Humboldt Foundation; the Science Committee, project no. 22rl-037 (Armenia); the Fonds pour la Formation \`a la Recherche dans l'Industrie et dans l'Agriculture (FRIA) and Fonds voor Wetenschappelijk Onderzoek contract No. 1228724N (Belgium); the Beijing Municipal Science \& Technology Commission, No. Z191100007219010, the Fundamental Research Funds for the Central Universities, the Ministry of Science and Technology of China under Grant No. 2023YFA1605804, the Natural Science Foundation of China under Grant No. 12535004, and USTC Research Funds of the Double First-Class Initiative No.\ YD2030002017 (China); the Ministry of Education, Youth and Sports (MEYS) of the Czech Republic; the Shota Rustaveli National Science Foundation (Georgia); the Deutsche Forschungsgemeinschaft (DFG), among others, under Germany's Excellence Strategy -- EXC 2121 ``Quantum Universe" -- 390833306, and under project number 400140256 - GRK2497; the Hellenic Foundation for Research and Innovation (HFRI), Project Number 2288 (Greece); the Hungarian Academy of Sciences, the New National Excellence Program - \'UNKP, the NKFIH research grants K 131991, K 138136, K 143460, K 143477, K 147557, K 146913, K 146914, K 147048, TKP2021-NKTA-64, and 2025-1.1.5-NEMZ\_KI-2025-00004, and MATE KKP and KKPCs Research Excellence and Flagship Research Groups grants (Hungary); the Council of Science and Industrial Research, India; ICSC -- National Research Center for High Performance Computing, Big Data and Quantum Computing, FAIR -- Future Artificial Intelligence Research, and CUP I53D23001070006 (Mission 4 Component 1), funded by the NextGenerationEU program, the Italian Ministry of University and Research (MUR) under Bando PRIN 2022 -- CUP I53C24002390006, PRIN PRIMULA 2022RBYK7T (Italy); the Latvian Council of Science; the Ministry of Science and Higher Education, project no. 2022/WK/14, and the National Science Center, contracts Opus 2021/41/B/ST2/01369, 2021/43/B/ST2/01552, 2023/49/B/ST2/03273, and the NAWA contract BPN/PPO/2021/1/00011 (Poland); the Funda\c{c}\~ao para a Ci\^encia e a Tecnologia (Portugal); the National Priorities Research Program by Qatar National Research Fund; MICIU/AEI/10.13039/501100011033, ERDF/EU, ``European Union NextGenerationEU/PRTR", projects PID2022-142604OB-C21, PID2022-139519OB-C21, PID2023-147706NB-I00, PID2023-148896NB-I00, PID2023-146983NB-I00, PID2023-147115NB-I00, PID2023-148418NB-C41, PID2023-148418NB-C42, PID2023-148418NB-C43, PID2023-148418NB-C44, PID2024-158190NB-C22, RYC2021-033305-I, RYC2024-048719-I, CNS2023-144781, CNS2024-154769 and Plan de Ciencia, Tecnolog{\'i}a e Innovaci{\'o}n de Asturias, Spain; the Chulalongkorn Academic into Its 2nd Century Project Advancement Project, the National Science, Research and Innovation Fund program IND\_FF\_68\_369\_2300\_097, and the Program Management Unit for Human Resources \& Institutional Development, Research and Innovation, grant B39G680009 (Thailand); the Eric \& Wendy Schmidt Fund for Strategic Innovation through the CERN Next Generation Triggers project under grant agreement number SIF-2023-004; the Kavli Foundation; the Nvidia Corporation; the SuperMicro Corporation; the Welch Foundation, contract C-1845; and the Weston Havens Foundation (USA).
\end{acknowledgments}\section*{Data availability} Release and preservation of data used by the CMS Collaboration as the basis for publications is guided by the  \href{https://doi.org/10.7483/OPENDATA.CMS.1BNU.8V1W}{CMS data preservation, re-use and open access policy}.

\bibliography{auto_generated}

@ARTICLE{PhysRevLett.19.1264,
	TITLE=	"A Model of Leptons",
	AUTHOR=	"Weinberg, S.",
	JOURNAL=	"Phys. Rev. Lett.",
	VOLUME=	"19",
	PAGES=	"1264",
	NUMPAGES=	"0",
	YEAR=	"1967",
	PUBLISHER=	"American Physical Society",
	DOI=	"10.1103/PhysRevLett.19.1264",
}

@ARTICLE{Salam:1968rm,
	AUTHOR=	"Salam, A.",
	TITLE=	"{Weak and electromagnetic interactions}",
	DOI=	"10.1142/9789812795915_0034",
	JOURNAL=	"Conf. Proc. C",
	VOLUME=	"680519",
	PAGES=	"367",
	YEAR=	"1968",
}

@ARTICLE{PhysRevLett.13.321,
	TITLE=	"Broken Symmetry and the Mass of Gauge Vector Mesons",
	AUTHOR=	"Englert, F. and Brout, R.",
	JOURNAL=	"Phys. Rev. Lett.",
	VOLUME=	"13",
	PAGES=	"321",
	NUMPAGES=	"0",
	YEAR=	"1964",
	PUBLISHER=	"American Physical Society",
	DOI=	"10.1103/PhysRevLett.13.321",
}

@ARTICLE{HIGGS1964132,
	AUTHOR=	"Higgs, Peter W.",
	TITLE=	"{Broken symmetries, massless particles and gauge fields}",
	DOI=	"10.1016/0031-9163(64)91136-9",
	JOURNAL=	"Phys. Lett.",
	VOLUME=	"12",
	PAGES=	"132",
	YEAR=	"1964",
}

@ARTICLE{PhysRevLett.13.508,
	TITLE=	"Broken Symmetries and the Masses of Gauge Bosons",
	AUTHOR=	"{Higgs}, P. W.",
	JOURNAL=	"Phys. Rev. Lett.",
	VOLUME=	"13",
	PAGES=	"508",
	NUMPAGES=	"0",
	YEAR=	"1964",
	PUBLISHER=	"American Physical Society",
	DOI=	"10.1103/PhysRevLett.13.508",
}

@ARTICLE{PhysRevLett.13.585,
	TITLE=	"Global Conservation Laws and Massless Particles",
	AUTHOR=	"Guralnik, G. S. and Hagen, C. R. and Kibble, T. W. B.",
	JOURNAL=	"Phys. Rev. Lett.",
	VOLUME=	"13",
	PAGES=	"585",
	NUMPAGES=	"0",
	YEAR=	"1964",
	PUBLISHER=	"American Physical Society",
	DOI=	"10.1103/PhysRevLett.13.585",
}

@ARTICLE{PhysRev.145.1156,
	TITLE=	"Spontaneous Symmetry Breakdown without Massless Bosons",
	AUTHOR=	"{Higgs}, P. W.",
	JOURNAL=	"Phys. Rev.",
	VOLUME=	"145",
	PAGES=	"1156",
	NUMPAGES=	"0",
	YEAR=	"1966",
	PUBLISHER=	"American Physical Society",
	DOI=	"10.1103/PhysRev.145.1156",
}

@ARTICLE{PhysRev.155.1554,
	TITLE=	"Symmetry breaking in non-{Abelian} gauge theories",
	AUTHOR=	"Kibble, T. W. B.",
	JOURNAL=	"Phys. Rev.",
	VOLUME=	"155",
	PAGES=	"1554",
	NUMPAGES=	"0",
	YEAR=	"1967",
	PUBLISHER=	"American Physical Society",
	DOI=	"10.1103/PhysRev.155.1554",
}

@ARTICLE{Aad_2012,
	AUTHOR=	"{ATLAS Collaboration}",
	TITLE=	"{Observation of a new particle in the search for the standard model Higgs boson with the ATLAS detector at the LHC}",
	EPRINT=	"1207.7214",
	ARCHIVEPREFIX=	"arXiv",
	PRIMARYCLASS=	"hep-ex",
	REPORTNUMBER=	"CERN-PH-EP-2012-218",
	DOI=	"10.1016/j.physletb.2012.08.020",
	JOURNAL=	"Phys. Lett. B",
	VOLUME=	"716",
	PAGES=	"1",
	YEAR=	"2012",
}

@ARTICLE{Chatrchyan_2012,
	AUTHOR=	"{CMS Collaboration}",
	TITLE=	"{Observation of a new boson at a mass of 125 GeV with the CMS experiment at the LHC}",
	EPRINT=	"1207.7235",
	ARCHIVEPREFIX=	"arXiv",
	PRIMARYCLASS=	"hep-ex",
	REPORTNUMBER=	"CMS-HIG-12-028, CERN-PH-EP-2012-220",
	DOI=	"10.1016/j.physletb.2012.08.021",
	JOURNAL=	"Phys. Lett. B",
	VOLUME=	"716",
	PAGES=	"30",
	YEAR=	"2012",
}

@ARTICLE{chatrchyan_observation_2013,
	AUTHOR=	"{CMS Collaboration}",
	TITLE=	"{Observation of a new boson with mass near 125 GeV in pp collisions at $\sqrt{s}$ = 7 and 8 TeV}",
	EPRINT=	"1303.4571",
	ARCHIVEPREFIX=	"arXiv",
	PRIMARYCLASS=	"hep-ex",
	REPORTNUMBER=	"CMS-HIG-12-036, CERN-PH-EP-2013-035",
	DOI=	"10.1007/JHEP06(2013)081",
	JOURNAL=	"JHEP",
	VOLUME=	"06",
	PAGES=	"081",
	YEAR=	"2013",
}

@ARTICLE{Grzadkowski_2010,
	AUTHOR=	"Grzadkowski, B. and Osland, P.",
	TITLE=	"Tempered two-{Higgs}-doublet Model",
	EPRINT=	"0910.4068",
	ARCHIVEPREFIX=	"arXiv",
	PRIMARYCLASS=	"hep-ph",
	REPORTNUMBER=	"IFT-09-10",
	DOI=	"10.1103/PhysRevD.82.125026",
	JOURNAL=	"Phys. Rev. D",
	VOLUME=	"82",
	PAGES=	"125026",
	YEAR=	"2010",
}

@ARTICLE{Drozd_2014,
	AUTHOR=	"Drozd, Aleksandra and Grzadkowski, Bohdan and Gunion, John F. and Jiang, Yun",
	TITLE=	"{Extending two-Higgs-doublet models by a singlet scalar field - the case for dark matter}",
	EPRINT=	"1408.2106",
	ARCHIVEPREFIX=	"arXiv",
	PRIMARYCLASS=	"hep-ph",
	DOI=	"10.1007/JHEP11(2014)105",
	JOURNAL=	"JHEP",
	VOLUME=	"11",
	PAGES=	"105",
	YEAR=	"2014",
}

@ARTICLE{CMS:2022dwd,
	AUTHOR=	"{CMS Collaboration}",
	TITLE=	"{A portrait of the Higgs boson by the CMS experiment ten years after the discovery.}",
	EPRINT=	"2207.00043",
	ARCHIVEPREFIX=	"arXiv",
	PRIMARYCLASS=	"hep-ex",
	REPORTNUMBER=	"CMS-HIG-22-001, CERN-EP-2022-039",
	DOI=	"10.1038/s41586-022-04892-x",
	JOURNAL=	"Nature",
	VOLUME=	"607",
	PAGES=	"60",
	YEAR=	"2022",
	NOTE=	"[Erratum: \DOI{10.1038/s41586-023-06164-8}]",
}

@ARTICLE{ATLAS:2022vkf,
	AUTHOR=	"{ATLAS Collaboration}",
	TITLE=	"{A detailed map of Higgs boson interactions by the ATLAS experiment ten years after the discovery}",
	EPRINT=	"2207.00092",
	ARCHIVEPREFIX=	"arXiv",
	PRIMARYCLASS=	"hep-ex",
	REPORTNUMBER=	"CERN-EP-2022-057",
	DOI=	"10.1038/s41586-022-04893-w",
	JOURNAL=	"Nature",
	VOLUME=	"607",
	PAGES=	"52",
	YEAR=	"2022",
	NOTE=	"[Erratum: \DOI{10.1038/s41586-023-06248-5}]",
}

@ARTICLE{ATLAS:2015hpr,
	AUTHOR=	"{ATLAS Collaboration}",
	TITLE=	"{Search for new light gauge bosons in Higgs boson decays to four-lepton final states in $\Pp\Pp$ collisions at $\sqrt{s}=8$ TeV with the ATLAS detector at the LHC}",
	EPRINT=	"1505.07645",
	ARCHIVEPREFIX=	"arXiv",
	PRIMARYCLASS=	"hep-ex",
	REPORTNUMBER=	"CERN-PH-EP-2015-111",
	DOI=	"10.1103/PhysRevD.92.092001",
	JOURNAL=	"Phys. Rev. D",
	VOLUME=	"92",
	PAGES=	"092001",
	YEAR=	"2015",
}

@ARTICLE{CMS:2012qms,
	AUTHOR=	"{CMS Collaboration}",
	TITLE=	"{Search for a non-standard-model Higgs boson decaying to a pair of new light bosons in four-muon final states}",
	EPRINT=	"1210.7619",
	ARCHIVEPREFIX=	"arXiv",
	PRIMARYCLASS=	"hep-ex",
	REPORTNUMBER=	"CMS-EXO-12-012, CERN-PH-EP-2012-292",
	DOI=	"10.1016/j.physletb.2013.09.009",
	JOURNAL=	"Phys. Lett. B",
	VOLUME=	"726",
	PAGES=	"564",
	YEAR=	"2013",
}

@ARTICLE{CMS:2015nay,
	AUTHOR=	"{CMS Collaboration}",
	TITLE=	"{A search for pair production of new light bosons decaying into muons}",
	EPRINT=	"1506.00424",
	ARCHIVEPREFIX=	"arXiv",
	PRIMARYCLASS=	"hep-ex",
	REPORTNUMBER=	"CMS-HIG-13-010, CERN-PH-EP-2015-116",
	DOI=	"10.1016/j.physletb.2015.10.067",
	JOURNAL=	"Phys. Lett. B",
	VOLUME=	"752",
	PAGES=	"146",
	YEAR=	"2016",
}

@ARTICLE{CMS:2017dmg,
	AUTHOR=	"{CMS Collaboration}",
	TITLE=	"{Search for light bosons in decays of the 125 GeV Higgs boson in proton-proton collisions at $ \sqrt{s}=8 $ TeV}",
	EPRINT=	"1701.02032",
	ARCHIVEPREFIX=	"arXiv",
	PRIMARYCLASS=	"hep-ex",
	REPORTNUMBER=	"CMS-HIG-16-015, CERN-EP-2016-292",
	DOI=	"10.1007/JHEP10(2017)076",
	JOURNAL=	"JHEP",
	VOLUME=	"10",
	PAGES=	"076",
	YEAR=	"2017",
}

@ARTICLE{CMS:2015twz,
	AUTHOR=	"{CMS Collaboration}",
	TITLE=	"{Search for a very light NMSSM Higgs boson produced in decays of the 125 GeV scalar boson and decaying into $\tau$ leptons in pp collisions at $\sqrt{s}=8$ TeV}",
	EPRINT=	"1510.06534",
	ARCHIVEPREFIX=	"arXiv",
	PRIMARYCLASS=	"hep-ex",
	REPORTNUMBER=	"CMS-HIG-14-019, CERN-PH-EP-2015-264",
	DOI=	"10.1007/JHEP01(2016)079",
	JOURNAL=	"JHEP",
	VOLUME=	"01",
	PAGES=	"079",
	YEAR=	"2016",
}

@ARTICLE{ATLAS:2015rsn,
	AUTHOR=	"{ATLAS Collaboration}",
	TITLE=	"{Search for new phenomena in events with at least three photons collected in $\Pp\Pp$ collisions at $\sqrt{s}$ = 8 TeV with the ATLAS detector}",
	EPRINT=	"1509.05051",
	ARCHIVEPREFIX=	"arXiv",
	PRIMARYCLASS=	"hep-ex",
	REPORTNUMBER=	"CERN-PH-EP-2015-187",
	DOI=	"10.1140/epjc/s10052-016-4034-8",
	JOURNAL=	"Eur. Phys. J. C",
	VOLUME=	"76",
	PAGES=	"210",
	YEAR=	"2016",
}

@ARTICLE{ATLAS:2018coo,
	AUTHOR=	"{ATLAS Collaboration}",
	TITLE=	"{Search for Higgs boson decays to beyond-the-standard-model light bosons in four-lepton events with the ATLAS detector at $\sqrt{s}=13\TeV$}",
	EPRINT=	"1802.03388",
	ARCHIVEPREFIX=	"arXiv",
	PRIMARYCLASS=	"hep-ex",
	REPORTNUMBER=	"CERN-EP-2017-293",
	DOI=	"10.1007/JHEP06(2018)166",
	JOURNAL=	"JHEP",
	VOLUME=	"06",
	PAGES=	"166",
	YEAR=	"2018",
}

@ARTICLE{CMS:2018jid,
	AUTHOR=	"{CMS Collaboration}",
	TITLE=	"{A search for pair production of new light bosons decaying into muons in proton-proton collisions at 13\TeV}",
	EPRINT=	"1812.00380",
	ARCHIVEPREFIX=	"arXiv",
	PRIMARYCLASS=	"hep-ex",
	REPORTNUMBER=	"CMS-HIG-18-003, CERN-EP-2018-288",
	DOI=	"10.1016/j.physletb.2019.07.013",
	JOURNAL=	"Phys. Lett. B",
	VOLUME=	"796",
	PAGES=	"131",
	YEAR=	"2019",
}

@ARTICLE{ATLAS:2015unc,
	AUTHOR=	"{ATLAS Collaboration}",
	TITLE=	"{Search for Higgs bosons decaying to aa in the $\mu\mu\tau\tau$ final state in $\Pp\Pp$ collisions at $\sqrt{s} = $ 8\TeV with the ATLAS experiment}",
	EPRINT=	"1505.01609",
	ARCHIVEPREFIX=	"arXiv",
	PRIMARYCLASS=	"hep-ex",
	REPORTNUMBER=	"CERN-PH-EP-2015-057",
	DOI=	"10.1103/PhysRevD.92.052002",
	JOURNAL=	"Phys. Rev. D",
	VOLUME=	"92",
	PAGES=	"052002",
	YEAR=	"2015",
}

@ARTICLE{CMS:2018qvj,
	AUTHOR=	"{CMS Collaboration}",
	TITLE=	"{Search for an exotic decay of the Higgs boson to a pair of light pseudoscalars in the final state of two muons and two $\tau$ leptons in proton-proton collisions at $ \sqrt{s}=13\TeV$}",
	EPRINT=	"1805.04865",
	ARCHIVEPREFIX=	"arXiv",
	PRIMARYCLASS=	"hep-ex",
	REPORTNUMBER=	"CMS-HIG-17-029, CERN-EP-2018-078",
	DOI=	"10.1007/JHEP11(2018)018",
	JOURNAL=	"JHEP",
	VOLUME=	"11",
	PAGES=	"018",
	YEAR=	"2018",
}

@ARTICLE{CMS:2020ffa,
	AUTHOR=	"{CMS Collaboration}",
	TITLE=	"{Search for a light pseudoscalar Higgs boson in the boosted $\mu\mu\tau\tau$ final state in proton-proton collisions at $\sqrt{s}=13\TeV$}",
	EPRINT=	"2005.08694",
	ARCHIVEPREFIX=	"arXiv",
	PRIMARYCLASS=	"hep-ex",
	REPORTNUMBER=	"CMS-HIG-18-024, CERN-EP-2020-061",
	DOI=	"10.1007/JHEP08(2020)139",
	JOURNAL=	"JHEP",
	VOLUME=	"08",
	PAGES=	"139",
	YEAR=	"2020",
}

@ARTICLE{ATLAS:2018emt,
	AUTHOR=	"{ATLAS Collaboration}",
	TITLE=	"{Search for Higgs boson decays into a pair of light bosons in the bb$\mu\mu$ final state in $\Pp\Pp$ collision at $\sqrt{s}=13\TeV$ with the ATLAS detector}",
	EPRINT=	"1807.00539",
	ARCHIVEPREFIX=	"arXiv",
	PRIMARYCLASS=	"hep-ex",
	REPORTNUMBER=	"CERN-EP-2018-153",
	DOI=	"10.1016/j.physletb.2018.10.073",
	JOURNAL=	"Phys. Lett. B",
	VOLUME=	"790",
	PAGES=	"1",
	YEAR=	"2019",
}

@ARTICLE{CMS:2024uru,
	AUTHOR=	"{CMS Collaboration}",
	TITLE=	"{Search for exotic decays of the Higgs boson to a pair of pseudoscalars in the $\mu\mu$bb and $\tau\tau$bb final states}",
	EPRINT=	"2402.13358",
	ARCHIVEPREFIX=	"arXiv",
	PRIMARYCLASS=	"hep-ex",
	REPORTNUMBER=	"CMS-HIG-22-007, CERN-EP-2023-284",
	DOI=	"10.1140/epjc/s10052-024-12727-4",
	JOURNAL=	"Eur. Phys. J. C",
	VOLUME=	"84",
	PAGES=	"493",
	YEAR=	"2024",
}

@ARTICLE{ATLAS:2018jnf,
	AUTHOR=	"{ATLAS Collaboration}",
	TITLE=	"{Search for Higgs boson decays into pairs of light (pseudo)scalar particles in the $\gamma\gamma \mathrm{jj}$ final state in $\Pp\Pp$ collisions at $\sqrt{s}=13\TeV$ with the ATLAS detector}",
	EPRINT=	"1803.11145",
	ARCHIVEPREFIX=	"arXiv",
	PRIMARYCLASS=	"hep-ex",
	REPORTNUMBER=	"CERN-EP-2017-295",
	DOI=	"10.1016/j.physletb.2018.06.011",
	JOURNAL=	"Phys. Lett. B",
	VOLUME=	"782",
	PAGES=	"750",
	YEAR=	"2018",
}

@ARTICLE{CMS:2019spf,
	AUTHOR=	"{CMS Collaboration}",
	TITLE=	"{Search for light pseudoscalar boson pairs produced from decays of the 125 GeV Higgs boson in final states with two muons and two nearby tracks in $\Pp\Pp$ collisions at $\sqrt{s}=$ 13 TeV}",
	EPRINT=	"1907.07235",
	ARCHIVEPREFIX=	"arXiv",
	PRIMARYCLASS=	"hep-ex",
	REPORTNUMBER=	"CMS-HIG-18-006, CERN-EP-2019-105",
	DOI=	"10.1016/j.physletb.2019.135087",
	JOURNAL=	"Phys. Lett. B",
	VOLUME=	"800",
	PAGES=	"135087",
	YEAR=	"2020",
}

@ARTICLE{CMS:2024zfv,
	AUTHOR=	"{CMS Collaboration}",
	TITLE=	"{Search for the decay of the Higgs boson to a pair of light pseudoscalar bosons in the final state with four bottom quarks in proton-proton collisions at $ \sqrt{\textrm{s}} $ = 13 TeV}",
	EPRINT=	"2403.10341",
	ARCHIVEPREFIX=	"arXiv",
	PRIMARYCLASS=	"hep-ex",
	REPORTNUMBER=	"CMS-HIG-18-026, CERN-EP-2024-028",
	DOI=	"10.1007/JHEP06(2024)097",
	JOURNAL=	"JHEP",
	VOLUME=	"06",
	PAGES=	"097",
	YEAR=	"2024",
}

@ARTICLE{CMS:2022xxa,
	AUTHOR=	"{CMS Collaboration}",
	TITLE=	"{Search for the exotic decay of the Higgs boson into two light pseudoscalars with four photons in the final state in proton-proton collisions at $ \sqrt{s} $ = 13 TeV}",
	EPRINT=	"2208.01469",
	ARCHIVEPREFIX=	"arXiv",
	PRIMARYCLASS=	"hep-ex",
	REPORTNUMBER=	"CMS-HIG-21-003, CERN-EP-2022-095",
	DOI=	"10.1007/JHEP07(2023)148",
	JOURNAL=	"JHEP",
	VOLUME=	"07",
	PAGES=	"148",
	YEAR=	"2023",
}

@ARTICLE{CMS:2022fyt,
	AUTHOR=	"{CMS Collaboration}",
	TITLE=	"{Search for exotic Higgs boson decays $\mathrm{H} \to \mathcal{A}\mathcal{A} \to 4\gamma$ with events containing two merged diphotons in proton-proton collisions at $\sqrt{s}$ = 13 TeV}",
	EPRINT=	"2209.06197",
	ARCHIVEPREFIX=	"arXiv",
	PRIMARYCLASS=	"hep-ex",
	REPORTNUMBER=	"CMS-HIG-21-016, CERN-EP-2022-151",
	DOI=	"10.1103/PhysRevLett.131.101801",
	JOURNAL=	"Phys. Rev. Lett.",
	VOLUME=	"131",
	PAGES=	"101801",
	YEAR=	"2023",
}

@UNPUBLISHED{CMS:2025hjt,
	AUTHOR=	"{CMS Collaboration}",
	TITLE=	"{Search for light pseudoscalar boson pairs produced from Higgs boson decays using the 4$\tau$ and 2$\mu$2$\tau$ final states in proton-proton collisions at $\sqrt{s}$ = 13 TeV}",
	EPRINT=	"2508.06947",
	ARCHIVEPREFIX=	"arXiv",
	PRIMARYCLASS=	"hep-ex",
	REPORTNUMBER=	"CMS-SUS-24-002, CERN-EP-2025-129",
	YEAR=	"2025",
	NOTE=	"Accepted by \textit{JHEP}",
}

@ARTICLE{CMS:2024jyb,
	AUTHOR=	"{CMS Collaboration}",
	TITLE=	"{Model-independent search for pair production of new bosons decaying into muons in proton-proton collisions at $\sqrt{s}$ = 13 TeV}",
	EPRINT=	"2407.20425",
	ARCHIVEPREFIX=	"arXiv",
	PRIMARYCLASS=	"hep-ex",
	REPORTNUMBER=	"CMS-HIG-21-004, CERN-EP-2024-199",
	DOI=	"10.1007/JHEP12(2024)172",
	JOURNAL=	"JHEP",
	VOLUME=	"12",
	PAGES=	"172",
	YEAR=	"2024",
}

@ARTICLE{ATLAS:2025qyn,
	AUTHOR=	"{ATLAS Collaboration}",
	TITLE=	"{Search for Higgs boson exotic decays into Lorentz-boosted light bosons in the four-{\ensuremath{\tau}} final state at $\sqrt{s}$ = 13 {TeV} with the ATLAS detector}",
	EPRINT=	"2503.05463",
	ARCHIVEPREFIX=	"arXiv",
	PRIMARYCLASS=	"hep-ex",
	REPORTNUMBER=	"CERN-EP-2025-038",
	DOI=	"10.1016/j.physletb.2025.139843",
	JOURNAL=	"Phys. Lett. B",
	VOLUME=	"870",
	PAGES=	"139843",
	YEAR=	"2025",
}

@ARTICLE{Branco_2012,
	AUTHOR=	"Branco, G. C. and Ferreira, P. M. and Lavoura, L. and Rebelo, M. N. and Sher, Marc and Silva, Joao P.",
	TITLE=	"{Theory and phenomenology of two-Higgs-doublet models}",
	EPRINT=	"1106.0034",
	ARCHIVEPREFIX=	"arXiv",
	PRIMARYCLASS=	"hep-ph",
	DOI=	"10.1016/j.physrep.2012.02.002",
	JOURNAL=	"Phys. Rept.",
	VOLUME=	"516",
	PAGES=	"1",
	YEAR=	"2012",
}

@ARTICLE{Fayet1975104,
	TITLE=	"Supergauge invariant extension of the {H}iggs mechanism and a model for the electron and its neutrino",
	JOURNAL=	"Nucl. Phys. B",
	VOLUME=	"90",
	PAGES=	"104",
	YEAR=	"1975",
	ISSN=	"0550-3213",
	DOI=	"10.1016/0550-3213(75)90636-7",
	AUTHOR=	"Pierre Fayet",
}

@ARTICLE{Kaul198236,
	TITLE=	"Cancellation of quadratically divergent mass corrections in globally supersymmetric spontaneously broken gauge theories",
	JOURNAL=	"Nucl. Phys. B",
	VOLUME=	"199",
	PAGES=	"36",
	YEAR=	"1982",
	ISSN=	"0550-3213",
	DOI=	"10.1016/0550-3213(82)90565-X",
	AUTHOR=	"Romesh K. Kaul and Parthasarathi Majumdar",
}

@ARTICLE{Barbieri1982343,
	TITLE=	"Gauge models with spontaneously broken local supersymmetry",
	JOURNAL=	"Phys. Lett. B",
	VOLUME=	"119",
	PAGES=	"343",
	YEAR=	"1982",
	ISSN=	"0370-2693",
	DOI=	"10.1016/0370-2693(82)90685-2",
	AUTHOR=	"R. Barbieri and S. Ferrara and C. A. Savoy",
}

@ARTICLE{Nilles1983346,
	TITLE=	"Weak interaction breakdown induced by supergravity",
	JOURNAL=	"Phys. Lett.",
	VOLUME=	"120",
	PAGES=	"346",
	YEAR=	"1983",
	ISSN=	"0370-2693",
	DOI=	"10.1016/0370-2693(83)90460-4",
	AUTHOR=	"H. P. Nilles and M. Srednicki and D. Wyler",
}

@ARTICLE{Frere198311,
	TITLE=	"Fermion masses and induction of the weak scale by supergravity",
	JOURNAL=	"Nucl. Phys. B",
	VOLUME=	"222",
	PAGES=	"11",
	YEAR=	"1983",
	ISSN=	"0550-3213",
	DOI=	"10.1016/0550-3213(83)90606-5",
	AUTHOR=	"J. M. Frere and D. R. T. Jones and S. Raby",
}

@ARTICLE{Derendinger1984307,
	TITLE=	"Quantum effects and {SU(2)$\times$U(1)} breaking in supergravity gauge theories",
	JOURNAL=	"Nucl. Phys. B",
	VOLUME=	"237",
	PAGES=	"307",
	YEAR=	"1984",
	ISSN=	"0550-3213",
	DOI=	"10.1016/0550-3213(84)90162-7",
	AUTHOR=	"J. P. Derendinger and C. A. Savoy",
}

@ARTICLE{Drees:1988fc,
	AUTHOR=	"Drees, Manuel",
	TITLE=	"Supersymmetric models with extended {H}iggs sector",
	JOURNAL=	"Int. J. Mod. Phys. A",
	VOLUME=	"4",
	PAGES=	"3635",
	DOI=	"10.1142/S0217751X89001448",
	YEAR=	"1989",
	REPORTNUMBER=	"MAD/PH/429",
	SLACCITATION=	"%%CITATION = IMPAE,A4,3635;%%",
}

@ARTICLE{Maniatis:2009re,
	AUTHOR=	"Maniatis, M.",
	TITLE=	"The next-to-minimal supersymmetric extension of the standard model reviewed",
	JOURNAL=	"Int. J. Mod. Phys. A",
	VOLUME=	"25",
	PAGES=	"3505",
	DOI=	"10.1142/S0217751X10049827",
	YEAR=	"2010",
	EPRINT=	"0906.0777",
	ARCHIVEPREFIX=	"arXiv",
	PRIMARYCLASS=	"hep-ph",
	REPORTNUMBER=	"HD-THEP-09-9",
	SLACCITATION=	"%%CITATION = ARXIV:0906.0777;%%",
}

@ARTICLE{Ellwanger:2009dp,
	AUTHOR=	"Ellwanger, Ulrich and Hugonie, Cyril and Teixeira, Ana M.",
	TITLE=	"The next-to-minimal supersymmetric standard model",
	JOURNAL=	"Phys. Rept.",
	VOLUME=	"496",
	PAGES=	"1",
	DOI=	"10.1016/j.physrep.2010.07.001",
	YEAR=	"2010",
	EPRINT=	"0910.1785",
	ARCHIVEPREFIX=	"arXiv",
	PRIMARYCLASS=	"hep-ph",
	REPORTNUMBER=	"LPT-ORSAY-09-76, CFTP-09-032, LPTA-09-066",
	SLACCITATION=	"%%CITATION = ARXIV:0910.1785;%%",
}

@ARTICLE{Kim:1979if,
	AUTHOR=	"Kim, Jihn E.",
	TITLE=	"{Weak interaction singlet and strong CP invariance}",
	REPORTNUMBER=	"UPR-0120T",
	DOI=	"10.1103/PhysRevLett.43.103",
	JOURNAL=	"Phys. Rev. Lett.",
	VOLUME=	"43",
	PAGES=	"103",
	YEAR=	"1979",
}

@ARTICLE{AxionsDM,
	AUTHOR=	"Leanne D Duffy and Karl van Bibber",
	TITLE=	"Axions as dark matter particles",
	JOURNAL=	"New J. Phys",
	VOLUME=	"11",
	PAGES=	"105008",
	YEAR=	"2009",
	DOI=	"10.1088/1367-2630/11/10/105008",
	EPRINT=	"hep-ph/0904.3346",
	ARCHIVEPREFIX=	"arXiv",
	PRIMARYCLASS=	"hep-ph",
	SLACCITATION=	"%%CITATION = ARXIV:0904.3346;%%",
}

@ARTICLE{ArkaniHamed:2008qn,
	AUTHOR=	"Arkani-Hamed, Nima and Finkbeiner, Douglas P. and Slatyer, Tracy R. and Weiner, Neal",
	TITLE=	"A theory of dark matter",
	JOURNAL=	"Phys. Rev. D",
	VOLUME=	"79",
	PAGES=	"015014",
	DOI=	"10.1103/PhysRevD.79.015014",
	YEAR=	"2009",
	EPRINT=	"0810.0713",
	ARCHIVEPREFIX=	"arXiv",
	PRIMARYCLASS=	"hep-ph",
	SLACCITATION=	"%%CITATION = ARXIV:0810.0713;%%",
}

@ARTICLE{Baumgart:2009tn,
	AUTHOR=	"Baumgart, Matthew and Cheung, Clifford and Ruderman, Joshua T. and Wang, Lian-Tao and Yavin, Itay",
	TITLE=	"{N}on-{A}belian dark sectors and their collider signatures",
	JOURNAL=	"JHEP",
	VOLUME=	"04",
	PAGES=	"014",
	DOI=	"10.1088/1126-6708/2009/04/014",
	YEAR=	"2009",
	EPRINT=	"0901.0283",
	ARCHIVEPREFIX=	"arXiv",
	PRIMARYCLASS=	"hep-ph",
	SLACCITATION=	"%%CITATION = ARXIV:0901.0283;%%",
}

@ARTICLE{Falkowski:2010cm,
	AUTHOR=	"Falkowski, Adam and Ruderman, Joshua T. and Volansky, Tomer and Zupan, Jure",
	TITLE=	"Hidden {H}iggs decaying to lepton jets",
	JOURNAL=	"JHEP",
	VOLUME=	"05",
	PAGES=	"077",
	DOI=	"10.1007/JHEP05(2010)077",
	YEAR=	"2010",
	EPRINT=	"1002.2952",
	ARCHIVEPREFIX=	"arXiv",
	PRIMARYCLASS=	"hep-ph",
	SLACCITATION=	"%%CITATION = ARXIV:1002.2952;%%",
}

@ARTICLE{Robens_2020,
	AUTHOR=	"Robens, Tania and Stefaniak, Tim and Wittbrodt, Jonas",
	TITLE=	"{Two-real-scalar-singlet extension of the SM: LHC phenomenology and benchmark scenarios}",
	EPRINT=	"1908.08554",
	ARCHIVEPREFIX=	"arXiv",
	PRIMARYCLASS=	"hep-ph",
	REPORTNUMBER=	"DESY-19-142, DESY 19-142",
	DOI=	"10.1140/epjc/s10052-020-7655-x",
	JOURNAL=	"Eur. Phys. J. C",
	VOLUME=	"80",
	PAGES=	"151",
	YEAR=	"2020",
}

@INPROCEEDINGS{Robens:2023oyz,
	AUTHOR=	"Robens, Tania",
	BOOKTITLE=	"2023 European Physical Society Conference on High Energy Physics (EPS-HEP)",
	TITLE=	"{TRSM benchmark planes - EPS-HEP2023 update}",
	YEAR=	"2024",
	PAGES=	"55",
	NOTE=	"{[PoS(EPS-HEP2023)055]}",
	EPRINT=	"2310.18045",
	ARCHIVEPREFIX=	"arXiv",
	PRIMARYCLASS=	"hep-ph",
	REPORTNUMBER=	"RBI-ThPhys-2023-43",
	DOI=	"10.22323/1.449.0055",
}

@ARTICLE{CMS:2022suh,
	AUTHOR=	"{CMS Collaboration}",
	TITLE=	"{Search for a massive scalar resonance decaying to a light scalar and a Higgs boson in the four b quarks final state with boosted topology}",
	EPRINT=	"2204.12413",
	ARCHIVEPREFIX=	"arXiv",
	PRIMARYCLASS=	"hep-ex",
	REPORTNUMBER=	"CMS-B2G-21-003, CERN-EP-2022-034",
	DOI=	"10.1016/j.physletb.2022.137392",
	JOURNAL=	"Phys. Lett. B",
	VOLUME=	"842",
	PAGES=	"137392",
	YEAR=	"2023",
}

@ARTICLE{CMS:2023boe,
	AUTHOR=	"{CMS Collaboration}",
	TITLE=	"{Search for a new resonance decaying into two spin-0 bosons in a final state with two photons and two bottom quarks in proton-proton collisions at $ \sqrt{s} $ = 13 TeV}",
	EPRINT=	"2310.01643",
	ARCHIVEPREFIX=	"arXiv",
	PRIMARYCLASS=	"hep-ex",
	REPORTNUMBER=	"CMS-HIG-21-011, CERN-EP-2023-132",
	DOI=	"10.1007/JHEP05(2024)316",
	JOURNAL=	"JHEP",
	VOLUME=	"05",
	PAGES=	"316",
	YEAR=	"2024",
}

@ARTICLE{CMS:2024phk,
	AUTHOR=	"{CMS Collaboration}",
	TITLE=	"{Searches for Higgs boson production through decays of heavy resonances}",
	EPRINT=	"2403.16926",
	ARCHIVEPREFIX=	"arXiv",
	PRIMARYCLASS=	"hep-ex",
	REPORTNUMBER=	"CMS-B2G-23-002, CERN-EP-2024-062",
	DOI=	"10.1016/j.physrep.2024.09.004",
	JOURNAL=	"Phys. Rept.",
	VOLUME=	"1115",
	PAGES=	"368",
	YEAR=	"2025",
}

@ARTICLE{ATLAS:2025rfm,
	AUTHOR=	"{ATLAS Collaboration}",
	TITLE=	"{Search for decays of the Higgs boson into scalar particles decaying into four or six b-quarks using pp collisions at $\sqrt{s}= 13\,\mathrm{TeV}$ with the ATLAS detector}",
	EPRINT=	"2507.01165",
	ARCHIVEPREFIX=	"arXiv",
	PRIMARYCLASS=	"hep-ex",
	REPORTNUMBER=	"CERN-EP-2025-121",
	DOI=	"10.1103/mzld-ldlt",
	JOURNAL=	"Phys. Rev. D",
	VOLUME=	"112",
	PAGES=	"072005",
	YEAR=	"2025",
}

@MISC{hepdata,
	HOWPUBLISHED=	"{HEPD}ata record for this analysis",
	DOI=	"10.17182/hepdata.172650",
	YEAR=	"2026",
}

@ARTICLE{CMS:2008xjf,
	AUTHOR=	"{CMS Collaboration}",
	TITLE=	"The {CMS} experiment at the {CERN} {LHC}",
	DOI=	"10.1088/1748-0221/3/08/S08004",
	JOURNAL=	"JINST",
	VOLUME=	"3",
	PAGES=	"S08004",
	YEAR=	"2008",
}

@ARTICLE{CMS:2023gfb,
	AUTHOR=	"{CMS Collaboration}",
	TITLE=	"{Development of the CMS detector for the CERN LHC Run 3}",
	EPRINT=	"2309.05466",
	ARCHIVEPREFIX=	"arXiv",
	PRIMARYCLASS=	"physics.ins-det",
	REPORTNUMBER=	"CMS-PRF-21-001, CERN-EP-2023-136",
	DOI=	"10.1088/1748-0221/19/05/P05064",
	JOURNAL=	"JINST",
	VOLUME=	"19",
	PAGES=	"P05064",
	YEAR=	"2024",
}

@ARTICLE{CMS:2020cmk,
	AUTHOR=	"{CMS Collaboration}",
	TITLE=	"{Performance of the CMS Level-1 trigger in proton-proton collisions at $\sqrt{s} = 13$\,TeV}",
	JOURNAL=	"JINST",
	VOLUME=	"15",
	PAGES=	"P10017",
	YEAR=	"2020",
	DOI=	"10.1088/1748-0221/15/10/P10017",
	EPRINT=	"2006.10165",
	ARCHIVEPREFIX=	"arXiv",
	PRIMARYCLASS=	"hep-ex",
	REPORTNUMBER=	"CMS-TRG-17-001, CERN-EP-2020-065",
}

@ARTICLE{CMS:2016ngn,
	AUTHOR=	"{CMS Collaboration}",
	TITLE=	"{The CMS trigger system}",
	JOURNAL=	"JINST",
	VOLUME=	"12",
	PAGES=	"P01020",
	DOI=	"10.1088/1748-0221/12/01/P01020",
	YEAR=	"2017",
	EPRINT=	"1609.02366",
	ARCHIVEPREFIX=	"arXiv",
	PRIMARYCLASS=	"physics.ins-det",
	REPORTNUMBER=	"CMS-TRG-12-001, CERN-EP-2016-160",
	SLACCITATION=	"%%CITATION = ARXIV:1609.02366;%%",
}

@ARTICLE{CMS:2024aqx,
	AUTHOR=	"{CMS Collaboration}",
	TITLE=	"Performance of the {CMS} high-level trigger during {LHC Run} 2",
	EPRINT=	"2410.17038",
	ARCHIVEPREFIX=	"arXiv",
	PRIMARYCLASS=	"physics.ins-det",
	REPORTNUMBER=	"CMS-TRG-19-001, CERN-EP-2024-259",
	DOI=	"10.1088/1748-0221/19/11/P11021",
	JOURNAL=	"JINST",
	VOLUME=	"19",
	PAGES=	"P11021",
	YEAR=	"2024",
}

@ARTICLE{CMS:2020uim,
	AUTHOR=	"{CMS Collaboration}",
	TITLE=	"Electron and photon reconstruction and identification with the {CMS} experiment at the {CERN} {LHC}",
	EPRINT=	"2012.06888",
	JOURNAL=	"JINST",
	VOLUME=	"16",
	PAGES=	"P05014",
	YEAR=	"2021",
	ARCHIVEPREFIX=	"arXiv",
	PRIMARYCLASS=	"hep-ex",
	REPORTNUMBER=	"CMS-EGM-17-001, CERN-EP-2020-219",
	DOI=	"10.1088/1748-0221/16/05/P05014",
}

@ARTICLE{CMS:2018rym,
	AUTHOR=	"{CMS Collaboration}",
	TITLE=	"{Performance of the CMS muon detector and muon reconstruction with proton-proton collisions at $\sqrt{s}=$ 13 TeV}",
	EPRINT=	"1804.04528",
	ARCHIVEPREFIX=	"arXiv",
	PRIMARYCLASS=	"physics.ins-det",
	REPORTNUMBER=	"CMS-MUO-16-001, CERN-EP-2018-058",
	DOI=	"10.1088/1748-0221/13/06/P06015",
	JOURNAL=	"JINST",
	VOLUME=	"13",
	PAGES=	"P06015",
	YEAR=	"2018",
}

@ARTICLE{CMS:2014pgm,
	AUTHOR=	"{CMS Collaboration}",
	TITLE=	"{Description and performance of track and primary-vertex reconstruction with the CMS tracker}",
	EPRINT=	"1405.6569",
	ARCHIVEPREFIX=	"arXiv",
	PRIMARYCLASS=	"physics.ins-det",
	REPORTNUMBER=	"CMS-TRK-11-001, CERN-PH-EP-2014-070",
	DOI=	"10.1088/1748-0221/9/10/P10009",
	JOURNAL=	"JINST",
	VOLUME=	"9",
	PAGES=	"P10009",
	YEAR=	"2014",
}

@ARTICLE{Sjostrand:2014zea,
	AUTHOR=	"Sj{\"o}strand, Torbj{\"o}rn and Ask, Stefan and Christiansen, Jesper R. and Corke, Richard and Desai, Nishita and Ilten, Philip and Mrenna, Stephen and Prestel, Stefan and Rasmussen, Christine O. and Skands, Peter Z.",
	TITLE=	"{An Introduction to PYTHIA 8.2}",
	JOURNAL=	"Comput. Phys. Commun.",
	VOLUME=	"191",
	YEAR=	"2015",
	PAGES=	"159",
	DOI=	"10.1016/j.cpc.2015.01.024",
	EPRINT=	"1410.3012",
	ARCHIVEPREFIX=	"arXiv",
	PRIMARYCLASS=	"hep-ph",
	REPORTNUMBER=	"LU-TP-14-36, MCNET-14-22, CERN-PH-TH-2014-190, FERMILAB-PUB-14-316-CD, DESY-14-178, SLAC-PUB-16122",
	SLACCITATION=	"%%CITATION = ARXIV:1410.3012;%%",
}

@ARTICLE{Agostinelli:2002hh,
	AUTHOR=	"Agostinelli, S. and others",
	COLLABORATION=	"GEANT4",
	TITLE=	"{\GEANTfour}---a simulation toolkit",
	JOURNAL=	"Nucl. Instrum. Meth. A",
	VOLUME=	"506",
	YEAR=	"2003",
	PAGES=	"250",
	DOI=	"10.1016/S0168-9002(03)01368-8",
	SLACCITATION=	"%%CITATION = NUIMA,A506,250;%%",
}

@ARTICLE{Alwall:2007fs,
	AUTHOR=	"Alwall, J. and H{\"o}che, S. and Krauss, F. and Lavesson, N. and L{\"o}nnblad, L. and Maltoni, F. and Mangano, M. L. and Moretti, M. and Papadopoulos, C. G. and Piccinini, F. and Schumann, S. and Treccani, M. and Winter, J. and Worek, M.",
	TITLE=	"{Comparative study of various algorithms for the merging of parton showers and matrix elements in hadronic collisions}",
	JOURNAL=	"Eur. Phys. J. C",
	VOLUME=	"53",
	YEAR=	"2008",
	PAGES=	"473",
	DOI=	"10.1140/epjc/s10052-007-0490-5",
	EPRINT=	"0706.2569",
	ARCHIVEPREFIX=	"arXiv",
	PRIMARYCLASS=	"hep-ph",
	REPORTNUMBER=	"SLAC-PUB-12604, CERN-PH-TH-2007-066, LU-TP-07-13, KA-TP-06-2007, DCPT-07-62, IPPP-07-31",
	SLACCITATION=	"%%CITATION = ARXIV:0706.2569;%%",
}

@ARTICLE{Alioli:2008tz,
	AUTHOR=	"Alioli, Simone and Nason, Paolo and Oleari, Carlo and Re, Emanuele",
	TITLE=	"{NLO Higgs} boson production via gluon fusion matched with shower in {POWHEG}",
	JOURNAL=	"JHEP",
	VOLUME=	"04",
	YEAR=	"2009",
	PAGES=	"002",
	DOI=	"10.1088/1126-6708/2009/04/002",
	EPRINT=	"0812.0578",
	ARCHIVEPREFIX=	"arXiv",
	PRIMARYCLASS=	"hep-ph",
	SLACCITATION=	"%%CITATION = ARXIV:0812.0578;%%",
}

@ARTICLE{Bagnaschi:2011tu,
	AUTHOR=	"Bagnaschi, E. and Degrassi, G. and Slavich, P. and Vicini, A.",
	TITLE=	"Higgs production via gluon fusion in the {POWHEG} approach in the {SM} and in the {MSSM}",
	JOURNAL=	"JHEP",
	VOLUME=	"02",
	YEAR=	"2012",
	PAGES=	"088",
	DOI=	"10.1007/JHEP02(2012)088",
	EPRINT=	"1111.2854",
	ARCHIVEPREFIX=	"arXiv",
	PRIMARYCLASS=	"hep-ph",
	REPORTNUMBER=	"RM3-TH-11-5, IFUM-990-FT",
	SLACCITATION=	"%%CITATION = ARXIV:1111.2854;%%",
}

@ARTICLE{Nason:2009ai,
	AUTHOR=	"Nason, Paolo and Oleari, Carlo",
	TITLE=	"{NLO Higgs} boson production via vector-boson fusion matched with shower in {POWHEG}",
	EPRINT=	"0911.5299",
	ARCHIVEPREFIX=	"arXiv",
	PRIMARYCLASS=	"hep-ph",
	DOI=	"10.1007/JHEP02(2010)037",
	JOURNAL=	"JHEP",
	VOLUME=	"02",
	PAGES=	"037",
	YEAR=	"2010",
}

@ARTICLE{Luisoni:2013cuh,
	AUTHOR=	"Luisoni, Gionata and Nason, Paolo and Oleari, Carlo and Tramontano, Francesco",
	TITLE=	"{$\PH\PWpm$/$\PH\PZ$} + 0 and 1 jet at {NLO} with the {POWHEG BOX} interfaced to {GoSam} and their merging within {MiNLO}",
	EPRINT=	"1306.2542",
	ARCHIVEPREFIX=	"arXiv",
	PRIMARYCLASS=	"hep-ph",
	DOI=	"10.1007/JHEP10(2013)083",
	JOURNAL=	"JHEP",
	VOLUME=	"10",
	PAGES=	"083",
	YEAR=	"2013",
}

@ARTICLE{Hartanto:2015uka,
	AUTHOR=	"Hartanto, Heribertus B. and Jager, Barbara and Reina, Laura and Wackeroth, Doreen",
	TITLE=	"{Higgs} boson production in association with top quarks in the {POWHEG BOX}",
	EPRINT=	"1501.04498",
	ARCHIVEPREFIX=	"arXiv",
	PRIMARYCLASS=	"hep-ph",
	DOI=	"10.1103/PhysRevD.91.094003",
	JOURNAL=	"Phys. Rev. D",
	VOLUME=	"91",
	PAGES=	"094003",
	YEAR=	"2015",
}

@TECHREPORT{deFlorian:2016spz,
	AUTHOR=	"D. de Florian and others",
	EDITOR=	"de Florian, D. and C. Grojean and F. Maltoni and C. Mariotti and A. Nikitenko and M. Pieri and P. Savard and M. Schumacher and R. Tanaka",
	TITLE=	"Handbook of {LHC} {H}iggs cross sections: 4. {D}eciphering the nature of the {H}iggs sector",
	DOI=	"10.23731/CYRM-2017-002",
	YEAR=	"2016",
	EPRINT=	"1610.07922",
	ARCHIVEPREFIX=	"arXiv",
	PRIMARYCLASS=	"hep-ph",
	TYPE=	"CERN Report",
	NUMBER=	"CERN-2017-002-M",
	REPORTNUMBER=	" CERN-2017-002-M",
	SLACCITATION=	"%%CITATION = ARXIV:1610.07922;%%",
}

@ARTICLE{embedding,
	AUTHOR=	"{CMS Collaboration}",
	TITLE=	"{An embedding technique to determine $\tau\tau$ backgrounds in proton-proton collision data}",
	EPRINT=	"1903.01216",
	ARCHIVEPREFIX=	"arXiv",
	PRIMARYCLASS=	"hep-ex",
	REPORTNUMBER=	"CMS-TAU-18-001, CERN-EP-2019-012",
	DOI=	"10.1088/1748-0221/14/06/P06032",
	JOURNAL=	"JINST",
	VOLUME=	"14",
	PAGES=	"P06032",
	YEAR=	"2019",
}

@ARTICLE{Nason:2004rx,
	AUTHOR=	"Nason, Paolo",
	TITLE=	"{A new method for combining NLO QCD with shower Monte Carlo algorithms}",
	EPRINT=	"hep-ph/0409146",
	ARCHIVEPREFIX=	"arXiv",
	REPORTNUMBER=	"BICOCCA-FT-04-11",
	DOI=	"10.1088/1126-6708/2004/11/040",
	JOURNAL=	"JHEP",
	VOLUME=	"11",
	PAGES=	"040",
	YEAR=	"2004",
}

@ARTICLE{Frixione:2007vw,
	AUTHOR=	"Frixione, Stefano and Nason, Paolo and Oleari, Carlo",
	TITLE=	"{Matching NLO QCD computations with parton shower simulations: the POWHEG method}",
	EPRINT=	"0709.2092",
	ARCHIVEPREFIX=	"arXiv",
	PRIMARYCLASS=	"hep-ph",
	REPORTNUMBER=	"BICOCCA-FT-07-9, GEF-TH-21-2007",
	DOI=	"10.1088/1126-6708/2007/11/070",
	JOURNAL=	"JHEP",
	VOLUME=	"11",
	PAGES=	"070",
	YEAR=	"2007",
}

@ARTICLE{Alioli:2010xd,
	AUTHOR=	"Alioli, Simone and Nason, Paolo and Oleari, Carlo and Re, Emanuele",
	TITLE=	"{A general framework for implementing NLO calculations in shower Monte Carlo programs: the POWHEG BOX}",
	JOURNAL=	"JHEP",
	VOLUME=	"06",
	YEAR=	"2010",
	PAGES=	"043",
	DOI=	"10.1007/JHEP06(2010)043",
	EPRINT=	"1002.2581",
	ARCHIVEPREFIX=	"arXiv",
	PRIMARYCLASS=	"hep-ph",
	REPORTNUMBER=	"DESY-10-018, SFB-CPP-10-22, IPPP-10-11, DCPT-10-22",
	SLACCITATION=	"%%CITATION = ARXIV:1002.2581;%%",
}

@ARTICLE{Alioli:2010xa,
	AUTHOR=	"Alioli, Simone and Hamilton, Keith and Nason, Paolo and Oleari, Carlo and Re, Emanuele",
	TITLE=	"Jet pair production in {POWHEG}",
	EPRINT=	"1012.3380",
	ARCHIVEPREFIX=	"arXiv",
	PRIMARYCLASS=	"hep-ph",
	REPORTNUMBER=	"DESY-10-233, SFB-CPP-10-128, IPPP-10-100, DCPT-10-200, MCNET-10-23",
	DOI=	"10.1007/JHEP04(2011)081",
	JOURNAL=	"JHEP",
	VOLUME=	"04",
	PAGES=	"081",
	YEAR=	"2011",
}

@ARTICLE{Czakon_2017,
	AUTHOR=	"Czakon, Michal and Heymes, David and Mitov, Alexander and Pagani, Davide and Tsinikos, Ioannis and Zaro, Marco",
	TITLE=	"{Top-pair production at the LHC through NNLO QCD and NLO EW}",
	EPRINT=	"1705.04105",
	ARCHIVEPREFIX=	"arXiv",
	PRIMARYCLASS=	"hep-ph",
	REPORTNUMBER=	"CAVENDISH-HEP-17-07, CP3-17-12, TUM-HEP-1084-17, TTK-17-15",
	DOI=	"10.1007/JHEP10(2017)186",
	JOURNAL=	"JHEP",
	VOLUME=	"10",
	PAGES=	"186",
	YEAR=	"2017",
}

@ARTICLE{Czakon:2011xx,
	AUTHOR=	"Czakon, Michal and Mitov, Alexander",
	TITLE=	"Top++: A Program for the Calculation of the Top-Pair Cross-Section at Hadron Colliders",
	EPRINT=	"1112.5675",
	ARCHIVEPREFIX=	"arXiv",
	PRIMARYCLASS=	"hep-ph",
	REPORTNUMBER=	"CERN-PH-TH-2011-303, TTK-11-58",
	DOI=	"10.1016/j.cpc.2014.06.021",
	JOURNAL=	"Comput. Phys. Commun.",
	VOLUME=	"185",
	PAGES=	"2930",
	YEAR=	"2014",
}

@ARTICLE{Ball:2014uwa,
	AUTHOR=	"Ball, Richard D. and others",
	TITLE=	"Parton distributions for the {LHC Run II}",
	COLLABORATION=	"NNPDF",
	JOURNAL=	"JHEP",
	VOLUME=	"04",
	YEAR=	"2015",
	PAGES=	"040",
	DOI=	"10.1007/JHEP04(2015)040",
	EPRINT=	"1410.8849",
	ARCHIVEPREFIX=	"arXiv",
	PRIMARYCLASS=	"hep-ph",
	SLACCITATION=	"%%CITATION = ARXIV:1410.8849;%%",
}

@ARTICLE{Martin:2009bu,
	AUTHOR=	"Martin, A. D. and Stirling, W. J. and Thorne, R. S. and Watt, G.",
	TITLE=	"{Uncertainties on $\alpha$(S) in global PDF analyses and implications for predicted hadronic cross sections}",
	EPRINT=	"0905.3531",
	ARCHIVEPREFIX=	"arXiv",
	PRIMARYCLASS=	"hep-ph",
	REPORTNUMBER=	"IPPP-09-33, DCPT-09-66, CAVENDISH-HEP-09-06",
	DOI=	"10.1140/epjc/s10052-009-1164-2",
	JOURNAL=	"Eur. Phys. J. C",
	VOLUME=	"64",
	PAGES=	"653",
	YEAR=	"2009",
}

@ARTICLE{Gao:2013xoa,
	AUTHOR=	"Gao, Jun and Guzzi, Marco and Huston, Joey and Lai, Hung-Liang and Li, Zhao and Nadolsky, Pavel and Pumplin, Jon and Stump, Daniel and Yuan, C. -P.",
	TITLE=	"{CT10 next-to-next-to-leading order global analysis of QCD}",
	EPRINT=	"1302.6246",
	ARCHIVEPREFIX=	"arXiv",
	PRIMARYCLASS=	"hep-ph",
	REPORTNUMBER=	"SMU-HEP-12-23",
	DOI=	"10.1103/PhysRevD.89.033009",
	JOURNAL=	"Phys. Rev. D",
	VOLUME=	"89",
	PAGES=	"033009",
	YEAR=	"2014",
}

@ARTICLE{Ball:2012cx,
	AUTHOR=	"Ball, Richard D. and Bertone, Valerio and Carrazza, Stefano and Deans, Christopher S. and Del Debbio, Luigi and Forte, Stefano and Guffanti, Alberto and Hartland, Nathan P. and Latorre, Jos{\'e} I. and Rojo, Juan and Ubiali, Maria",
	TITLE=	"{Parton distributions with LHC data}",
	EPRINT=	"1207.1303",
	ARCHIVEPREFIX=	"arXiv",
	PRIMARYCLASS=	"hep-ph",
	REPORTNUMBER=	"EDINBURGH-2012-08, IFUM-FT-997, FR-PHENO-2012-014, RWTH-TTK-12-25, CERN-PH-TH-2012-037, SFB-CPP-12-47",
	DOI=	"10.1016/j.nuclphysb.2012.10.003",
	JOURNAL=	"Nucl. Phys. B",
	VOLUME=	"867",
	PAGES=	"244",
	YEAR=	"2013",
}

@ARTICLE{Campbell:2020fhf,
	AUTHOR=	"Campbell, John and Neumann, Tobias and Sullivan, Zack",
	TITLE=	"{Single-top-quark production in the $t$-channel at NNLO}",
	EPRINT=	"2012.01574",
	ARCHIVEPREFIX=	"arXiv",
	PRIMARYCLASS=	"hep-ph",
	REPORTNUMBER=	"FERMILAB-PUB-20-608-T, IIT-CAPP-20-05",
	DOI=	"10.1007/JHEP02(2021)040",
	JOURNAL=	"JHEP",
	VOLUME=	"02",
	PAGES=	"040",
	YEAR=	"2021",
}

@ARTICLE{PDF4LHCWorkingGroup:2022cjn,
	AUTHOR=	"Ball, Richard D. and others",
	COLLABORATION=	"PDF4LHC Working Group",
	TITLE=	"{The PDF4LHC21 combination of global PDF fits for the LHC Run III}",
	EPRINT=	"2203.05506",
	ARCHIVEPREFIX=	"arXiv",
	PRIMARYCLASS=	"hep-ph",
	REPORTNUMBER=	"Edinburgh 2021/31, FERMILAB-PUB-22-121-QIS-SCD-T, MSUHEP-22-010, Nikhef 2021-033, SMU-HEP-22-01",
	DOI=	"10.1088/1361-6471/ac7216",
	JOURNAL=	"J. Phys. G",
	VOLUME=	"49",
	PAGES=	"080501",
	YEAR=	"2022",
}

@ARTICLE{Melnikov:2006kv,
	AUTHOR=	"Melnikov, Kirill and Petriello, Frank",
	TITLE=	"{Electroweak gauge boson production at hadron colliders through $O(\alpha_s^2)$}",
	EPRINT=	"hep-ph/0609070",
	ARCHIVEPREFIX=	"arXiv",
	REPORTNUMBER=	"UH-511-1092-06",
	DOI=	"10.1103/PhysRevD.74.114017",
	JOURNAL=	"Phys. Rev. D",
	VOLUME=	"74",
	PAGES=	"114017",
	YEAR=	"2006",
}

@ARTICLE{Martin:2009iq,
	AUTHOR=	"Martin, A. D. and Stirling, W. J. and Thorne, R. S. and Watt, G.",
	TITLE=	"{Parton distributions for the LHC}",
	EPRINT=	"0901.0002",
	ARCHIVEPREFIX=	"arXiv",
	PRIMARYCLASS=	"hep-ph",
	REPORTNUMBER=	"IPPP-08-95, DCPT-08-190, CAVENDISH-HEP-08-16",
	DOI=	"10.1140/epjc/s10052-009-1072-5",
	JOURNAL=	"Eur. Phys. J. C",
	VOLUME=	"63",
	PAGES=	"189",
	YEAR=	"2009",
}

@ARTICLE{CMS:2022kdi,
	AUTHOR=	"{CMS Collaboration}",
	TITLE=	"{Measurements of Higgs boson production in the decay channel with a pair of $\tau $ leptons in proton{\textendash}proton collisions at $\sqrt{s}=13$ TeV}",
	EPRINT=	"2204.12957",
	ARCHIVEPREFIX=	"arXiv",
	PRIMARYCLASS=	"hep-ex",
	REPORTNUMBER=	"CMS-HIG-19-010, CERN-EP-2022-027",
	DOI=	"10.1140/epjc/s10052-023-11452-8",
	JOURNAL=	"Eur. Phys. J. C",
	VOLUME=	"83",
	PAGES=	"562",
	YEAR=	"2023",
}

@ARTICLE{CMS:2017yfk,
	AUTHOR=	"{CMS Collaboration}",
	TITLE=	"Particle-flow reconstruction and global event description with the {CMS} detector",
	JOURNAL=	"JINST",
	VOLUME=	"12",
	YEAR=	"2017",
	PAGES=	"P10003",
	DOI=	"10.1088/1748-0221/12/10/P10003",
	EPRINT=	"1706.04965",
	ARCHIVEPREFIX=	"arXiv",
	PRIMARYCLASS=	"physics.ins-det",
	REPORTNUMBER=	"CMS-PRF-14-001, CERN-EP-2017-110",
	SLACCITATION=	"%%CITATION = ARXIV:1706.04965;%%",
}

@ARTICLE{Cacciari:2008gp,
	AUTHOR=	"Cacciari, Matteo and Salam, Gavin P. and Soyez, Gregory",
	TITLE=	"{The anti-\kt jet clustering algorithm}",
	JOURNAL=	"JHEP",
	VOLUME=	"04",
	YEAR=	"2008",
	PAGES=	"063",
	DOI=	"10.1088/1126-6708/2008/04/063",
	EPRINT=	"0802.1189",
	ARCHIVEPREFIX=	"arXiv",
	PRIMARYCLASS=	"hep-ex",
}

@ARTICLE{Cacciari:2011ma,
	AUTHOR=	"Cacciari, Matteo and Salam, Gavin P. and Soyez, Gregory",
	TITLE=	"{FastJet user manual}",
	JOURNAL=	"Eur. Phys. J. C",
	VOLUME=	"72",
	PAGES=	"1896",
	DOI=	"10.1140/epjc/s10052-012-1896-2",
	YEAR=	"2012",
	EPRINT=	"1111.6097",
	ARCHIVEPREFIX=	"arXiv",
	PRIMARYCLASS=	"hep-ph",
	REPORTNUMBER=	"CERN-PH-TH-2011-297",
	SLACCITATION=	"%%CITATION = ARXIV:1111.6097;%%",
}

@ARTICLE{CMS:2020ebo,
	AUTHOR=	"{CMS Collaboration}",
	TITLE=	"{Pileup mitigation at CMS in 13\TeV data}",
	EPRINT=	"2003.00503",
	ARCHIVEPREFIX=	"arXiv",
	PRIMARYCLASS=	"hep-ex",
	REPORTNUMBER=	"CMS-JME-18-001, CERN-EP-2020-017",
	DOI=	"10.1088/1748-0221/15/09/P09018",
	JOURNAL=	"JINST",
	VOLUME=	"15",
	PAGES=	"P09018",
	YEAR=	"2020",
}

@ARTICLE{CMS:2016lmd,
	AUTHOR=	"{CMS Collaboration}",
	TITLE=	"{Jet energy scale and resolution in the CMS experiment in pp collisions at 8 TeV}",
	JOURNAL=	"JINST",
	VOLUME=	"12",
	YEAR=	"2017",
	PAGES=	"P02014",
	DOI=	"10.1088/1748-0221/12/02/P02014",
	EPRINT=	"1607.03663",
	ARCHIVEPREFIX=	"arXiv",
	PRIMARYCLASS=	"hep-ex",
	REPORTNUMBER=	"CMS-JME-13-004, CERN-PH-EP-2015-305",
	SLACCITATION=	"%%CITATION = ARXIV:1607.03663;%%",
}

@ARTICLE{Bols:2020bkb,
	AUTHOR=	"Bols, Emil and Kieseler, Jan and Verzetti, Mauro and Stoye, Markus and Stakia, Anna",
	TITLE=	"Jet Flavour Classification Using \textsc{DeepJet}",
	EPRINT=	"2008.10519",
	ARCHIVEPREFIX=	"arXiv",
	PRIMARYCLASS=	"hep-ex",
	DOI=	"10.1088/1748-0221/15/12/P12012",
	JOURNAL=	"JINST",
	VOLUME=	"15",
	PAGES=	"P12012",
	YEAR=	"2020",
}

@ARTICLE{Sirunyan:2017ezt,
	AUTHOR=	"{CMS Collaboration}",
	TITLE=	"{Identification of heavy-flavour jets with the CMS detector in {\Pp}{\Pp} collisions at 13 TeV}",
	JOURNAL=	"JINST",
	VOLUME=	"13",
	YEAR=	"2018",
	PAGES=	"P05011",
	DOI=	"10.1088/1748-0221/13/05/P05011",
	EPRINT=	"1712.07158",
	ARCHIVEPREFIX=	"arXiv",
	PRIMARYCLASS=	"physics.ins-det",
	REPORTNUMBER=	"CMS-BTV-16-002, CERN-EP-2017-326",
	SLACCITATION=	"%%CITATION = ARXIV:1712.07158;%%",
}

@ARTICLE{CMS:2019ctu,
	AUTHOR=	"{CMS Collaboration}",
	TITLE=	"Performance of missing transverse momentum reconstruction in proton-proton collisions at $\sqrt{s} = 13$\,{TeV} using the {CMS} detector",
	JOURNAL=	"JINST",
	VOLUME=	"14",
	YEAR=	"2019",
	PAGES=	"P07004",
	DOI=	"10.1088/1748-0221/14/07/P07004",
	EPRINT=	"1903.06078",
	ARCHIVEPREFIX=	"arXiv",
	PRIMARYCLASS=	"hep-ex",
	REPORTNUMBER=	"CMS-JME-17-001, CERN-EP-2018-335",
	SLACCITATION=	"%%CITATION = ARXIV:1903.06078;%%",
}

@ARTICLE{CMS:2018jrd,
	AUTHOR=	"{CMS Collaboration}",
	TITLE=	"{Performance of reconstruction and identification of $\tau$ leptons decaying to hadrons and $\nu_\tau$ in pp collisions at $\sqrt{s}=$ 13 TeV}",
	EPRINT=	"1809.02816",
	ARCHIVEPREFIX=	"arXiv",
	PRIMARYCLASS=	"hep-ex",
	REPORTNUMBER=	"CMS-TAU-16-003, CERN-EP-2018-229",
	DOI=	"10.1088/1748-0221/13/10/P10005",
	JOURNAL=	"JINST",
	VOLUME=	"13",
	PAGES=	"P10005",
	YEAR=	"2018",
}

@ARTICLE{CMS:2022prd,
	AUTHOR=	"{CMS Collaboration}",
	TITLE=	"{Identification of hadronic tau lepton decays using a deep neural network}",
	EPRINT=	"2201.08458",
	ARCHIVEPREFIX=	"arXiv",
	PRIMARYCLASS=	"hep-ex",
	REPORTNUMBER=	"CMS-TAU-20-001, CERN-EP-2021-257",
	DOI=	"10.1088/1748-0221/17/07/P07023",
	JOURNAL=	"JINST",
	VOLUME=	"17",
	PAGES=	"P07023",
	YEAR=	"2022",
}

@ARTICLE{svfit,
	AUTHOR=	"Bianchini, L. and Calpas, B. and Conway, J. and Fowlie, A. and Marzola, L. and Perrini, L. and Veelken, C.",
	TITLE=	"Reconstruction of the {Higgs} mass in events with {Higgs} bosons decaying into a pair of tau leptons using matrix element technique",
	EPRINT=	"1603.05910",
	ARCHIVEPREFIX=	"arXiv",
	PRIMARYCLASS=	"hep-ex",
	DOI=	"10.1016/j.nima.2017.05.001",
	JOURNAL=	"Nucl. Instrum. Meth. A",
	VOLUME=	"862",
	PAGES=	"54",
	YEAR=	"2017",
}

@ARTICLE{Kalinowski:2025vci,
	AUTHOR=	"Kalinowski, Artur and Matyszkiewicz, Wiktor",
	TITLE=	"{Efficient tau-pair invariant mass reconstruction with simplified matrix element techniques}",
	EPRINT=	"2509.26069",
	ARCHIVEPREFIX=	"arXiv",
	PRIMARYCLASS=	"hep-ph",
	DOI=	"10.1016/j.nima.2026.171318",
	JOURNAL=	"Nucl. Instrum. Meth. A",
	VOLUME=	"1086",
	PAGES=	"171318",
	YEAR=	"2026",
}

@INPROCEEDINGS{Chen:2016:XGBoost,
	AUTHOR=	"Chen, Tianqi and Guestrin, Carlos",
	TITLE=	"XGBoost: A Scalable Tree Boosting System",
	BOOKTITLE=	"Proceedings of the 22nd ACM SIGKDD International Conference on Knowledge Discovery and Data Mining",
	YEAR=	"2016",
	PAGES=	"785",
}

@ARTICLE{2016bbtautau,
	AUTHOR=	"{CMS Collaboration}",
	TITLE=	"{Search for an exotic decay of the Higgs boson to a pair of light pseudoscalars in the final state with two b quarks and two $\tau$ leptons in proton-proton collisions at $\sqrt{s}=$ 13 TeV}",
	EPRINT=	"1805.10191",
	ARCHIVEPREFIX=	"arXiv",
	PRIMARYCLASS=	"hep-ex",
	REPORTNUMBER=	"CMS-HIG-17-024, CERN-EP-2018-089",
	DOI=	"10.1016/j.physletb.2018.08.057",
	JOURNAL=	"Phys. Lett. B",
	VOLUME=	"785",
	PAGES=	"462",
	YEAR=	"2018",
}

@ARTICLE{Abulencia:2005kq,
	AUTHOR=	"Abulencia, A. and others",
	TITLE=	"Search for Neutral {Higgs} Bosons of the Minimal Supersymmetric Standard Model Decaying to $\tau$ Pairs in {$\Pp\bar{\Pp}$} Collisions at {$\sqrt{s} = 1.96\TeV$}",
	COLLABORATION=	"CDF",
	JOURNAL=	"Phys. Rev. Lett.",
	VOLUME=	"96",
	YEAR=	"2006",
	PAGES=	"011802",
	DOI=	"10.1103/PhysRevLett.96.011802",
	EPRINT=	"hep-ex/0508051",
	ARCHIVEPREFIX=	"arXiv",
	PRIMARYCLASS=	"hep-ex",
	REPORTNUMBER=	"FERMILAB-PUB-05-374-E",
	SLACCITATION=	"%%CITATION = HEP-EX/0508051;%%",
}

@ARTICLE{CMS-LUM-17-003,
	AUTHOR=	"{CMS Collaboration}",
	TITLE=	"Precision luminosity measurement in proton-proton collisions at $\sqrt{s} =$ 13 {TeV} in 2015 and 2016 at {CMS}",
	EPRINT=	"2104.01927",
	ARCHIVEPREFIX=	"arXiv",
	PRIMARYCLASS=	"hep-ex",
	REPORTNUMBER=	"CMS-LUM-17-003, CERN-EP-2021-033",
	YEAR=	"2021",
	DOI=	"10.1140/epjc/s10052-021-09538-2",
	JOURNAL=	"Eur. Phys. J. C",
	VOLUME=	"81",
	PAGES=	"800",
}

@TECHREPORT{CMS-PAS-LUM-17-004,
	TITLE=	"{CMS} luminosity measurement for the 2017 data-taking period at $\sqrt{s}$ = 13 {TeV}",
	AUTHOR=	"{CMS Collaboration}",
	URL=	"https://cds.cern.ch/record/2621960/",
	TYPE=	"CMS Physics Analysis Summary",
	NUMBER=	"CMS-PAS-LUM-17-004",
	YEAR=	"2018",
}

@TECHREPORT{CMS-PAS-LUM-18-002,
	TITLE=	"{CMS} luminosity measurement for the 2018 data-taking period at $\sqrt{s}$ = 13 {TeV}",
	AUTHOR=	"{CMS Collaboration}",
	URL=	"https://cds.cern.ch/record/2676164/",
	TYPE=	"CMS Physics Analysis Summary",
	NUMBER=	"CMS-PAS-LUM-18-002",
	YEAR=	"2019",
}

@ARTICLE{Barlow:249779,
	AUTHOR=	"Barlow, Roger J. and Beeston, Christine",
	TITLE=	"{Fitting using finite Monte Carlo samples}",
	REPORTNUMBER=	"MAN-HEP-93-1",
	DOI=	"10.1016/0010-4655(93)90005-W",
	JOURNAL=	"Comput. Phys. Commun.",
	VOLUME=	"77",
	PAGES=	"219",
	YEAR=	"1993",
}

@ARTICLE{CMS:2024onh,
	AUTHOR=	"{CMS Collaboration}",
	TITLE=	"The {CMS} Statistical Analysis and Combination Tool: \textsc{Combine}",
	EPRINT=	"2404.06614",
	ARCHIVEPREFIX=	"arXiv",
	PRIMARYCLASS=	"physics.data-an",
	REPORTNUMBER=	"CMS-CAT-23-001, CERN-EP-2024-078",
	DOI=	"10.1007/s41781-024-00121-4",
	JOURNAL=	"Comput. Softw. Big Sci.",
	VOLUME=	"8",
	PAGES=	"19",
	YEAR=	"2024",
}

@ARTICLE{Cowan:2010js,
	AUTHOR=	"Cowan, Glen and Cranmer, Kyle and Gross, Eilam and Vitells, Ofer",
	TITLE=	"Asymptotic formulae for likelihood-based tests of new physics",
	JOURNAL=	"Eur. Phys. J. C",
	VOLUME=	"71",
	YEAR=	"2011",
	PAGES=	"1554",
	DOI=	"10.1140/epjc/s10052-011-1554-0",
	NOTE=	"[Erratum: \DOI{10.1140/epjc/s10052-013-2501-z}]",
	EPRINT=	"1007.1727",
	ARCHIVEPREFIX=	"arXiv",
	PRIMARYCLASS=	"physics.data-an",
	SLACCITATION=	"%%CITATION = ARXIV:1007.1727;%%",
}

\appendix
\numberwithin{table}{section}
\numberwithin{figure}{section}
\section{Cut-based categorization and results}
\label{appendix}
The cut-based strategy for signal categorization follows Ref.~\cite{2016bbtautau}, which considers the signal process $\PH \to \Pa \Pa \to 2\PGt 2\PQb$ and defines one CR and three SRs targeting specific ranges of \Pa masses: a low- (SRL), medium- (SRM), and high-mass (SRH) SR.
The SR categories are based on the visible invariant mass of the combined leading \PQb-tagged jet and \PGt{}\PGt candidate, $m^{\text{vis}}(\PGt\PGt\PQb_1)$, the transverse invariant mass between the \ptvecmiss and the leading constituent of the \PGt{}\PGt, and the alignment variable \dzeta.
The thresholds for the definitions of the categories are shown in Table~\ref{tab:cut-based-categories}, applied on preselected events with at least one \PQb-tagged jet.

\begin{table*}[ht!]
        \centering
        \topcaption{Definitions of cut-based event categories in the three channels. The selection boundaries are indicated in \GeV.}
        \begin{tabular}{lcccc}
                Cut & SRL & SRM & SRH & CR \\
                \hline
                & \multicolumn{4}{c}{\emu} \\
                $m^{\text{vis}}(\PGt\PGt\PQb_1)$ & $<$65 & $\in[65,80]$ & $\in[80,95]$ & $>$95  \\
                $\mT(\Pe, \ptmiss)$ & $<$40  & $<$40  & $<$40  & $<$40   \\
                $\mT(\PGm, \ptmiss)$ & $<$40  & $<$40  & $<$40  & $<$40  \\
                \dzeta & ${>}-30$  & ${>}-30$  & ${>}-30$  & ${>}-30$  \\
                [\cmsTabSkip]
                & \multicolumn{4}{c}{\etau} \\
                $m^{\text{vis}}(\PGt\PGt\PQb_1)$ & $<$80 & $\in[80,100]$  & $\in[100,120]$  & $>$120  \\
                $\mT(\Pe, \ptmiss)$ & $<$40  & $<$50  & $<$50  & $<$40   \\
                $\mT(\tauh, \ptmiss)$ & $<$60  & $<$60  & $<$60  & $<$60  \\
                [\cmsTabSkip]
                & \multicolumn{4}{c}{\mutau} \\
                $m^{\text{vis}}(\PGt\PGt\PQb_1)$ & $<$75  & $\in[75,95]$  & $\in[95,115]$  & $>$115  \\
                $\mT(\PGm, \ptmiss)$ & $<$40  & $<$50  & $<$50  & $<$40   \\
                $\mT(\tauh, \ptmiss)$ & $<$60  & $<$60  & $<$60  & $<$60  \\
                \dzeta & \NA & $<0$ & \NA  & \NA \\
        \end{tabular}
        \label{tab:cut-based-categories}
\end{table*}

\begin{figure}[ht!]
    \centering
        \includegraphics[width=0.42\textwidth]{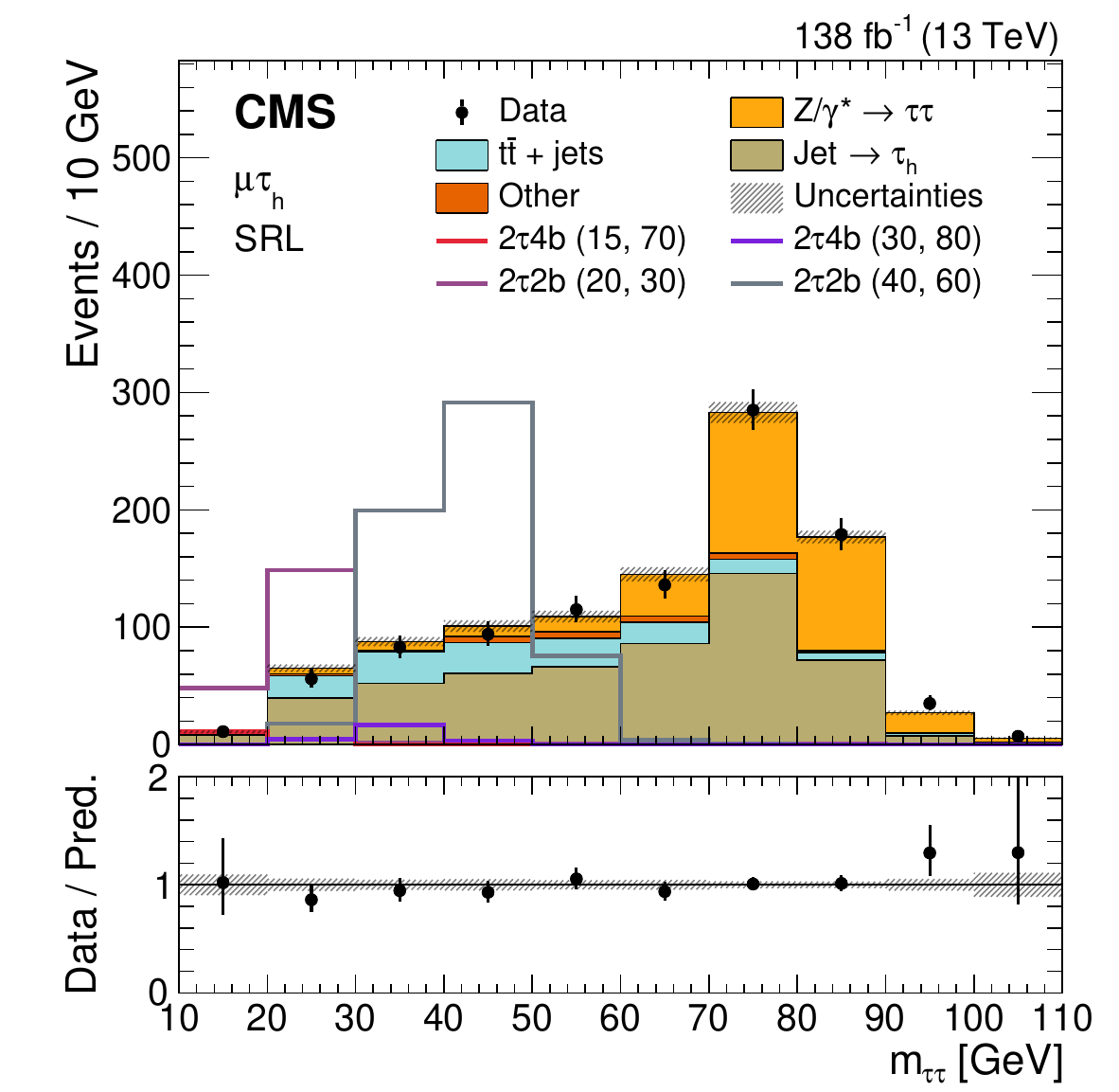}
        \includegraphics[width=0.42\textwidth]{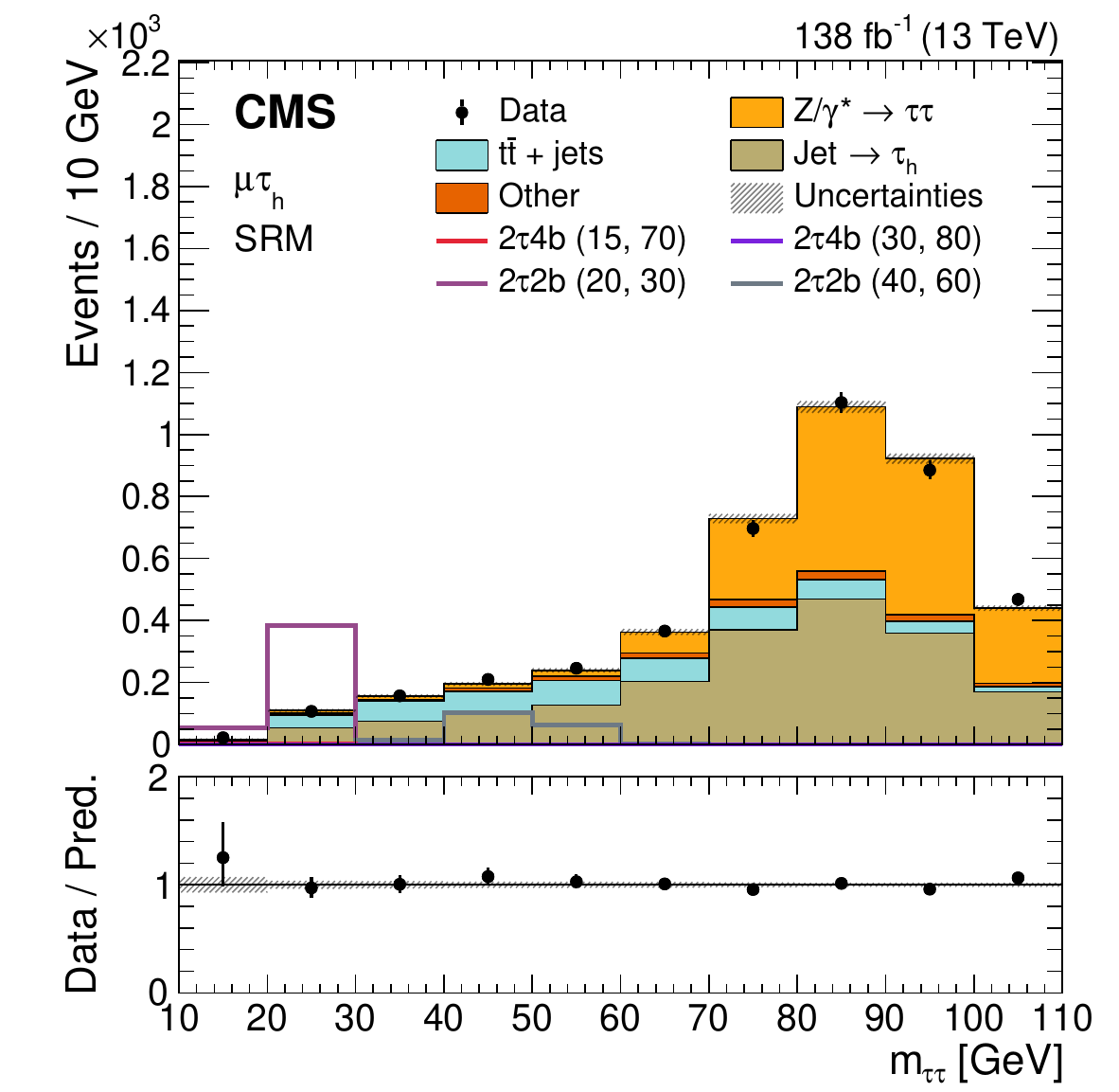} \\
        \includegraphics[width=0.42\textwidth]{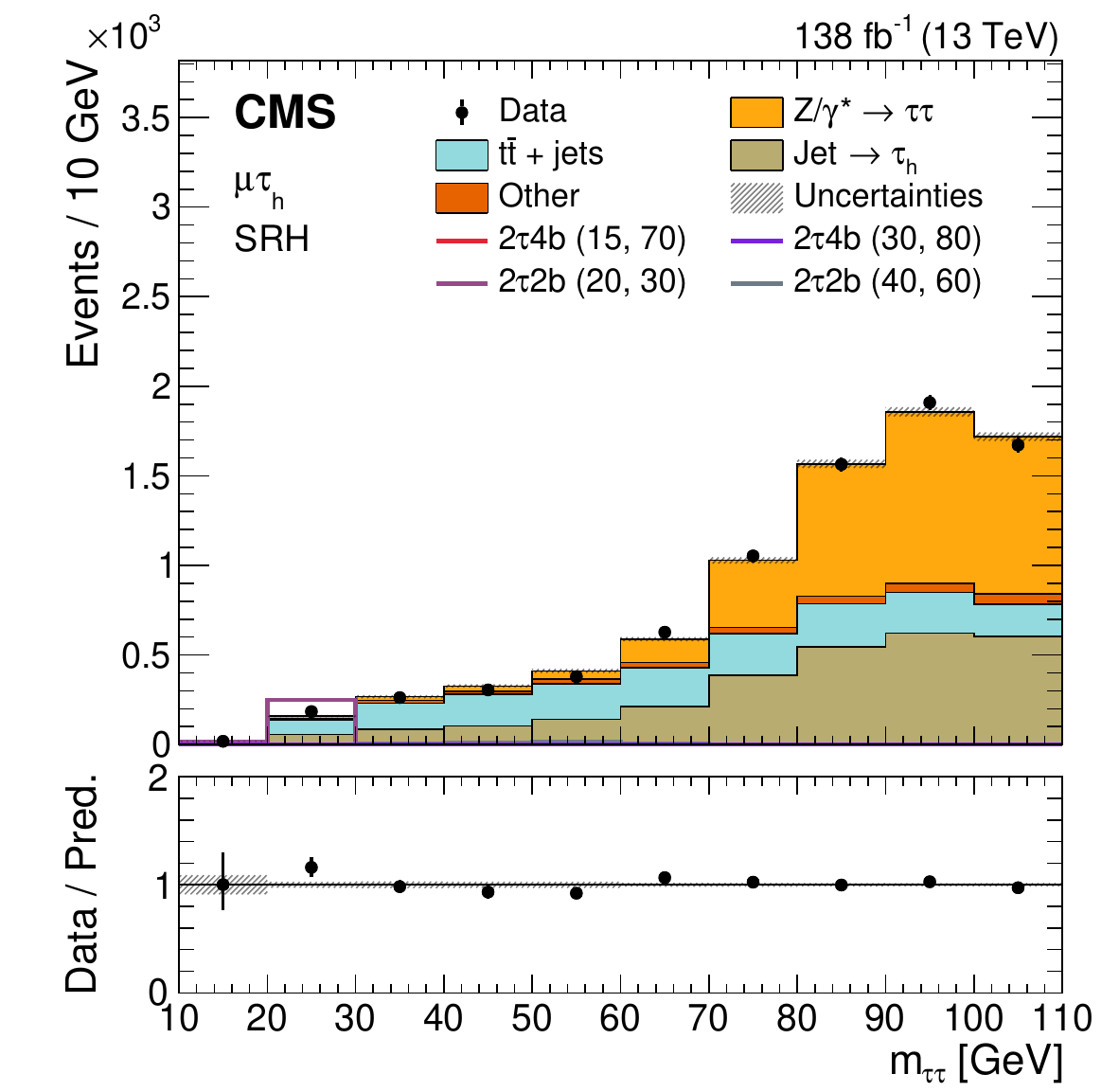}
    
    \caption{Background only, post-fit \mtt distributions for the \mutau channel, in events with at least one \PQb-tagged jet: SRL (upper left), SRM (upper right), and SRH (lower). The data are shown by the markers with vertical bars and various backgrounds by the colored histograms. The total systematic uncertainty is shown by the hatched area. The colored open histograms display the predicted signal distribution for two cascade decays and two non-cascade decays, with four different values of \Paa and \Pab masses, for an assumed branching fraction of 100\%. The lower plot of each panel gives the ratio of the data to the sum of the predicted number of background events. The vertical bars display the statistical uncertainty in the ratio.}
    \label{fig:results_postfit_mutau}
\end{figure}

\begin{figure}[ht!]
    \centering
        \includegraphics[width=0.42\textwidth]{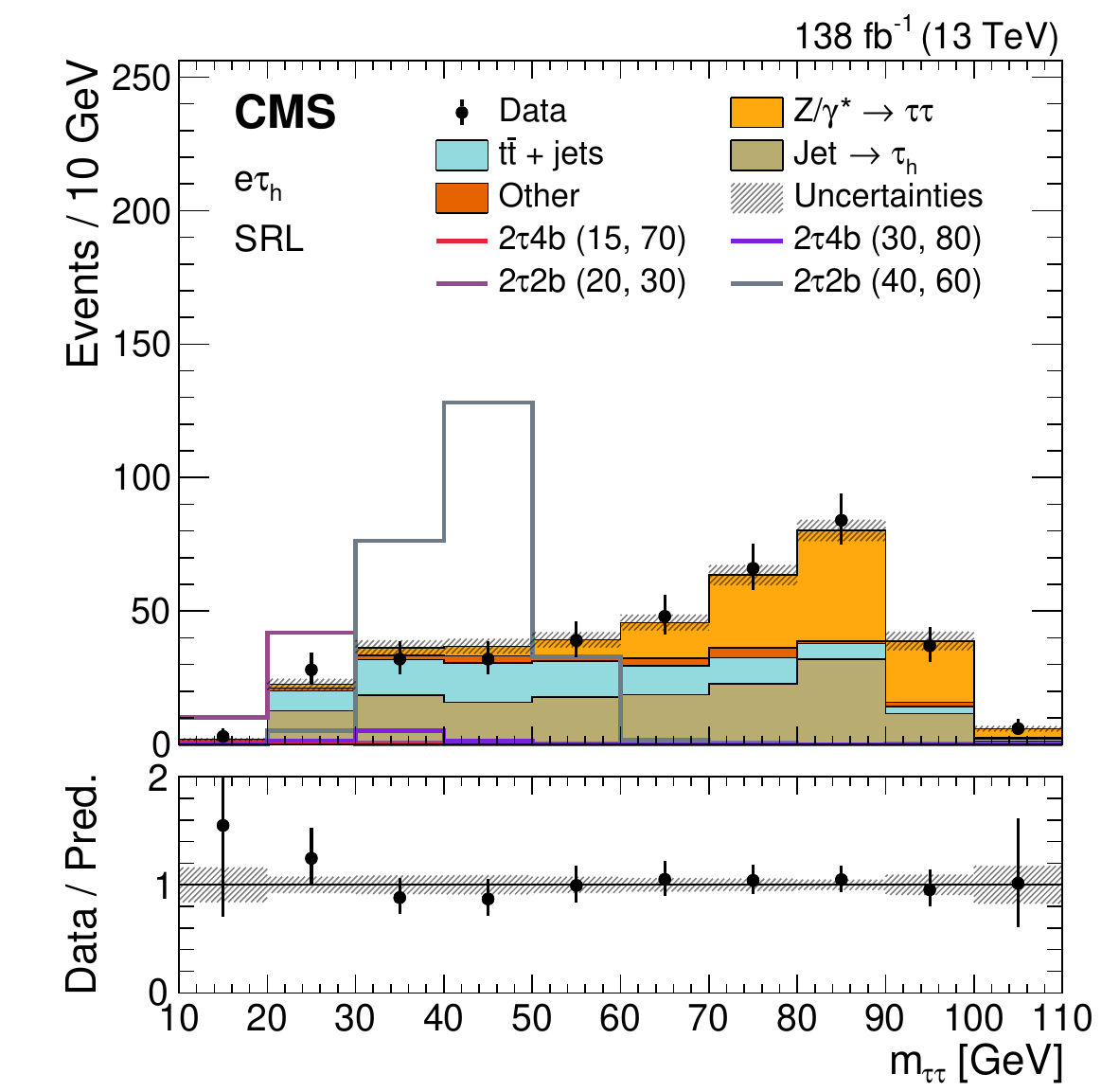}
        \includegraphics[width=0.42\textwidth]{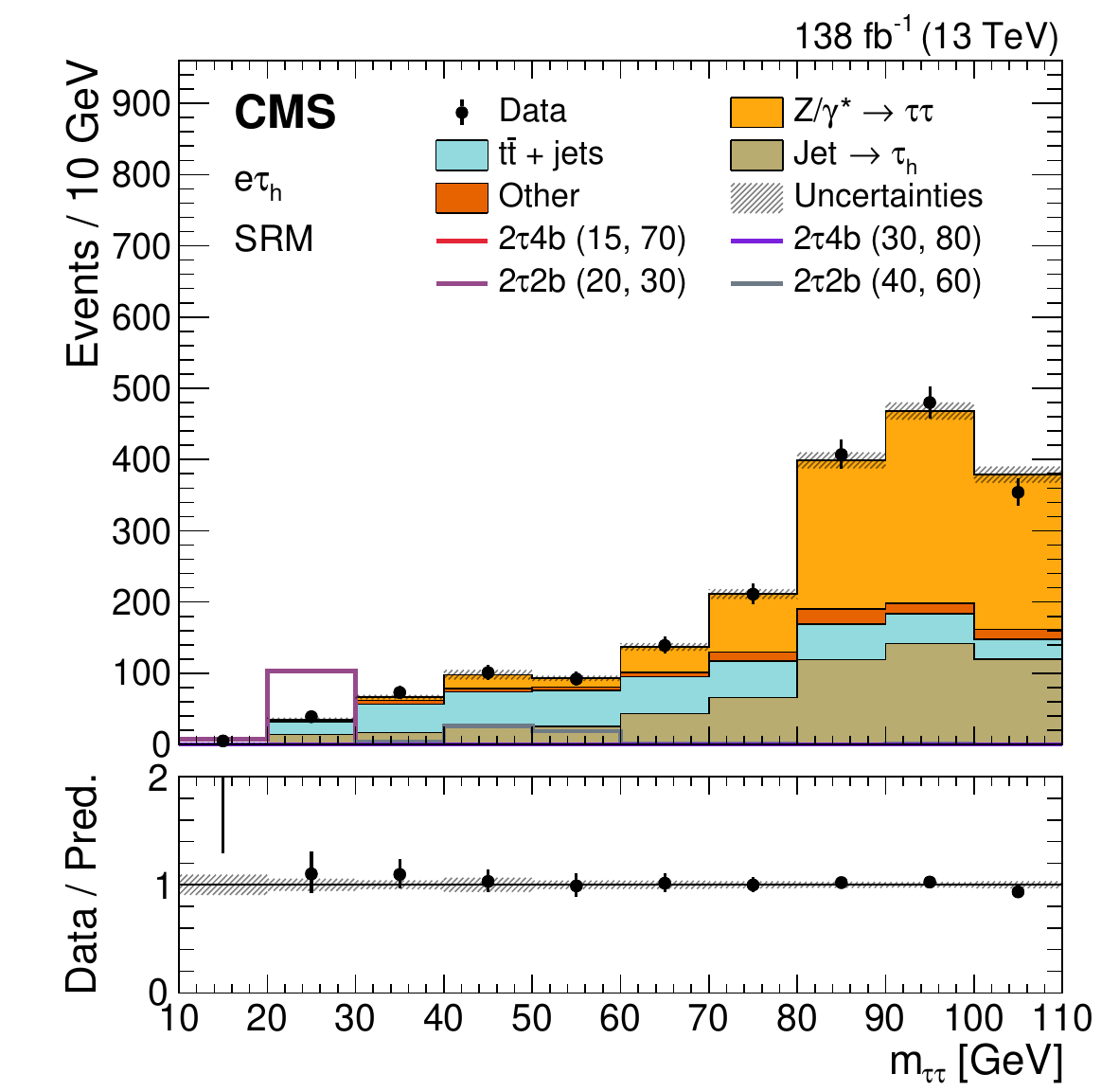} \\
        \includegraphics[width=0.42\textwidth]{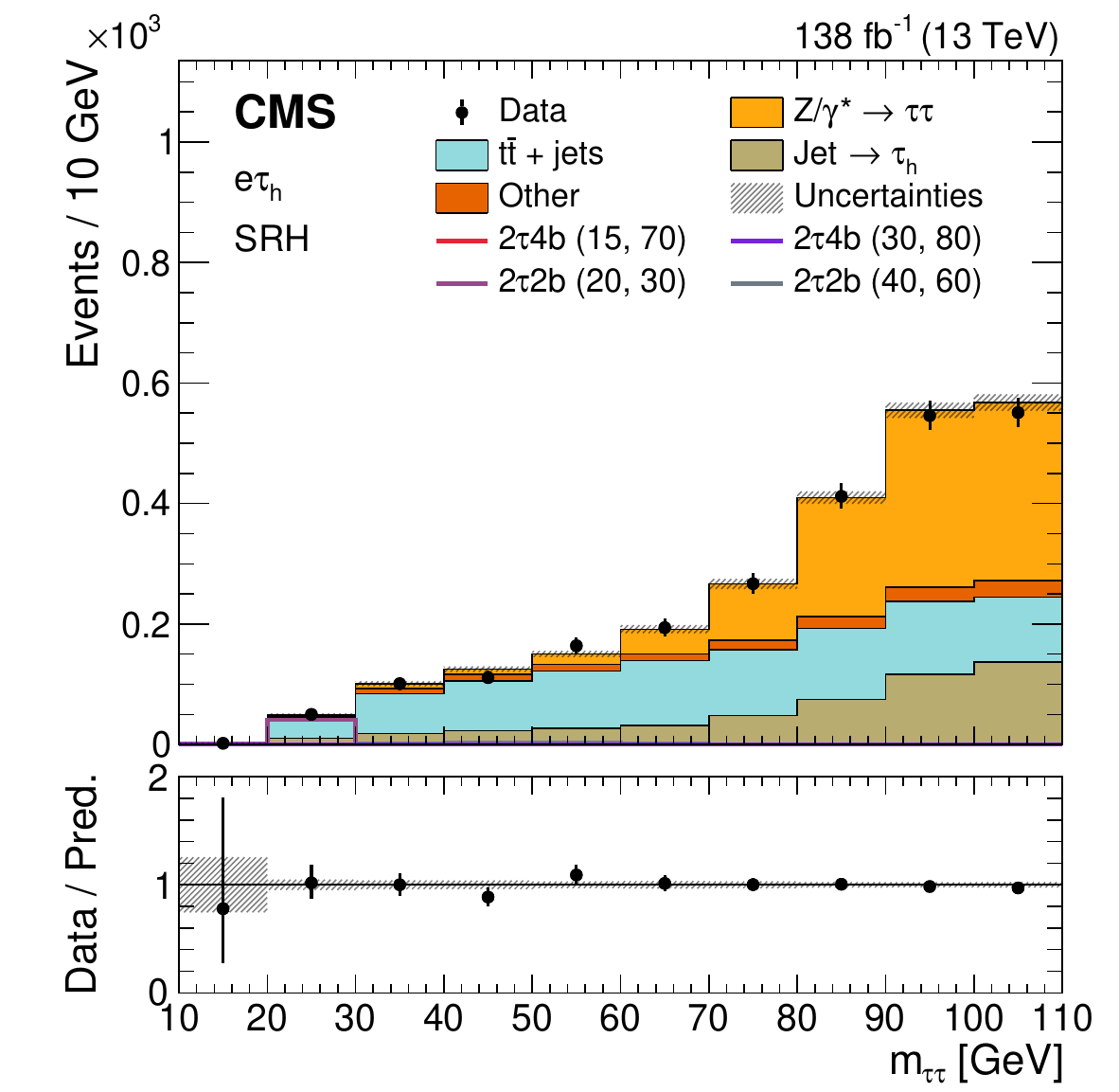}
    
    \caption{Background only, post-fit \mtt distributions for the \etau channel, in events with at least one \PQb-tagged jet: SRL (upper left), SRM (upper right), and SRH (lower). The data are shown by the markers with vertical bars and various backgrounds by the colored histograms. The total systematic uncertainty is shown by the hatched area. The colored open histograms display the predicted signal distribution for two cascade decays and two non-cascade decays, with four different values of \Paa and \Pab masses, for an assumed branching fraction of 100\%. The lower plot of each panel gives the ratio of the data to the sum of the predicted number of background events. The vertical bars display the statistical uncertainty in the ratio.}
    \label{fig:results_postfit_etau}
\end{figure}

\begin{figure}[ht!]
    \centering
        \includegraphics[width=0.42\textwidth]{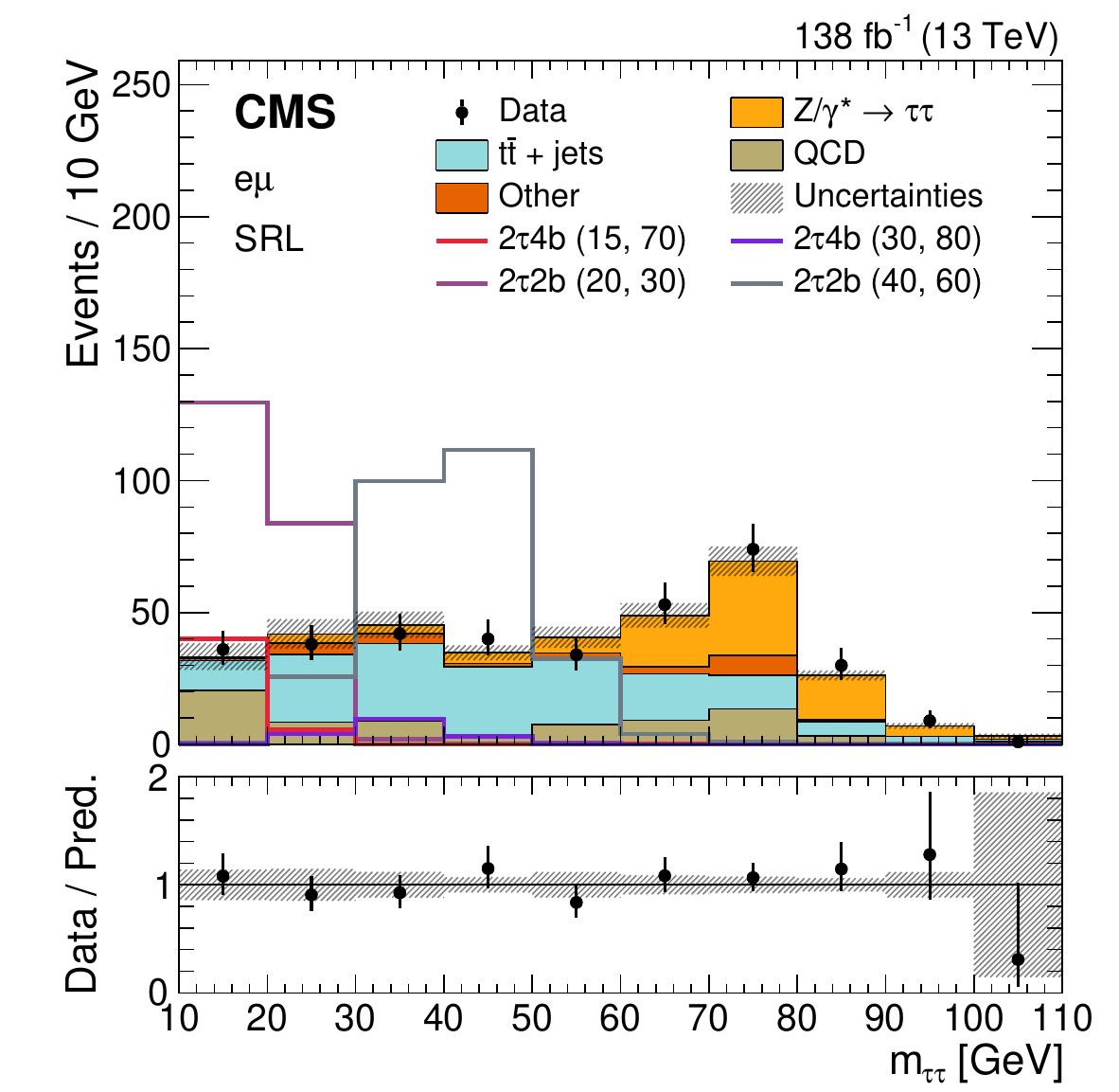}
        \includegraphics[width=0.42\textwidth]{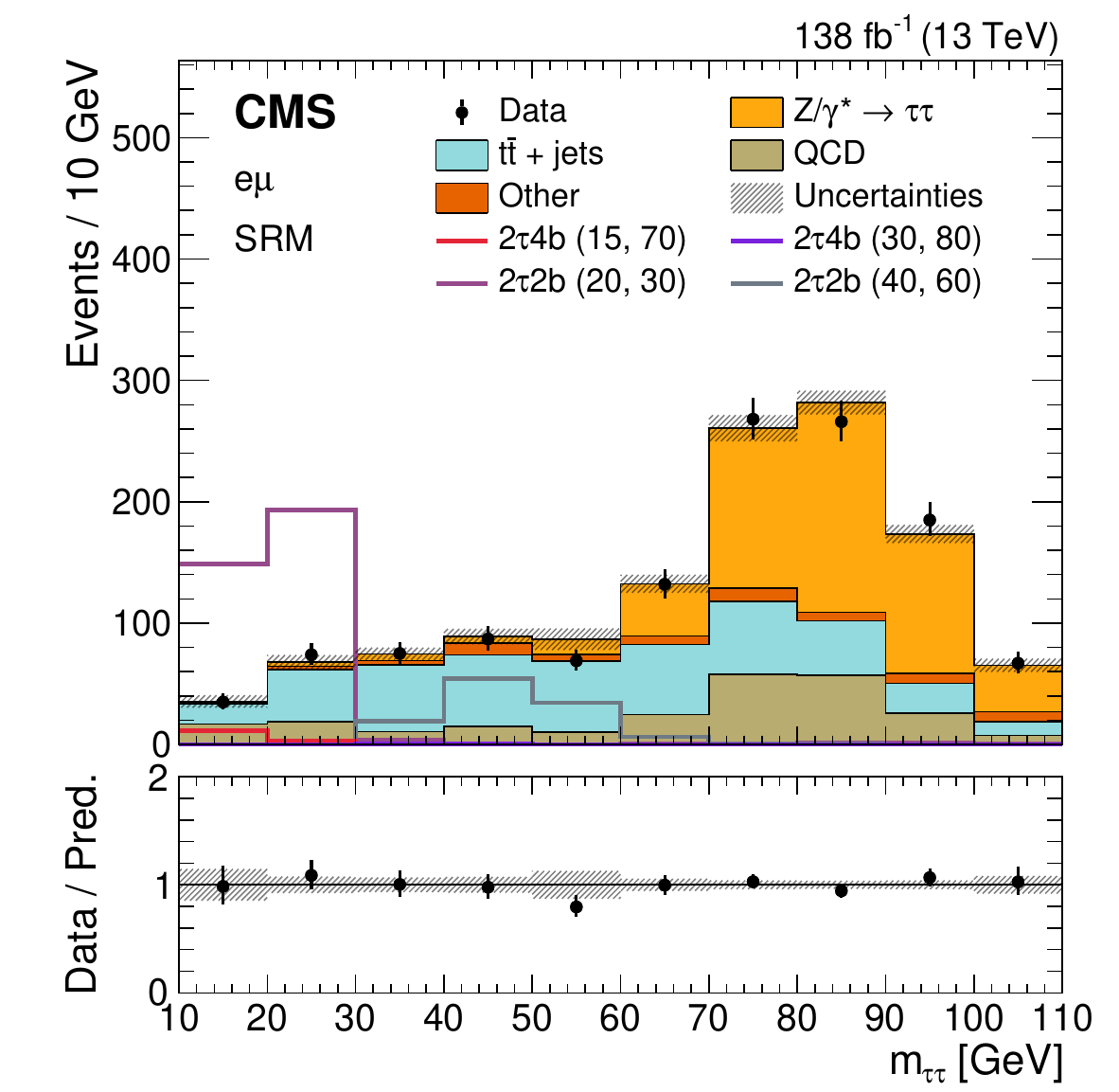} \\
        \includegraphics[width=0.42\textwidth]{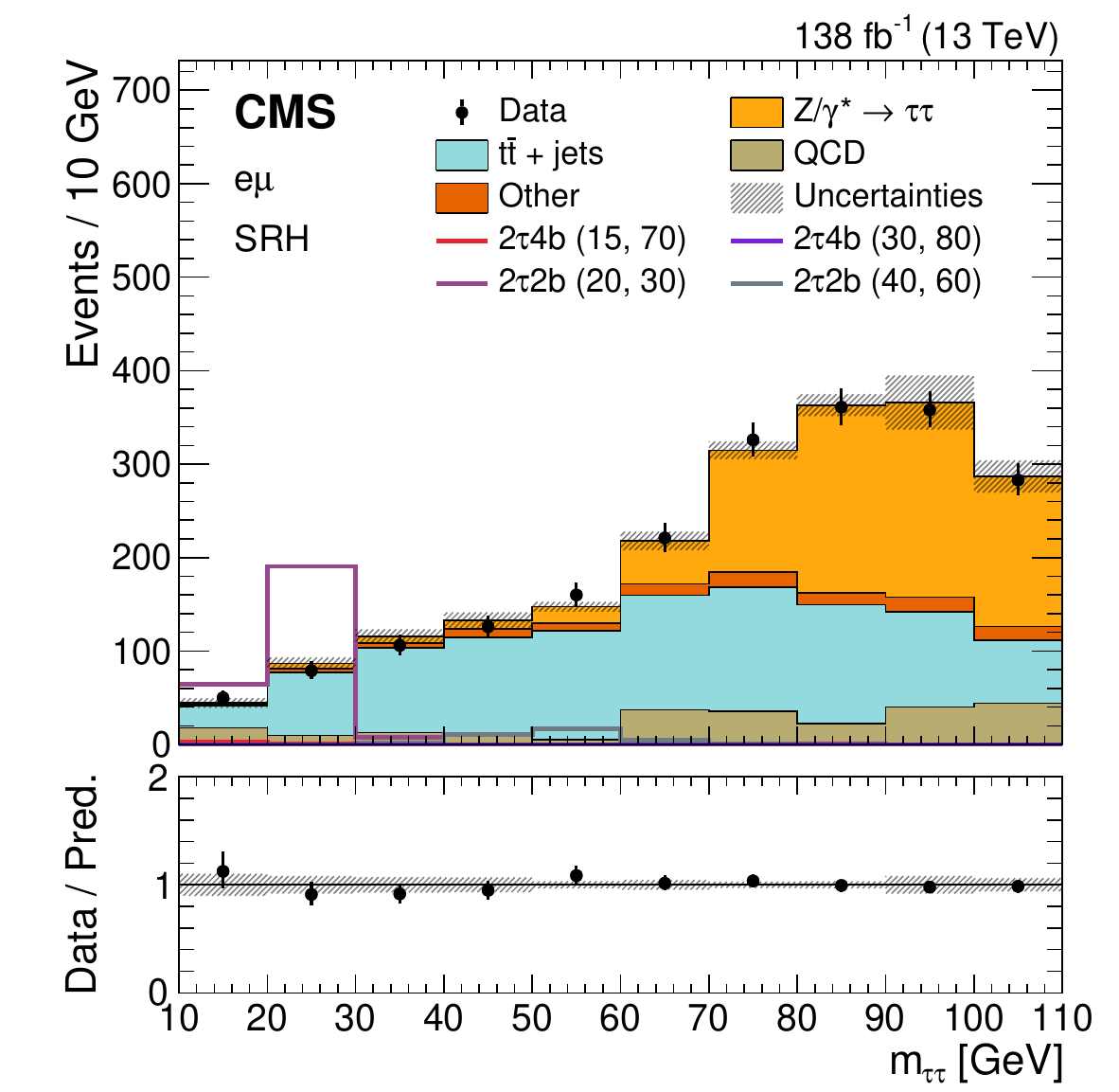}
    
    \caption{Background only, post-fit \mtt distributions for the \emu channel, in events with at least one \PQb-tagged jet: SRL (upper left), SRM (upper right), and SRH (lower). The data are shown by the markers with vertical bars and various backgrounds by the colored histograms. The total systematic uncertainty is shown by the hatched area. The colored open histograms display the predicted signal distribution for two cascade decays and two non-cascade decays, with four different values of \Paa and \Pab masses, for an assumed branching fraction of 100\%. The lower plot of each panel gives the ratio of the data to the sum of the predicted number of background events. The vertical bars display the statistical uncertainty in the ratio.}
    \label{fig:results_postfit_emu}
\end{figure}

Figures~\ref{fig:results_postfit_mutau}--\ref{fig:results_postfit_emu} show the background-only post-fit distributions of the \mtt variable in the cut-based SR categories.
Representative signal distributions are overlaid assuming a 100\% branching fraction for the Higgs boson decay into $2\PGt4\PQb$ for cascade mass hypotheses (15, 70) and (30, 80)\GeV and into $2\PGt2\PQb$ for non-cascade mass hypotheses (20, 30) and (40, 60)\GeV.

\begin{figure}[ht!]
    \centering
        \includegraphics[width=\textwidth]{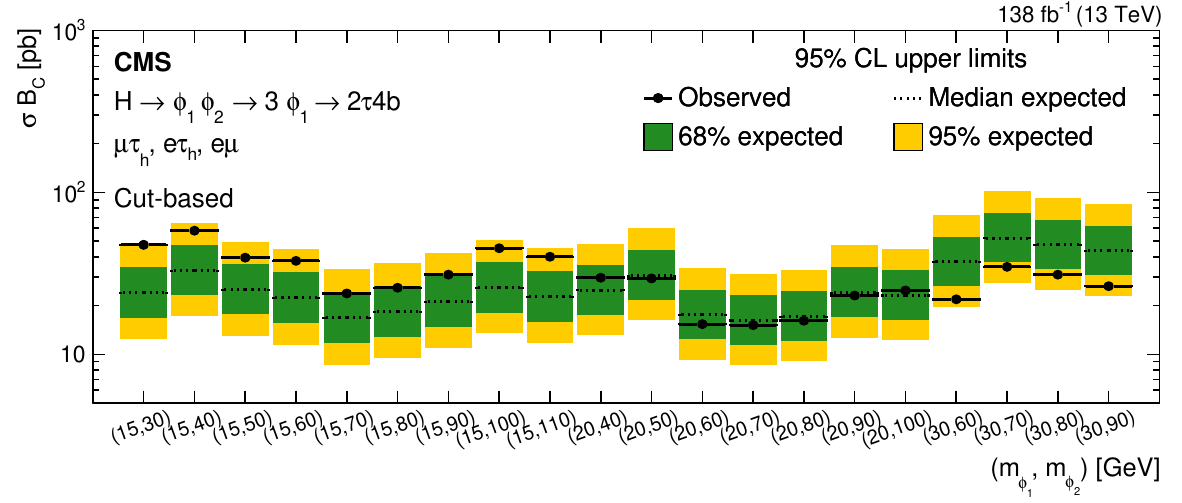}
    
    \caption{The observed (points) and median expected (dotted line) 95\% \CL upper limits on $\sigma \mathrm{B_{C}}$ for the cascade scenario using the cut-based event categorization and the fit to the \mtt distribution, for different mass hypotheses (\maa, \mab). The horizontal bars on the points are for better legibility only. The green and yellow regions show the 68 and 95\% expected range for the median value, respectively.}
    \label{fig:limits_cascade_run2}
\end{figure}

\begin{figure}[ht!]
    \centering
        \includegraphics[width=\textwidth]{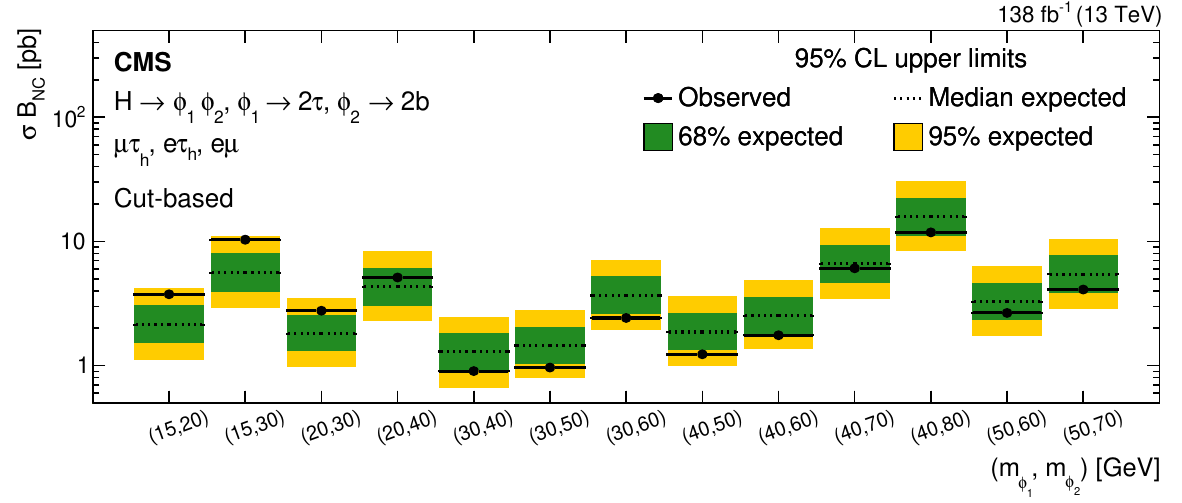}
    
    \caption{The observed (points) and median expected (dotted line) 95\% \CL upper limits on $\sigma \mathrm{B_{NC}}$ for the non-cascade scenario using the cut-based event categorization and the fit to the \mtt distribution, for different mass hypotheses (\maa, \mab). The horizontal bars on the points are for better legibility only. The green and yellow regions show the 68 and 95\% expected range for the median value, respectively.}
    \label{fig:limits_noncascade_run2}
\end{figure}

For the cut-based SRs, the 95\% \CL upper limits on the product $\sigma \mathrm{B_{C}}$ for all considered mass hypotheses in the cascade scenario are plotted in Fig.~\ref{fig:limits_cascade_run2}.
The 95\% \CL upper limits on the product $\sigma \mathrm{B_{NC}}$ for the non-cascade scenario are displayed in Fig.~\ref{fig:limits_noncascade_run2}.

\begin{figure}[ht!]
    \centering
        \includegraphics[width=0.49\textwidth]{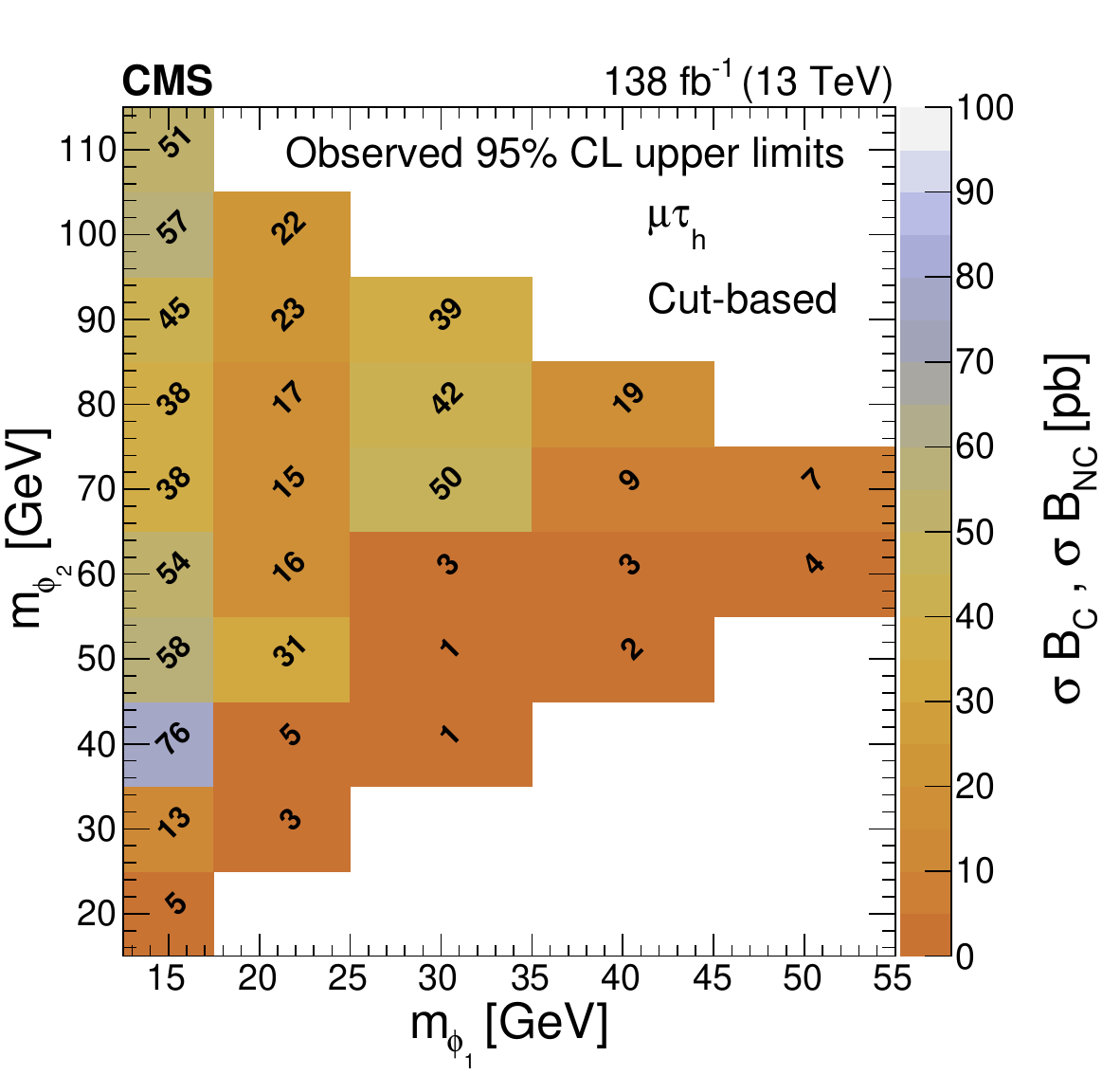}
        \includegraphics[width=0.49\textwidth]{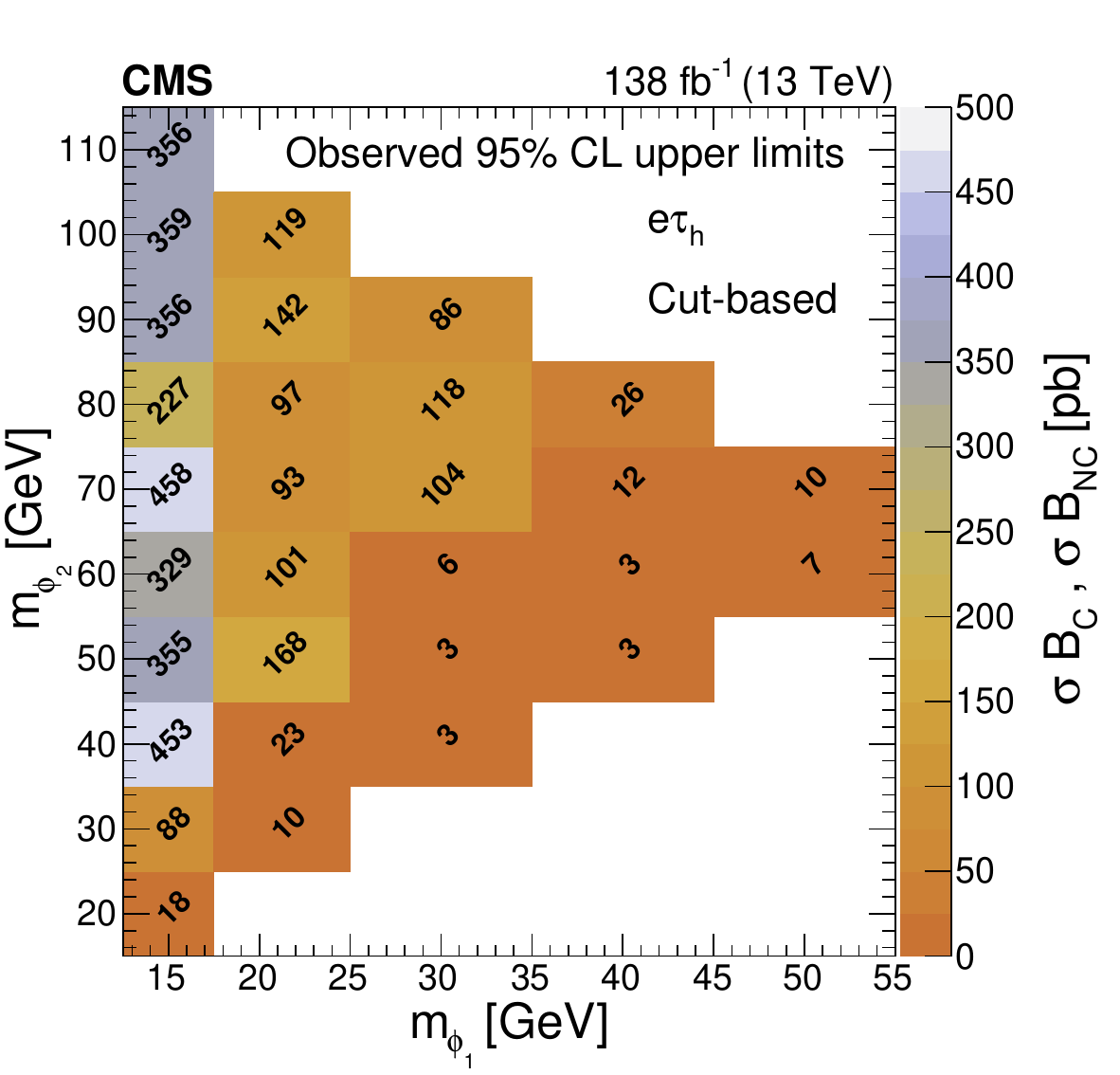} \\
        \includegraphics[width=0.49\textwidth]{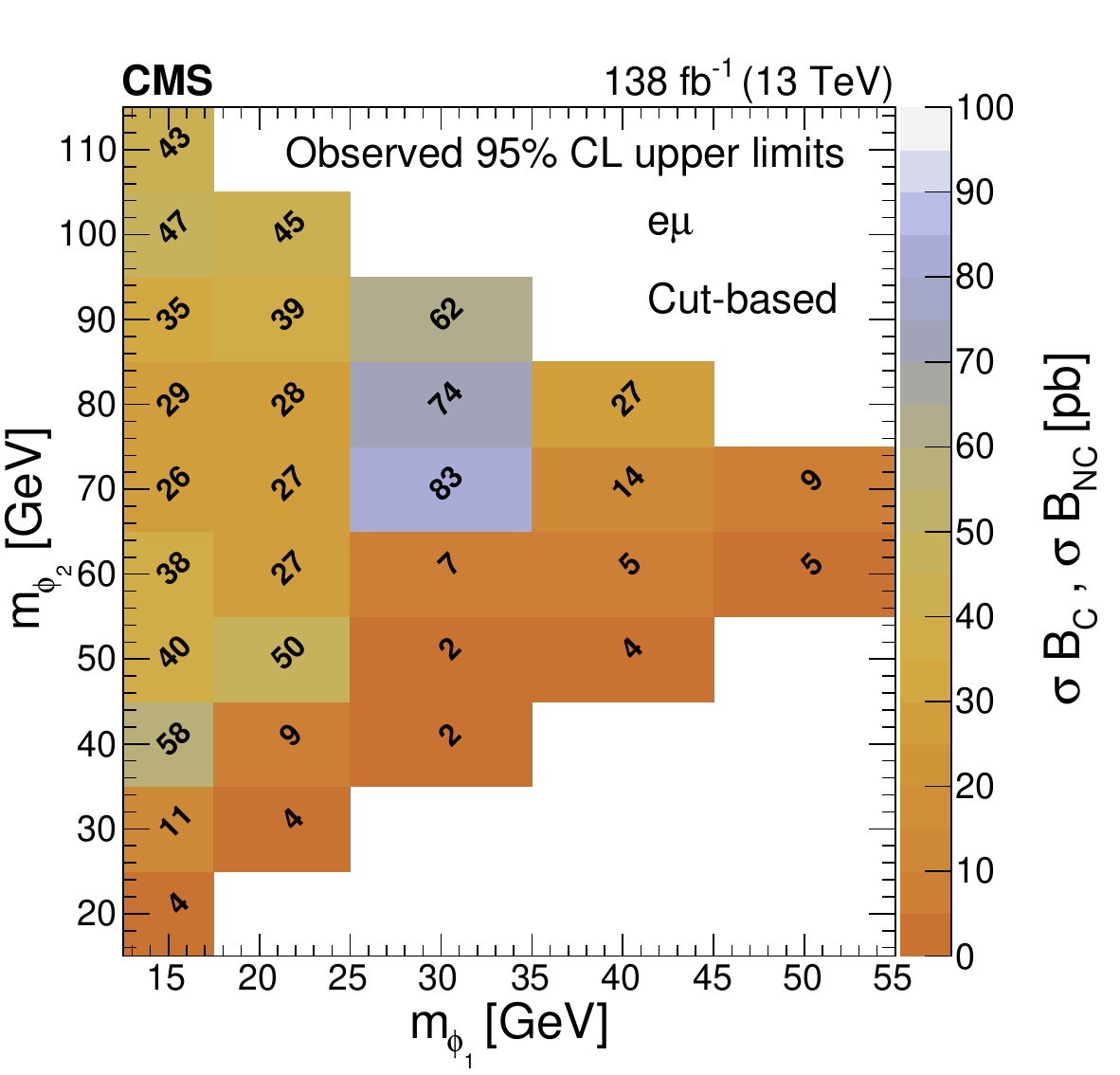}
        \includegraphics[width=0.49\textwidth]{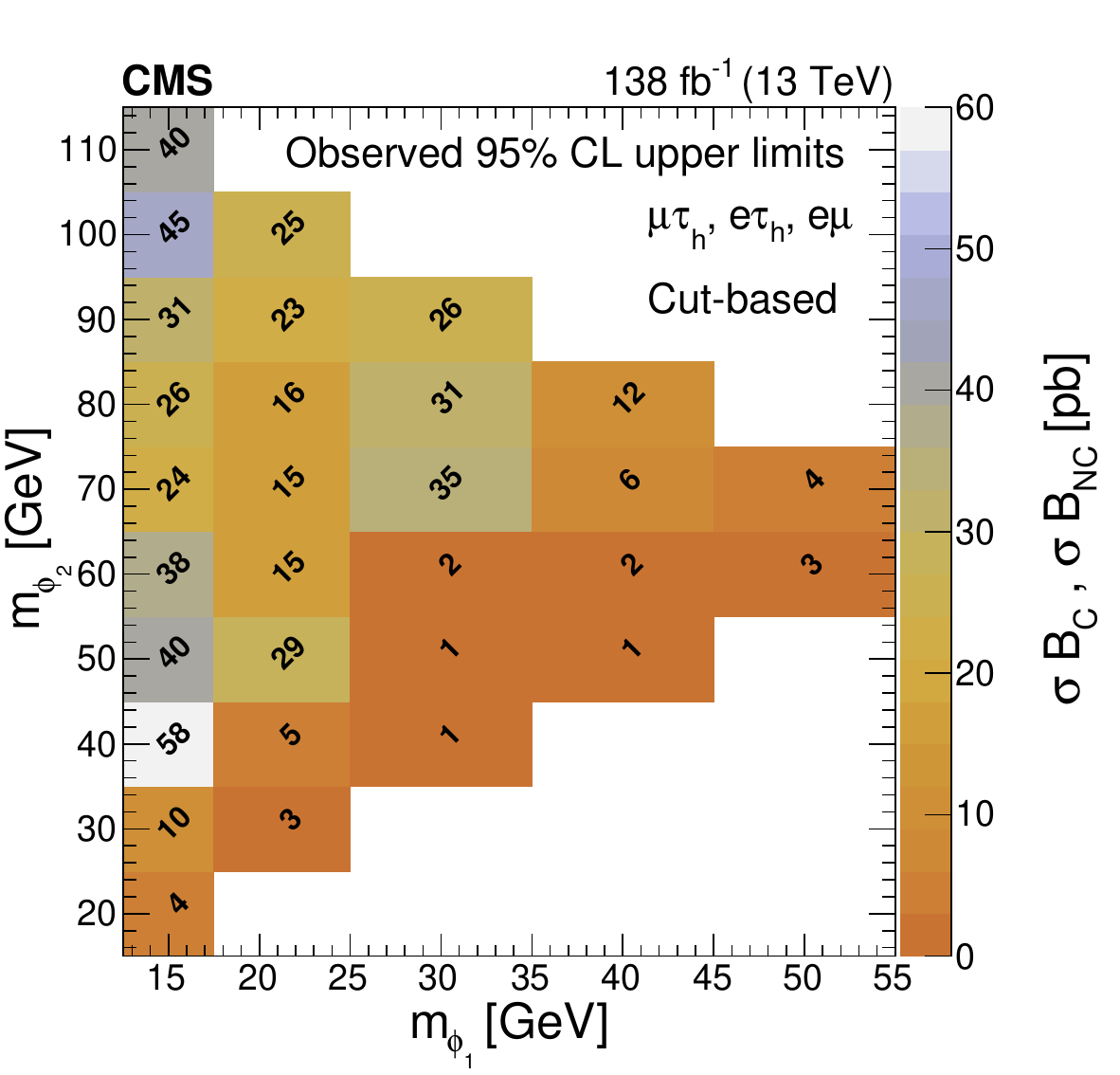}
    
    \caption{The 95\% \CL upper limits on the products $\sigma \mathrm{B_{C}}$ and $\sigma \mathrm{B_{NC}}$ obtained using the cut-based event categorization, as a function of the scalar masses $\maa$ and $\mab$. For the (\Paa, \Pab) mass hypotheses (15, 30), (20, 40), and (30, 60)\GeV, only the non-cascade limits are shown. For all other mass hypotheses, either cascade or non-cascade limits are presented, depending on whether the cascade decay is kinematically allowed ($m_{\Pab} \geq 2 m_{\Paa}$). The numbers displayed in the plot are in \unit{pb}.}
    \label{fig:limits_cutbased}
\end{figure}

Observed 95\% \CL upper limits on the products $\sigma \mathrm{B_{C}}$ and $\sigma \mathrm{B_{NC}}$ obtained from cut-based SRs are shown in Fig.~\ref{fig:limits_cutbased}, found from the combination of the three individual channels.

\cleardoublepage \section{The CMS Collaboration \label{app:collab}}\begin{sloppypar}\hyphenpenalty=5000\widowpenalty=500\clubpenalty=5000\cmsinstitute{Yerevan Physics Institute, Yerevan, Armenia}
{\tolerance=6000
A.~Gevorgyan\cmsorcid{0000-0003-2751-9489}, A.~Hayrapetyan, V.~Makarenko\cmsorcid{0000-0002-8406-8605}, A.~Tumasyan\cmsAuthorMark{1}\cmsorcid{0009-0000-0684-6742}
\par}
\cmsinstitute{Marietta Blau Institute for Particle Physics, Vienna, Austria}
{\tolerance=6000
W.~Adam\cmsorcid{0000-0001-9099-4341}, L.~Benato\cmsorcid{0000-0001-5135-7489}, T.~Bergauer\cmsorcid{0000-0002-5786-0293}, M.~Dragicevic\cmsorcid{0000-0003-1967-6783}, S.~Gundacker\cmsorcid{0000-0003-2087-3266}, A.K.~Guven\cmsorcid{0009-0004-5670-5138}, P.S.~Hussain\cmsorcid{0000-0002-4825-5278}, M.~Jeitler\cmsAuthorMark{2}\cmsorcid{0000-0002-5141-9560}, N.~Krammer\cmsorcid{0000-0002-0548-0985}, A.~Li\cmsorcid{0000-0002-4547-116X}, D.~Liko\cmsorcid{0000-0002-3380-473X}, M.~Matthewman, J.~Schieck\cmsAuthorMark{2}\cmsorcid{0000-0002-1058-8093}, R.~Sch\"{o}fbeck\cmsAuthorMark{2}\cmsorcid{0000-0002-2332-8784}, M.~Shooshtari\cmsorcid{0009-0004-8882-4887}, M.~Sonawane\cmsorcid{0000-0003-0510-7010}, N.~Van~Den~Bossche\cmsorcid{0000-0003-2973-4991}, W.~Waltenberger\cmsorcid{0000-0002-6215-7228}, C.-E.~Wulz\cmsAuthorMark{2}\cmsorcid{0000-0001-9226-5812}
% ...
\par}
\vskip\cmsinstskip
\dag:~Deceased\\
$^{1}$Also at Yerevan State University, Yerevan, Armenia\\
$^{2}$Also at TU Wien, Vienna, Austria\\
% ...
\end{sloppypar}
%%% END EDITABLE REGION %%%
% skeleton_end
\end{document}